\documentclass[11pt, reqno,preprint]{article}
\usepackage{jheppub}
\usepackage{epsfig}
\usepackage{amssymb}
\usepackage{amsmath}
\usepackage{mathrsfs}
\usepackage{hyperref}
\usepackage{multirow}
\usepackage[normalem]{ulem}
\usepackage{graphicx}
\usepackage{textcomp}
\usepackage{subcaption}

\usepackage{soul,xcolor}
\setstcolor{red} % Change font color of strikethrough

\def\be{\begin{equation}}
\def\ee{\end{equation}}
\def\ba{\begin{eqnarray}}
\def\ea{\end{eqnarray}}
\def\nl{\nonumber \\}
\def\a{\alpha}
\def\ab{\overline{\alpha}}
\def\zb{\overline{z}}
\def\j{J}

\def\Df{\Delta_\phi}
\def\OPE{\lambda^2}  %easy-to-change symbol for OPE coefficients
\def\Konishi{K}

\newif{\ifremarks}
\remarkstrue
\ifremarks
\newcommand{\remarkfc}[1]{{\renewcommand{\bfdefault}{b}\color[RGB]{0,0,150}{\textbf{#1}}}}
\fi
\providecommand{\remarkfc}[1]{\ignorespaces}

\newcommand{\beq}{\begin{equation}}
\newcommand{\eeq}{\end{equation}}
\newcommand{\bba}{\begin{align}}
\newcommand{\eea}{\end{align}}

\newcommand{\mS}{\mathrm{s}}
\newcommand{\mT}{\mathrm{t}}
\newcommand{\mU}{\mathrm{u}}

\def\OO{\mathcal{O}}
\def\cH{\mathcal{H}}
\def\cO{\mathcal{O}}
\def\cP{\mathcal{P}}
\def\dDisc{{\rm dDisc}}
\def\D{\Delta}

\title{Bootstrapping $\mathcal{N}=4$ sYM correlators using integrability}
\author{Simon Caron-Huot$^1$, Frank Coronado$^1$, Anh-Khoi Trinh$^1$, Zahra Zahraee$^1$}
\affiliation{
${}^1$Department of Physics, McGill University, 3600 Rue University, Montr\'eal, H3A 2T8, QC Canada
}
\emailAdd{schuot@physics.mcgill.ca}
\emailAdd{fcidrogo@gmail.com}
\emailAdd{anh-khoi.trinh@mail.mcgill.ca}
\emailAdd{zr.zahraee@physics.mcgill.ca}

\abstract{How much spectral information is needed to determine the correlation functions of a conformal theory?
We study this question in the context of planar supersymmetric Yang-Mills theory,
where integrability techniques accurately determine the single-trace spectrum at finite 't Hooft coupling. 
Corresponding OPE coefficients are constrained by dispersive sum rules, which implement crossing symmetry.
Focusing on correlators of four stress-tensor multiplets, we construct combinations of sum rules
which determine one-loop correlators, and we study a numerical bootstrap problem that
nonperturbatively bounds planar OPE coefficients. 
We observe interesting cusps at the location of physical operators,
and we obtain a nontrivial upper bound on the OPE coefficient of the Konishi operator outside the perturbative regime.
}

\begin{document}

\maketitle

\section{Introduction}

The study of $\mathcal{N}=4$ supersymmetric Yang-Mills theory (sYM) has driven advancements in key areas of theoretical physics, 
including insights into the AdS/CFT correspondence through
the development of scattering amplitudes, correlation functions, supersymmetric localization, %positive geometry/amplituhedron, 
integrability, and the conformal bootstrap.
Some of these techniques offer exact results in special subsectors,
while others are perturbative approximations which often exploit exact results as boundary conditions.
It is interesting to ask to what extent the full theory is nonperturbatively determined by exact subsectors.

%%The closely related integrand for correlation function exhibits a rich structure revealed to 10 loops in \cite{Bourjaily:2016evz},
%%and direct integration has provided conformal data up to 5 loops \cite{goncalves}. %, providing key comparisons with integrability \cite{}.
In the planar 't Hooft limit $\mathcal{N}=4$ sYM, integrability has led to impressive advancements for the computation of the spectrum of local operators, some correlations functions and scattering amplitudes at finite %'t Hooft 
coupling in various kinematical limits \cite{Beisert:2010jr,Basso:2013vsa,Sever:2020jjx}. This boundary data was instrumental
in recent perturbative scattering amplitudes and form factors %in $\mathcal{N}=4$ sYM
%have been analytically computed
which reached a record high loop order (8 loops) \cite{Dixon:2022rse} and high multiplicity (up to seven-points) \cite{Caron-Huot:2019vjl,Dixon:2020cnr}.
In the strong coupling limit, most efforts have focused on correlators of $\tfrac{1}{2}$-BPS operators,
which is primarily driven by string theory, supergravity, conformal bootstrap, and supersymmetric localization \cite{Binder:2019jwn,Chester:2021aun} techniques.
Stringy corrections \cite{Costa:2012cb,Goncalves:2014ffa,Alday:2018pdi} and non-planar corrections \cite{Huang:2021xws,Drummond:2022dxw} have been computed for four points,
while the highest multiplicity correlator is the tree-level five-point function \cite{Goncalves:2019znr},
obtained by exploiting all existing constraints.
%Further simplifications occur in the planar limit, where the theory exhibits a still mysterious 10-dimensional conformal symmetry \cite{Caron-Huot:2018kta,Caron-Huot:2021usw}.
Although still out of reach for most observables,
these developments suggest that more quantities will eventually be computed at finite coupling in this theory.

%Despite this wealth of options, the theory remains unresolved nonperturbatively even in the planar limit, and nonpertubative techniques remain scarce.

A promising nonperturbative avenue originates from the integrability literature:
in the planar limit, the spectral problem is completely solved as a result of the Quantum Spectral Curve (QSC) \cite{Gromov:2009tv,Gromov:2013pga,Gromov:2014caa}, which governs the spectrum of
single-trace operators at arbitrary values of the coupling; see ref.~\cite{Gromov:2017blm} for a recent pedagogical review.
At the moment the QSC only provides spectrum of single-trace operators, and therefore dynamics,
encoded through OPE coefficients for example, remain elusive at finite coupling.
Nonetheless, there are encouraging developments with respect to the hexagon \cite{Basso:2015zoa,Fleury:2016ykk,Coronado:2018ypq,Coronado:2018cxj} and octagon \cite{Bargheer:2019kxb,Bargheer:2019exp} formalisms, which may overcome this challenge;
the hexagon and octagon formalisms reinterpret the Feynman expansion as the scattering of magnons and they are readily computable in the large R-charge limit.
Unfortunately, away from this limit, the complexity of such computations grow exponentially.

An independent nonperturbative framework is the numerical conformal bootstrap \cite{Rattazzi:2008pe,El-Showk:2012cjh,Poland:2018epd}.
The latter combines unitarity, crossing symmetry, and other nonperturbative constraints to resolve the spectrum of a generic conformal field theory and to bound its OPE coefficients; see \cite{Poland:2022qrs} for a recent review of numerical bootstrap results. In the context of $\mathcal{N}=4$ sYM, this method provides a handle into energy-momentum correlation functions
for finite gauge groups SU($N_c$)  \cite{Beem:2016wfs}, and,
when combined with exact results from localization, arbitrary 't Hooft coupling \cite{Chester:2021aun}.

%This is achieved by constructing extremal linear functionals \cite{El-Showk:2012vjm}, which are positive above a twist gap.
%, and exhibits double-zeros on exchanged operators in the spectrum.
%The construction of such functionals has driven numerous advancements in the study CFTs nonperturbatively; see \cite{Poland:2022qrs} for a recent review of numerical bootstrap results.
%In parallel, the analytic bootstrap 
%while analytic bootstrap tools such as the Lorentzian inversion formula \cite{Caron-Huot:2017vep} have enabled users to obtain closed-form expressions for the OPE coefficients and anomalous dimensions in CFTs.

%Given these complementary methods, it behooves us to combine them to solve $\mathcal{N}=4$ sYM in the planar limit.
It was recently proposed that integrability and bootstrap techniques could be combined to solve $\mathcal{N}=4$ sYM in the planar limit \cite{Cavaglia:2021bnz,Cavaglia:2022qpg}. 
The authors introduced an approach (``Bootstrability") to study the 1D defect CFT defined by inserting local operators
along a $\tfrac{1}{2}$-BPS Wilson line in $\mathcal{N}=4$ sYM.
Taking exact spectral data as input from integrability, they used bootstrap techniques to derive
tight bounds on the OPE coefficients of the first few lightest operator in the spectrum.

\begin{figure}[t]
\centering
\includegraphics[width=0.95\textwidth]{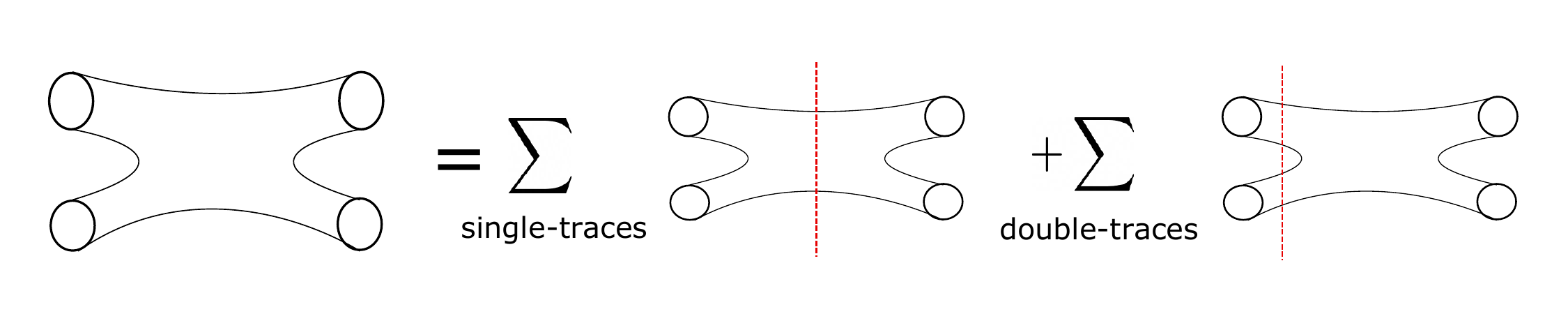}
\caption{A connected correlator in the planar limit: its operator product expansion receives contributions from
single and double trace operators. The latter's OPE data scale at large-$N_c$ like $\lambda_{\OO \OO [\OO\OO]} \sim 1 + O(1/N_c^2)$, where the ``1'' yields the disconnected correlator and the $O(1/N_c^2)$ corrections enter the figure.}
\label{fig:four-point cut}
\end{figure}

In this paper, we tackle a similar problem for a fully four-dimensional correlator involving four stress tensors
in the planar limit of $\mathcal{N}=4$ sYM theory, 
meaning the $N_c\to\infty$ of a SU($N_c$) gauge theory with fixed 't Hooft coupling $\lambda=g_{\rm YM}^2N_c$.
We will use spectral information from integrability, together with suitable nonperturbative sum rules on the spectrum,
to constrain OPE coefficients.
%the stress tensor multiplet is the lightest $\tfrac{1}{2}$-BPS operator in the spectrum.
In the planar limit, both single-trace and double-trace operators are exchanged as illustrated in fig.~\ref{fig:four-point cut}.
The spectrum of single-trace operators can be calculated precisely owing to the QSC.
Our goal will be to use the numerical bootstrap to
bound the OPE coefficient of the lightest unprotected single-trace operators in the spectrum, the so-called Konishi operator.

At strong coupling, the Konishi operator is dual to a genuine massive string mode.
Its properties have been studied extensively over the years.
Its scaling dimension has been computed perturbatively in both weak and strong 't Hooft coupling
limits \cite{Eden:2004ua,Vallilo:2011fj}, results which are now exactly connected at intermediate coupling
by the QSC \cite{Gromov:2009tv,Gromov:2015wca,Hegedus:2016eop}.
However, less is known about the Konishi operator's OPE coefficient.
It has been computed up to five loops in the weak coupling limit \cite{Georgoudis:2017meq},
and is known to the leading order at strong coupling through its connection with the flat-space
Veneziano-Shapiro amplitude \cite{Costa:2012cb,Minahan:2014usa,Goncalves:2014ffa}.
Can it be bootstrapped at finite coupling, given spectral information from the QSC?
%This paper aims to bridge the gap between perturbative results at weak and strong couplings for the Konishi operator's OPE coefficient.
%String theory 3-point vertex references: \cite{Bargheer:2013faa}

While our work is similar in spirit to refs.~\cite{Cavaglia:2021bnz,Cavaglia:2022qpg},  
the passage from $D=1$ defects to $D=4$ correlators presents significant challenges.
A main one is that an infinite number of double-trace operators enter the OPE, polluting it with
undesirable operators about which we have no spectral information.
Worse, the numerical bootstrap leverages the positivity of OPE coefficients,
but $O(1/N_c^2)$ corrections to OPE coefficients control planar correlators (see fig.~\ref{fig:four-point cut}) and
do not have definite signs.
%Constructing extremal functionals satisfying such positivity conditions can be expensive computationally.
These hurdles are overcome by the recently introduced dispersive CFT functionals
\cite{Penedones:2019tng,Caron-Huot:2020adz,Caron-Huot:2021enk,Trinh:2021mll},
which decouple the double-traces; planar correlators are reconstructed from single-trace data only \cite{Carmi:2019cub}.
This work constitutes the first systematic application of dispersive functionals to a numerical bootstrap problem.
A second challenge is that the single-trace operators entering the OPE are more numerous than in 1D,
being now labelled by dimension and spin,
but in practice only a finite number of dimensions can be computed from the QSC.

%These relations have led to the construction of 
%dispersive CFT numerical functionals, which are key to obtaining nonperturbative sum rules at finite coupling.
%Not only do they kill the contribution of double-trace operators, they further possess tunable positivity properties that improves the convergence of the numerical bootstrap.

This paper is organized as follows.
In section~\ref{sec:setup}, we describe our setup by detailing properties of the stress tensor multiplet correlator, the decoupling of the double-twist sector, and aspects of integrability relevant for our bootstrap algorithm.
In section~\ref{sec:functionals}, we discuss the dispersive functionals used in this paper; 
details of their construction and numerical evaluation are described in the appendix.
Section~\ref{sec:bootstrap} contains our primary results -- bounds on the OPE coefficient of the Konishi operator -- obtained from the numerical bootstrap. Finally, we summarize our findings and discuss future work in section~\ref{sec:discussion}.
Three appendices review integrability formulas; detail our construction of functionals which solve the 1-loop problem;
and detail efficient formulas for numerically evaluating of dispersive functional.

\section{Setup and methods} \label{sec:setup}

\subsection{Stress tensor multiplet correlators} \label{sec:Stress tensor}

We consider the simplest half-BPS operator in the sYM theory, which transforms
as a symmetric traceless tensor so(6)$_R$. Its supermultiplet notably contains the stress tensor.
Working in index-free notation, this operator can be viewed as function of spacetime coordinates $x$ and
a null 6-vector $y$:
\be
 \OO(x,y) \propto {\rm Tr}\left[ (y{\cdot}\phi(x))^2\right]\,. \label{def O}
\ee
We use the canonical normalization $\langle \OO(x_1,y_1) \OO(x_2,y_2)\rangle = (y_{12}^2/x_{12}^2)^2$.
Due to conformal symmetry, the four-point correlator factors through spacetime and R-charge cross-ratios,
\begin{align}
u= \frac{x_{12}^{2}x_{34}^2}{x_{13}^2x_{24}^2} = z \bar{z}\,, \qquad &v = \frac{x_{23}^2x_{14}^2}{x_{13}^2x_{24}^2} = (1-z)(1-\bar{z})\,, \label{cross-ratios u,v} \\
\sigma= \frac{y_{12}^2 y_{34}^2}{y_{13}^2 y_{24}^2} = \alpha \bar{\alpha}\,, \qquad& \tau = \frac{y_{23}^2 y_{14}^2}{y_{13}^2 y_{24}^2} = (1-\alpha)(1-\bar{\alpha}) \label{cross-ratios sigma tau}\,.
\end{align}
Furthermore, superconformal Ward identities constrain the dependence on R-charge vectors. Namely, they imply that the correlator with $z=\alpha$ is protected and does not depend on the coupling.
This allows to separate the free theory limit ($g^2\to 0$) from the dynamical part $\cH$ of the correlator \cite{Dolan:2004mu,Nirschl:2004pa}:
\begin{align}
\frac{x_{13}^4x_{24}^4}{y_{13}^4y_{24}^4} \langle \OO(x_1,y_1)\cdots \OO(x_4,y_4)\rangle
&=1+\frac{\sigma^2}{u^2}+\frac{\tau^2}{v^2} +\frac{1}{c}\left(\frac{\sigma}{u}+\frac{\tau}{v}+\frac{\sigma\tau}{u v}\right)
\nl &\quad + \frac{1}{c}(z-\a) (z - \ab) (\zb- \a) (\zb - \ab)\cH(z,\zb)\,, \label{G ansatz}
\end{align}
where $c\equiv \frac{N_c^2-1}{4}$, and $\cH$ is independent of R-symmetry cross-ratios $\a,\ab$, which only appear in its prefactor.
We are interested in the planar limit, where $\cH$ is $N_c$-independent.

The function ${\cal H}$ effectively behaves like a correlator of four scalar primaries with $\Delta=4$.
It enjoys the following properties:
\begin{enumerate}
\item Crossing:
\be\cH(u,v) = \cH(v,u) = u^{-4}\cH(\tfrac{1}{u},\tfrac{v}{u}). \label{crossing Eucl}
\ee
\item Operator Product Expansion: it can be expressed as a sum of short (protected) and long (unprotected) multiplets,
the latter being labelled by their dimension and spin $(\Delta,J)$:
\be
 \cH(u,v) = %\frac{1}{u^2} +
 %\frac{u^2-1}{u^2v^2}+\frac{1}{c}\frac{u-1}{u^2v}+
% \cH_{\rm short}(u,v)+
\cH^{\rm protected}(u,v) + \sum_{(\Delta,\j)\ \rm long} \OPE_{\Delta,J} G^{N=4}_{\Delta,J}(u,v), \qquad
 G^{N=4}_{\Delta,\j}\equiv u^{-4}G_{\Delta+4,\j}\,.
\label{OPE}
\ee
The protected part, equal to minus the $g\to 0$ limit of the sum, will be discussed below.
\item Regge limit: as $z,\zb\to\infty$ with fixed ratio $z/\zb$, $\cH\sim z^{\j_*-4}$ where $z$ is the Regge intercept.
Nonperturbatively in $N_c$ we have $\j_*\leq 1$, ie. the quantity $uv\cH$ is bounded.
In the planar limit this bound becomes trivial because of the overall $1/c$, but
the bound of chaos still ensures $\j_*\leq 2$, which is saturated at infinite 't Hooft coupling.
We will assume that at finite coupling, $\j_*<2$ strictly.
\item % I used that R goes to (z-a)()()()/(uv), up to powers of y_{13}^2 and x_{13}^2.
Weak and Strong limits: in our conventions,
at extreme values of the 't Hooft coupling ($g^2\equiv\frac{g^2_{\rm YM}N_c}{16\pi^2}$) \cite{Dolan:2004iy},
\be\begin{aligned} \label{known H}
\lim_{g^2\to 0} \cH(u,v) &=-2g^2 \frac{F_1(u,v)}{uv}+O(g^4)\,,
\\
\lim_{g^2\to \infty} \cH(u,v) &= \cH^{\rm strong} (u,v) +O(1/g) \quad\mbox{with}\quad \cH^{\rm strong}= -\bar{D}_{2,4,2,2}\,.
\end{aligned}
\ee
\end{enumerate}
In the above, we use the standard conformal block
\be
 G_{\Delta,\j}(u,v)=\frac{z\zb}{\zb-z}\left[
 k_{\frac{\Delta-\j-2}{2}}(z)k_{\frac{\Delta+\j}{2}}(\zb)-
 k_{\frac{\Delta+\j}{2}}(z)k_{\frac{\Delta-\j-2}{2}}(\zb)\right]
\ee
with $k_h(z)=z^h{}_2F_1(h,h,2h,z)$.  Furthermore, $F_1$ is the box integral and $\bar{D}_{2,4,2,2}$ is a derivative of it (see \cite{Arutyunov:2002fh}):
\begin{align}
 F_1(u,v)&\equiv \frac{2{\rm Li}_2(z)-2{\rm Li}_2(\zb)+\log(z\zb)(\log(1-z)-\log(1-\zb))}{z-\zb}\,,
\\
\bar{D}_{2,4,2,2}  &= \partial_u\partial_v(1+u\partial_u+v\partial_v)F_1(u,v)\,.
\end{align}
Note that we have factored the large-$N_c$ scaling of the single-trace OPE coefficients into $\OPE$
so that $\lambda$ does not depend on $N_c$:
\be
 \OPE_{\Delta,\j}\Big|_{\rm canonical} = \frac{1}{c}\OPE_{\Delta,\j}.  \label{OPE scaling}
\ee
As an abuse of notation we will still refer to $\lambda$ as OPE coefficients.

In the planar limit $c\to\infty$, the OPE \eqref{OPE} receives contributions only from single- and double-trace operators.
Our main goal will be to constrain the single-trace coefficients $\OPE_{\Delta,\j}$
given input about the single-trace spectrum from integrability.

\subsection{Decoupling double-traces}

For our purposes, the double-trace contribution to the OPE is a nuisance.
Since double-traces enter already in the disconnected correlator ($\sim c^0$ terms in eq.~\eqref{G ansatz}),
their contributions to $\cH$ represent $1/c$ corrections to coefficients and scaling dimensions
that do not have definite signs.
To formulate a nonperturbative bootstrap in the planar limit, it is crucial to project out all double-traces.

This is naturally achieved by taking a double-discontinuity of the correlator. For $z,\zb<0$, let:
\be
 \dDisc_s \cH(z,\zb) \equiv \cH(z_\curvearrowleft,\zb_{\scalebox{1}[-1]{${}_\curvearrowleft$}})
 -\tfrac12\cH(z_\curvearrowleft,\zb_\curvearrowleft) - \tfrac12\cH(z_{\scalebox{1}[-1]{${}_\curvearrowleft$}},\zb_{\scalebox{1}[-1]{${}_\curvearrowleft$}})\,, 
\ee
where the arrows denote analytic continuation paths starting from the Euclidean region with $0<z,\zb<1$.
The first term is simply the Euclidean correlator (the path maintains $\zb=z^*$), which enjoys the usual OPE,
while for the other terms the analytic continuation simply adds phases, so the OPE \eqref{OPE} yields
\be
 \dDisc_s \cH(z,\zb) = \dDisc_s \cH^{\rm protected}(z,\zb) +  \sum_{(\Delta,\j)\ \rm long} 2\sin^2\left(\pi \tfrac{\Delta-J}{2}\right)\OPE_{\Delta,J} G^{N=4}_{\Delta,J}(z,\zb)\,.
\label{dDiscOPE}
\ee
The crucial point is the trigonometric factor, which has double-zeroes at the position of double trace operators,
$\Delta{-}J=4{+}2m+O(1/c)$ with $m \in \mathbb{N}$.
Thus, in the planar limit, the above sum is saturated by single-trace operators.

The protected double-discontinuity is simple to describe: only operators of twist exactly two, from the stress-tensor multiplet, contribute. Taking the singular terms in eq.~(2.31) of \cite{Beem:2016wfs}, we find
\be
 \dDisc_s \cH^{\rm protected}(u,v)  = f^{(0)}(v)\dDisc_s \frac{1}{u}\quad\mbox{with}\quad f^{(0)}(v)\equiv \frac{v^2-1-2v\log v}{v(1-v)^3}\,.
\label{f0}
\ee
Note that the double-discontinuity of $1/u$ is a nonvanishing singular distribution near $u=0$ \cite{Caron-Huot:2017vep}.
A simple check is that this is precisely the double-discontinuity of the strong coupling result \eqref{known H}:
\be
 \lim_{u\to 0} \cH^{\rm strong}(u,v) = \frac{f^{(0)}(v)}{u}  + \mbox{(terms with vanishing dDisc${}_s$)}.
\ee
This happens because in the supergravity limit
all non-protected single-traces become heavy and decouple from \eqref{dDiscOPE}.

The double-discontinuity kills double-traces but is not crossing symmetric since it picks a specific channel (above, the $s$-channel).
How do we get crossing equations? 
The nontrivial fact is that conformal correlators are uniquely determined by their double-discontinuity
and Regge limits.
Concretely, they can be reconstructed through dispersive integrals \cite{Carmi:2019cub}:
%This will allow us to formulate crossing equation as a constraint on $\dDisc\cH$.
%This will give rise to integral constraints on $\dDisc\cH$. %, and thus to directly constrain single-trace data.
%This reconstruction is performed by a dispersive integral:
\be
\cH(u,v) = \int_{s} du'dv' K(u,v; u',v')\ \dDisc[\cH(u',v')]\,. \label{disp}
\ee
The kernel $K$ is recorded in eq.~\eqref{K} but won't be important for the present discussion. The integration region lies inside $s$-channel kinematics $z',\zb'\leq 0$,
where the OPE \eqref{dDiscOPE} converges. (The integration region is further restricted by step-functions and delta-functions inside $K$.)
By defining ``Polyakov-Regge block'' as the dispersive transform of a single $s$-channel block,
\be \label{PR coords}
  \cP_{u,v}^{N=4}[\Delta,\j]\equiv \int_{s}du'dv' K(u,v;u',v')\ \dDisc_s[G^{N=4}_{\Delta,\j}(u',v')]\,,
\ee
the correlator can thus be expressed as
\be
 \cH(u,v) = \cH^{\rm strong}(u,v) + \sum_{(\Delta,\j)\ \rm long} \OPE_{\Delta,\j} \cP^{N=4}_{u,v}[\Delta,\j].  \label{PR OPE}
\ee
The crucial point is that only \emph{single-traces} enter this sum in the planar limit, since $\cP$ inherits the double zeroes from dDisc.
The protected contribution is simply $\cH^{\rm strong}(u,v)$ because of the decoupling just mentioned;
we verified this numerically from the formulas in appendix.

The above is valid for any Euclidean $u,v$, namely, $u$ and $v$ which come from complex-conjugate cross-ratios $\zb=z^*$. This condition can be stated as:
\be
\mbox{Euclidean region: } u,v>0 \mbox{ real} \quad\mbox{and}\quad 4uv\geq (1{-}u{-}v)^2.
\ee

The dispersive representation manifests $u\leftrightarrow v$ crossing symmetry, which correspond to the $s \leftrightarrow t$-channel crossing equation: $\cP^{N=4}_{\Delta,\j}(u,v)=\cP^{N=4}_{\Delta,\j}(v,u)$.
To get crossing relations, the idea is that the second relation in \eqref{crossing Eucl} is nontrivial, and amounts to an infinite number of constraints:
\be
 \boxed{0 = \sum_{(\Delta,\j)\ \rm long} \OPE_{\Delta,\j} X_{u,v}[\Delta,\j]
 \quad\mbox{with}\quad X_{u,v}\equiv \cP_{u,v} - u^{-4}\cP_{1/u,v/u}\,,
% \Big( u^2\cP^{N=4}_{\Delta,\j}(u,v)-u^{-2}\cP^{N=4}_{\Delta,\j}(\tfrac{1}{u},\tfrac{v}{u})\Big)
  \quad\mbox{for $(u,v)$ Euclidean}.}
 \label{Xuv}
\ee
This statement of crossing symmetry involves only single-trace data in the planar limit.

It is not the most general statement yet, because the (unsubtracted) dispersion relation \eqref{disp} only relied on
the Regge behavior $\j_*< 4$. (The threshold is shifted by four compared with the usual threshold
of an unsubtracted dispersion relation due to supersymmetry and the factors in \eqref{G ansatz}.)
But since we expect $\j_*<2$ at finite 't Hooft coupling, more is true: \emph{anti-subtracted} dispersion relations converge. 
As explained in \cite{Caron-Huot:2020adz} and reviewed in section \ref{sec:Mellin functionals},
the difference between different subtraction schemes are ``dispersive sum rules'' characterized by their patterns of zeroes on double-twist operators.
Here we are not allowed to cancel any double-trace zero, so there is only a one-parameter family of extra constraints.
We can take it to be the $B_{2,v}$ sum rule in eq.~(4.39) of \cite{Caron-Huot:2020adz} applied to $u'v'\cH$.
Dividing it by $v$, we will call it simply the $B_v$ functional: 
\be
 B_v[f(u',v')] = \int\limits_v^\infty dv'\!\! \int\limits_0^{(\sqrt{v'}-\sqrt{v})^2} du' \frac{v'-u'}{\pi^2v\sqrt{v^2-2(u'+v')v+(u'-v')^2}} \dDisc_s[f(u',v')]. \label{Bv}
\ee
It can be proved directly that $B_v[\cH]=0$, essentially by deforming the integration contour from the $s$-channel to the $t$-channel double-discontinuity,
and exploiting $u'{\leftrightarrow}v'$ antisymmetry of the integrand \cite{Caron-Huot:2020adz}. For a generic correlator
the contour deformation would pick a contribution from $u$-channel identity, but this is absent for $\cH$.
The integral against $(u')^\delta$ becomes singular for $u'\to 0$ if $\delta\leq -1$, but can be defined by analytic continuation in $\delta$.
One finds in this way that when acting on twist-two exchanges, $B_{v}$ simply returns the coefficient of $1/u$ \cite{Caron-Huot:2020adz},
so $B_v[\cH^{\rm protected}]$ gives
\be
 B_v^{\rm protected} = f^{(0)}(v)
\ee
with $f^{(0)}(v)$ in \eqref{f0}. Therefore, the $B_v$ sum rules take the form:
\be
 \boxed{ 0 = B_v^{\rm protected}+\sum_{(\Delta,\j)\ \rm long} \OPE_{\Delta,\j} B_v[\Delta,\j]\qquad (v>0\ {\rm real}).} \label{Bv sum}
\ee
Here and below we use the notation $B_v[\Delta,\j]\equiv B_v[G^{N=4}_{\Delta,\j}]$ for the action of a functional on a block.
The salient feature of these sum rules is the protected contribution, which will provide an absolute normalization
to OPE coefficients; it will play a similar role in our analysis as the identity operator in numerical bootstrap studies.
At strong coupling, it can be interpreted as a relation between protected graviton exchanges and heavy string modes.
It is crucial for its validity that the 't Hooft coupling is finite, so the Regge intercept is strictly less than 2.

The crossing relation \eqref{Xuv} and $B_v$ sum rule \eqref{Bv} will be our main tool: they
exhaust the constraints on single-trace data coming from crossing symmetry and good Regge behavior.
Formulas for their efficient numerical evaluation are detailed in appendix~\ref{app:efficient eval of funcs}.
Following the bootstrap method, the key idea will be to exploit positivity of the unknowns $\OPE_{\Delta,\j}$.

\subsection{Input from integrability: Spectrum from quantum spectral curve} \label{sec:integrability and QSC}

Operators in $\mathcal{N}=4$ SYM can be identified through their charges under the global symmetries, the conformal group $SO(4,2)$\,:\,$\{\Delta,J_{1},J_{2}\}$ and the R-symmetry group $SO(6)$:\,$\{r_{1},\,r_{2},\,r_{3}\}$.
However, the real ``fingerprint" of a (single-trace) operator is its set of charges under the infinite family of symmetries that make the theory integrable.  This fingerprint is encoded in a QSC \cite{Gromov:2013pga,Gromov:2015wca}. 
The latter is composed of a set of 8 functions: $P_{a}(u)$ and $Q_{j}(u)$ with indices $a,j\in\{1,2,3,4\}$, which depend on the spectral parameter $u$.\footnote{
 We recognize the overload of the letter ``u'', which represents in turn the cross-ratio $u$,
 the spectral parameter $u$, and below, the Mellin-Mandelstam variable $\mU$.
 We hope that no confusion will appear from the context.}
For each operator, there is a unique set $\{P_{a},Q_{j}\}$.
In particular, the global charges are recovered in their large-$u$ asymptotics:
\beq
P_{a}(u) \underset{u\to\infty}{\sim} u^{M_{a}}\qquad\text{and}\qquad Q_{j}(u) \underset{u\to\infty}{\sim} u^{\hat{M}_{j}}.
\label{PQ asympt}
\eeq
Since we look at a single correlation function, of the stress-tensor multiplet, all the operators we are interested in
have the same R-charges, and spacetime charges of the form $(\Delta,J_{1}=\j+2,\,J_{2}=0,\, r_i=0)$,
corresponding to the exponents:
\bba
M_{a}=\big\{-2,-1,0,1\big\},\qquad
\hat{M}_{j} =\big\{\tfrac{\Delta-\j}{2},\,\tfrac{\Delta+\j+2}{2},\,\tfrac{-\Delta-\j-4}{2},\,\tfrac{-\Delta + \j-2}{2}\big\}.
\end{align}

In this paper we use the quantum spectral curve to determine the scaling dimensions for a few values of the coupling $g$
for the first few primary operators in the leading and sub-leading Regge trajectories.
Specifically, we consider the operators with the following identification at weak coupling:
\bba\label{eq:trajectories}
\text{leading trajectory: }&\Delta = 2+J+O(g^2)\quad \text{for}\quad J=0,2,4,6,8,10,\nonumber\\
\text{subleading trajectory: }&\Delta = 4+ J+O(g^2)\quad \text{for}\quad J=0\,.
\end{align}
Additionally, we will use the Asymptotic Bethe Ansatz to study the leading and subleading trajectory at asymptotically large spins.

%To achieve this, we will find the QSC, i.e. the functions $\{P_a,Q_i\}$, associated with each operator in \eqref{eq:trajectories}.  In section \ref{sec:QSCequations}, we review the analytic properties and closed system of equations satisfied by $\{P_a, Q_i\}$.  In section \ref{sec:QSCsolutions}, we use the algorithm of \cite{Gromov:2015wca} to find numercal solutions of the QSC with high precision. We report on the results of \cite{Gromov:2015wca,Hegedus:2016eop} on the leading Regge trajectory and extend them by a few higher spins. We also present the solution for the first spinless operator in the subleading Regge trajectory.  Finally in section \ref{sec:largeSpin} we summarize previous results on large spin operators, the tail of the Regge trajectories, which we use with the low spin results to interpolate and find approximate scaling dimensions for operators with intermediate values of spin $J$.

\begin{figure}[t]
\centering
    \includegraphics[width=1\linewidth]{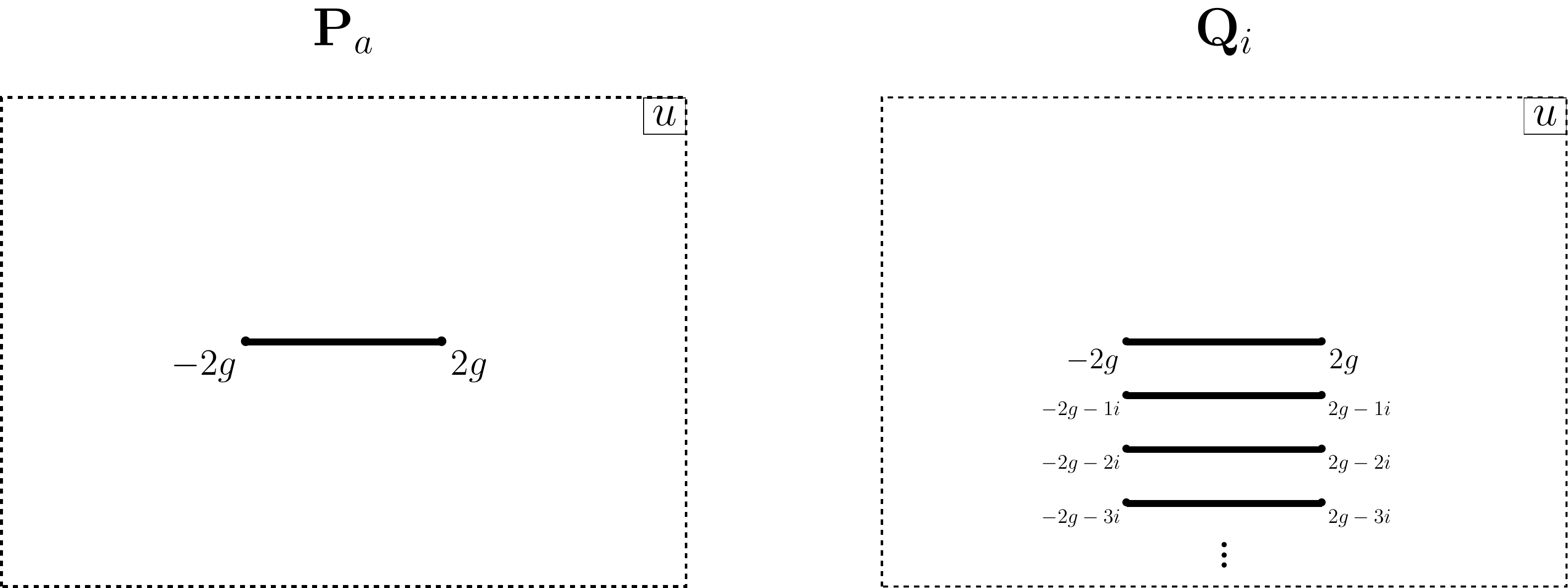}
  \caption{Analytic structure of $P_{a}$ and $Q_{j}$. The $P_{a}$ have a single square-root type branch cut at $u\in[-2g,2g]$, while the $Q_j$ have an infinite ladder of short branch cuts in the lower half-plane.  Alternatively,
the principal sheet of $Q_j$ could be defined so it is an analytic function outside two long cuts on the real axis.
  }
  \label{fig:PQ-cuts}
\end{figure}

We now briefly review how the QSC is solved to find the dimension $\Delta$ of a given operator.
The starting point is the analytic properties of the $P_a$, which is analytic outside a short cut $[-2g,2g]$,
where it has a square-root branch point, see figure \ref{fig:PQ-cuts}.
This allows to parametrize it as
\beq
P_{a}(u) \,=\,x^{M_{a}}\,\sum_{n=0}^{\infty}\frac{c_{a,n}}{x^{2n}} , \label{P ansatz}
\eeq
where the Zhukovsky variable is
\be
 x(u)= \frac{u+\sqrt{u-2g}\sqrt{u+2g}}{2g}
\ee
The series proceeds in even powers of $1/x$ due to the left-right symmetry of our operators \cite{Gromov:2013pga}.
The QSC equations allow us to gauge-fix $c_{4,1}=0$.
This parametrization converges in a neighborhood of the cut on the second sheet, where the continued function
is obtained by a simple replacement $x\mapsto 1/x$:
\beq \label{Ptilde}
\tilde{P}_{a}(u) \,=\,x^{-M_{a}}\,\sum_{n=0}^{\infty}\,c_{a,n}\,x^{2n} \,.
\eeq

Given $P_a(u)$, the $Q_j(u)$ are obtained by solving a finite difference equation known as P-Q system.
It involves an intermediate function $Q_{a|j}(u)$ which satisfies
\be
Q_{a|j}(u+i/2) - Q_{a|j}(u-i/2) =P_a(u)\,Q_j(u) \label{QQ recursion}
\ee
where
\be
Q_{j}(u) = P_{a}(u) Q_{b|j}(u + i/2)\,\chi^{ab} \label{Q from QP}
\ee
with $\chi^{ab} =(-1)^{a}\delta_{a,5-b}$ is a constant antisymmetric matrix.
Given $P_a(u)$, the two preceding equations give a homogeneous finite difference equation
which can be solved for $Q_{a|j}(u)$, subject to the boundary condition
$Q_{a|j}\underset{u\to\infty}{\sim} u^{M_{a}+\hat{M}_j+1}$ at large imaginary $u$.
This determines $Q_{a|j}$ as an analytic function in the upper-half-plane, with a sequence of short
cuts in the lower-half-plane starting at $u=-\frac{i}{2}+[-2g,2g]$.

To close the equations, one uses eq.~\eqref{Q from QP} together with the values of $Q_{a|j}(u)$ at $u\in \frac{i}{2}+[-2g,2g]$
to evaluate $Q_j(u)$ for $u\in [-2g,2g]$ along the real axis.
On the sheet shown in fig.~\ref{fig:PQ-cuts}, the function $Q_j(u)$ has an infinite series of short cuts in the lower-half-plane.
A crucial requirement is that if one were to go through the first short cut at $[-2g,2g]$, one would find a function
$\tilde{Q}_j(u)$ that is analytic in the lower-half-plane.
For the symmetrical operators that we consider in  \eqref{eq:trajectories},
we use the gluing conditions in (5.13) of \cite{Gromov:2017blm} (with $\beta=\gamma=0$ therein),
\be\begin{aligned}
 \tilde{Q}_1(u)&= \alpha\,\bar{Q}_3(u) \quad\text{and}\quad \tilde{Q}_2(u)&= -\alpha^*\,\bar{Q}_4(u)\label{glueing} 
\end{aligned}\ee
where $\alpha$ is a constant (only g-dependent), ${}^*$ is complex conjugation,
and the continuation $\tilde{Q}$ is obtained simply by using $\tilde{P}_a$ instead of $P_a$ in \eqref{Q from QP};
$\bar{Q}_j(u)\equiv Q_j(u^*)^*$.
These can be viewed as relations between analytic functions in the lower-half-plane, which can be analytically
from there.

For numerical implementation, following \cite{Gromov:2017blm} we take \eqref{glueing} to the real axis
and define the following ratios:\footnote{The identity $\tilde{Q}_{j}(u-i0) = Q_{j}(u+i0)$ is helpful to show the equivalence with \eqref{glueing}, both slightly below ($\alpha_{13}, \alpha_{24}$) and slightly above ($\alpha_{31}, \alpha_{42}$) the real cut.}
\beq
\alpha_{13}(u) {\equiv} \frac{Q_{1}(u{+}i 0)}{Q_{3}(u{+}i0)^*},\quad\alpha_{31}(u) {\equiv} \frac{\tilde{Q}_{1}(u{+}i0)}{\tilde{Q}_{3}(u{+}i0)^*},\quad\alpha_{24}(u) {\equiv} -\frac{Q_{2}(u{+}i 0)^*}{Q_{4}(u{+}i0)},\quad\alpha_{24}(u) {\equiv} -\frac{\tilde{Q}_{2}(u{+}i0)^*}{\tilde{Q}_{4}(u{+}i0)} \,. \label{eq:gluing}
\eeq
These four quantities should be constant ($u$-independent) and equal to each other.
We demand that this holds for a set of $N_{\rm cut}$ sampling points $u_{m}\in [-2g,2g]$.
These gluing conditions determine the parameters $c_{a,n}$ in \eqref{P ansatz},
from which one reads off the scaling dimension from the exponents in \eqref{PQ asympt},
which are themselves related to the constants $c_{a,0}$.

To solve these conditions we follow the numerical algorithm based on the multivariate Newton's method
described in section 6 of \cite{Gromov:2017blm}, as used initially in \cite{Gromov:2015wca}.
This requires to start with a seed of approximate values for the $c_{a,n}$ in eq.~\eqref{P ansatz}.
One then solve the difference equation \eqref{QQ recursion}-\eqref{Q from QP} to evaluate the four ratios $\alpha_{jk}(u)$ at a discrete set of $m$ points along the cut.
Since all $\alpha$'s should be equal and constant, the variance should vanish:\footnote{
It was observed that the $\alpha_{13}$ and $\alpha_{31}$ gluing conditions suffice to determine leading-twist operators \cite{Gromov:2015wca}.  For subleading trajectories, we observe that the four terms in \eqref{S QSC} are essential
to lift unconstrained directions in parameter space and ensure numerical stability.}
\be \label{S QSC}
S =  \sum_{m=1}^{N_{\rm cut}} \big|\alpha_{13}(u_{m})\,-\,\bar{\alpha}\big|^{2}+\big|\alpha_{31}(u_{m})\,-\,\bar{\alpha}\big|^{2}+ \big|\alpha_{24}(u_{m})\,-\,\bar{\alpha}\big|^{2}+\big|\alpha_{42}(u_{m})\,-\,\bar{\alpha}\big|^{2} % \mbox{with} \quad\bar{\alpha} \equiv \frac1{N_{\rm cut}} \sum_{m=1}^{N_{\rm cut}} \alpha (u_{m})\,.
\ee
where $\bar{\alpha}$ is the mean value of the list $\{\alpha_{13}(u_{m}),\,\alpha_{31}(u_{m}),\alpha_{24}(u_{m}),\,\alpha_{42}(u_{m})\}$. The multivariate Newton method minimizes $S$ iteratively as
\be \{c_{a,n}\}\longrightarrow \{c_{a,n}\} + \{\delta c_{a,n}\}_{\text{Newton Shift}}
\ee
where the changes in parameters are estimated from the derivatives $\frac{\partial\alpha_{jk}(u_m)}{\partial c_{a,n}}$
(estimated numerically by varying each $c_{a,n}$ by a small amount $\epsilon$), similar to \cite{Gromov:2017blm}.
By iterating the algorithm several times starting from an adequate seed,
the parameters $\{c_{a,n}\}$ converge to a value which solves the QSC equations with high accuracy.
The algorithm and the parameters involved at each step are summarized in table~\ref{tab:QSCalgorithm}.

\begin{table}[t]\begin{center}
\begin{tabular}{l|lccccccc}
 & \textbf{Description} \\\hline \\
\textbf{$P_{a}$ Ansatz} & Start with a guess $\{c_{a,n}\}_{\rm seed}$ in
\eqref{P ansatz} truncated at order $n_{\rm max}={\color{blue}N_p}$.\\ 
& \\[2pt]\hline \textbf{\multirow{6}{*}{\!\!\!\begin{tabular}{l} Glueing \\ conditions \end{tabular}}} &  \\
& Find the series $Q_{a,j}^{\rm large}=u^{M_{a}+\tilde{M}_{j}+1}\sum\limits_{n=0}^{{\color{blue}N_{q}}}\frac{b_{a,j,n}}{u^{2n}}$   by solving  \eqref{QQ recursion} given $\{c_{a,n}\}$.
\\[2pt]
& Iterate \eqref{PQ asympt} to find
$Q_{a,j}(u_m+\tfrac{i}{2})$ starting from $Q_{a,j}^{\rm large}(u_m+\tfrac{i}{2}+ i{\color{blue}u_{\infty}})$.
 \\[2pt]
 & Use \eqref{Q from QP} to find $Q_j(u_m)$, $\tilde{Q}_j(u_m)$ and $\alpha_{ij}(u_m)$ at ${\color{blue}N_{\rm cut}}$ points on the cut.
\\[2pt] \\\hline & \\
\multirow{2}{*}{\textbf{Update $\{c_{a,n}\}$}} &
Evaluate the errors $\alpha_{ij}(u_m)$ for parameters that differ by ${\color{blue}\epsilon}$, \\ &
and use Newton's method to update $\{c_{a,n}\}$. 
\\ 
\\[2pt]\hline 
\end{tabular}
\caption{Summary of steps and parameters in numerical algorithm for QSC\label{tab:QSCalgorithm}.
The last step is repeated until coefficients $\{c_{a,n}\}$ are found which minimize the error $\sum_{m=1}^{N_{\rm cut}} |F(u_m)|^2$
for the chosen parameters $\color{blue}N_p$, $\color{blue}N_q$, $\color{blue}u_\infty$ and $\color{blue}N_{\rm cut}$.
We normally choose those such that the dominant error is from $\color{blue}N_p$.
}
\end{center}\end{table}

\begin{figure}[t]
\centering
    \includegraphics[width=.75\linewidth]{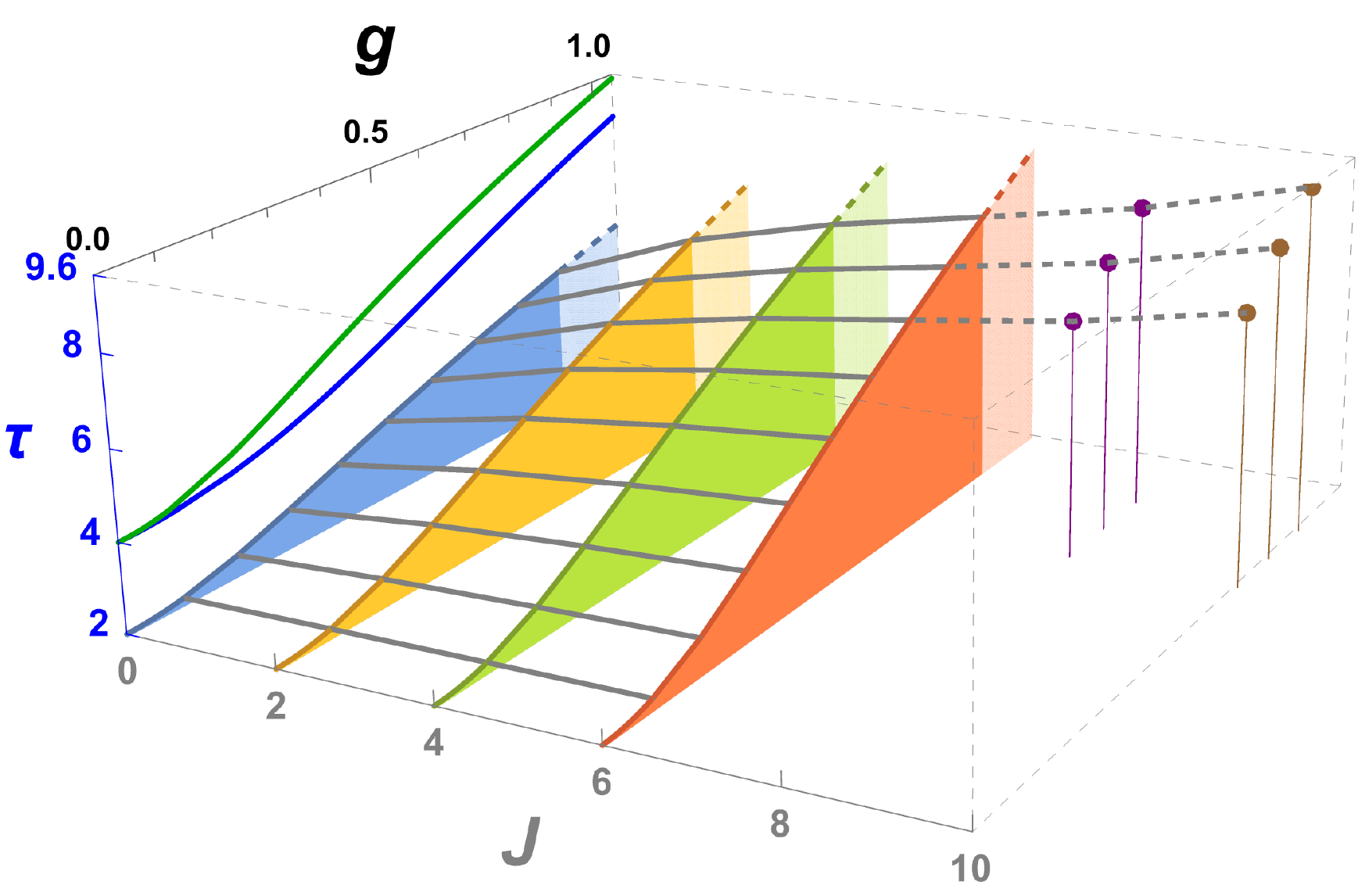}
  \caption{Scaling dimension for operators along the leading Regge trajectory at various values of the coupling,
   and for the first two subleading spin-0 operators.}
   \label{fig:Konishi&HigherSpin3D}
 \end{figure}

We benefit from the fact that the algorithm has been extensively applied already to low-lying operators.
Results for the Konishi operator will be described below.
Later in \cite{Hegedus:2016eop}, the same algorithm was used to produce data for higher spin operators $J=2,4,6$,
for a large range of coupling values $g\in [0.1,\,8]$.
In figure \ref{fig:Konishi&HigherSpin3D} we show data provided by the ancillary of \cite{Hegedus:2016eop}.
We have used this data to test our own code.

For our purposes, we extended this database to include higher spin operators $J=8,10,\cdots$ in the leading Regge trajectory and the lightest operator of the first sub-leading trajectory. For this, we need to find good seeds to start the numerical algorithm. We use three complementary ways to find them:
\begin{itemize}
\item In \cite{Marboe:2018ugv} we can find a database of solutions to the QSC at weak coupling. The solutions are presented as series expansions in $g$, see for instance \eqref{eq:WeakTwist4Spin0}. We use these results as seeds when the coupling is small $g<0.25$,
which is the typical radius of convergence of perturbation theory for many quantities. 
\item After generating a list of data for small coupling we can extrapolate to generate seeds for $g>0.25$. After using the numerical algorithm to refine them, we can extrapolate again to larger and larger values of $g$.
\item We can also extrapolate on the spin $J$ and move along a Regge trajectory for a fix value of the coupling. This is possible after producing a database for a few spins, such as the one provided in \cite{Hegedus:2016eop}.
\end{itemize}

The operators on the subleading trajectory have been less studied but we can obtain seeds
using the code from \cite{Marboe:2018ugv} which solves the QSC as a weak coupling series.
There are two nearly degenerate operators at spin 0 and twist near 4, whose scaling dimensions
are respectively:
\ba\label{eq:WeakTwist4Spin0}
\Delta \,&=\, 4+ (13 - \sqrt{41})\,g^2+ 2.413866\,g^4 +\mathcal{O}(g^6)\nonumber \\
\Delta' \,&=\, 4+ (13 + \sqrt{41})\,g^2- 94.41387\,g^4 +\mathcal{O}(g^6)\nonumber.
\ea
%{4.000000 + 6.596876 g^2 + 2.413866 g^4 - 32.34966 g^6 - 
%  57.09669 g^8 + 6580.235 g^10 - 161182.8 g^12, 
% 4.000000 + 19.40312 g^2 - 94.41387 g^4 + 774.8497 g^6 - 
%  8118.781 g^8 + 94925.63 g^10 - 1.181326*10^6 g^12}
The seed corresponding to the first operator is recorded in \eqref{twist 4 seed}.
We observe using the quantum spectral curve that the spacing between the two operators continues to increase with coupling
in the range of interest in this paper, so they do not cross.
To confirm that these operators are indeed the ones corresponding to our problem, we constructed their eigenfunction
and leading order structure constants as discussed in \eqref{eigen}:
\beq
C_{\Delta}^2 = \frac{3}{640}\left(7+\frac{27}{\sqrt{41}}\right)+O(g^2),\qquad
C_{\Delta'}^2 = \frac{3}{640}\left(7-\frac{27}{\sqrt{41}}\right)+O(g^2)\,.  \label{C4}
\eeq
We compared this data with the leading-logarithm terms in our four point correlator expanded to three loops
using \cite{Drummond:2013nda} (minus the contribution from twist-two operators):
%\bba
%C_{\Delta}^2 u^{\Delta/2}+C_{\Delta'}^2 u^{\Delta'/2} \,&=\, \left(\frac{2}{15}+\mathcal{O}(g^2)\right) + \left( \frac{14}{15}g^{2}+\mathcal{O}(g^4)\right)\,\frac{\log u}{2} \nonumber\\
%&\qquad + \left(\frac{36}{5} g^4 + \mathcal{O}(g^6)\right)\,\frac{(\log u)^2}{8} \nonumber\\
%&\qquad + \left(\frac{1016}{15} g^6 + \mathcal{O}(g^8)\right)\,\frac{(\log u)^3}{48} +\cdots
%\end{align}
\be
\def\pl{\,\,{+}\,\,}
\begin{array}{rl@{\pl}l@{\pl}l@{\pl}l@{\pl}l}\displaystyle
C_{\Delta}^2 u^{\frac{\Delta}{2}}+C_{\Delta'}^2 u^{\frac{\Delta'}{2}}\big|_{\rm leading\ log}\!=
& \frac{2}{15} & \frac{7}{15}g^2\log u & \frac{9}{10}g^4\log^2 u & \frac{23}{10}g^6\log^3 u  & O(g^8\log^4u),
\\
\left(\mathcal{H}-{\rm twist\ 2}\right)\big|_{\rm leading\ log}\!=
&0& \frac{-1}{5}g^2\log u & \frac{9}{10}g^4\log^2 u & \frac{23}{10}g^6\log^3 u  &O(g^8\log^4u,z,\zb).
\end{array}
\ee
%\bba
%C_{\Delta}^2 u^{\Delta/2}+C_{\Delta'}^2 u^{\Delta'/2}\big|_{\rm leading\ log} &= \frac{2}{15} &+ \frac{7}{15}g^2\log u
%&+ \frac{9}{10}g^4\log^2 u &+ \frac{23}{10}g^6\log^3 u  &+ O(g^8\log^4u) , \\
%\left(\mathcal{H}-{\rm twist\ 2}\right)\big|_{\rm leading\ log} &= 0 &-\frac{1}{5}g^2\log u &+ \frac{9}{10}g^4\log^2 u &+
% \frac{23}{10}g^6\log^3 u &+ O(z,\zb,g^8\log^4u)\,.
%\end{align}
The first two terms are not expected to match due to double-trace contributions, but the perfect
agreement of the $\log^2u$ and $\log^3u$ terms nontrivially confirms that we correctly identified the exchanged operator.

\begin{table}[t]\begin{center}
\begin{tabular}{|c|c|c|c|c|}\hline
 & $g=0.1$ & $g=0.2$ & $g=0.3$\\\hline
$\tau(\j=0)$ &2.115506378 & 2.418859881&  2.826948662  \\[2pt]\hline 
$\tau(\j=2)$ & 2.160267638 & 2.580161632 & 3.144804548  \\[2pt]\hline 
$\tau(\j=4)$ & 2.188431616 & 2.681905193 & 3.346021685   \\[2pt]\hline 
$\tau(\j=6)$ & 2.209027779 & 2.756495396 & 3.493970284  \\[2pt]\hline 
$\tau(\j=8)$ & 2.225274740 & 2.815455912&  3.611191520 \\[2pt]\hline 
$\tau(\j=10)$ & 2.238693733  & 2.864235300 & 3.708355788   \\[2pt]\hline
$\Delta\tau(\j=0)$ & 1.950671369 & 1.846969572 & 1.768329035 \\[2pt]\hline\hline
$\Gamma_{\rm cusp}$ & 0.03877086865 & 0.1433749321 & 0.291663365\\[2pt]\hline
$2\Gamma_{\rm virtual}$ &  $\,-0.00267496013  $& $-0.03534213604$ & $-0.139719169$ \\[2pt]\hline 
$\Delta\tau(\j{\to}\infty)$ & 1.895980997  & 1.645777840 & 1.351463213  \\[2pt]\hline 
\end{tabular}
\caption{Spectral data for the leading Regge trajectory and gap to the first subleading trajectory, used for numerical bootstrap.
These include the twists obtained from the Quantum Spectrum Curve,
as well as large-spin asymptotics from the asymptotic Bethe Ansatz.
\label{tab:data}}
\end{center}\end{table}

At large spin, we use that operators in the leading Regge trajectory have a universal anomalous dimension with logarithmic scaling:
\begin{align} 
\tau(\j)_{\text{twist-2},\, J\to \infty} \,&=
 2+ 2\Gamma_{\text{cusp}}(g)\,\log \big(J e^{\gamma_E}\big) \,+\, 2\Gamma_{\text{virtual}}(g) + O(\log^\#\!(J)/J)\\
 &= 2+ 2\Gamma_{\text{cusp}}(g)\,\log \left(\big(J+\tfrac{\tau+3}{2}\big) e^{\gamma_E}\right) \,+\, 2\Gamma_{\text{virtual}}(g)+ O(\log^\#\!(J)/J^2).
 \label{large-spin}
\end{align}
The cusp anomalous dimension, which also controls the UV divergences of lightlike cusped Wilson loops,
and virtual anomalous dimension are computed by the formulas recorded in \eqref{cusp}-\eqref{virtual}.
We use the second version of the formula, in which twist is expressed as a function of the conformal Casimir,
and which can be solved iteratively for $\tau$.
It is more accurate since it automatically removes $1/\j$ corrections \cite{Freyhult:2009my,Basso:2006nk,Alday:2015eya}.
In addition, we can compute the gap between the leading and first subleading trajectory at large spin
by adding excitations over the so-called GKP string, which represents the reference state corresponding to large spin operators:
\beq
\tau(\j)_{\text{twist-4},\, J\to \infty} \,=\, 2+2\Gamma_{\text{cusp}}(g)\,\log \big(J e^{\gamma_E}\big) \,+\, 2\Gamma_{\text{virtual}}(g) + \Delta\tau_{J\to\infty}(g). \label{twist4}
\eeq
The lightest R-singlet excitations is a pair of zero-momentum scalars,
$\Delta\tau_{J\to\infty}(g) = 2 E_{\phi}(g)$, whose energy is calculated from eq.~\eqref{Ephi}.
The raw data we use from integrability are summarized in table \ref{tab:data}.

%{{-0.001337480065 + 1.895980997 gap +  0.03877086865 LogJ}, 
%{-0.01767106802 + 1.645777840 gap +  0.1433749321 LogJ},
% {-0.06985958451 + 1.351463213 gap +    0.2916633658 LogJ}}

\subsection{Properties of the Konishi operator at weak and strong coupling} \label{ssec:Konishi}

\begin{figure}[t]
\centering
    \includegraphics[width=.75\linewidth]{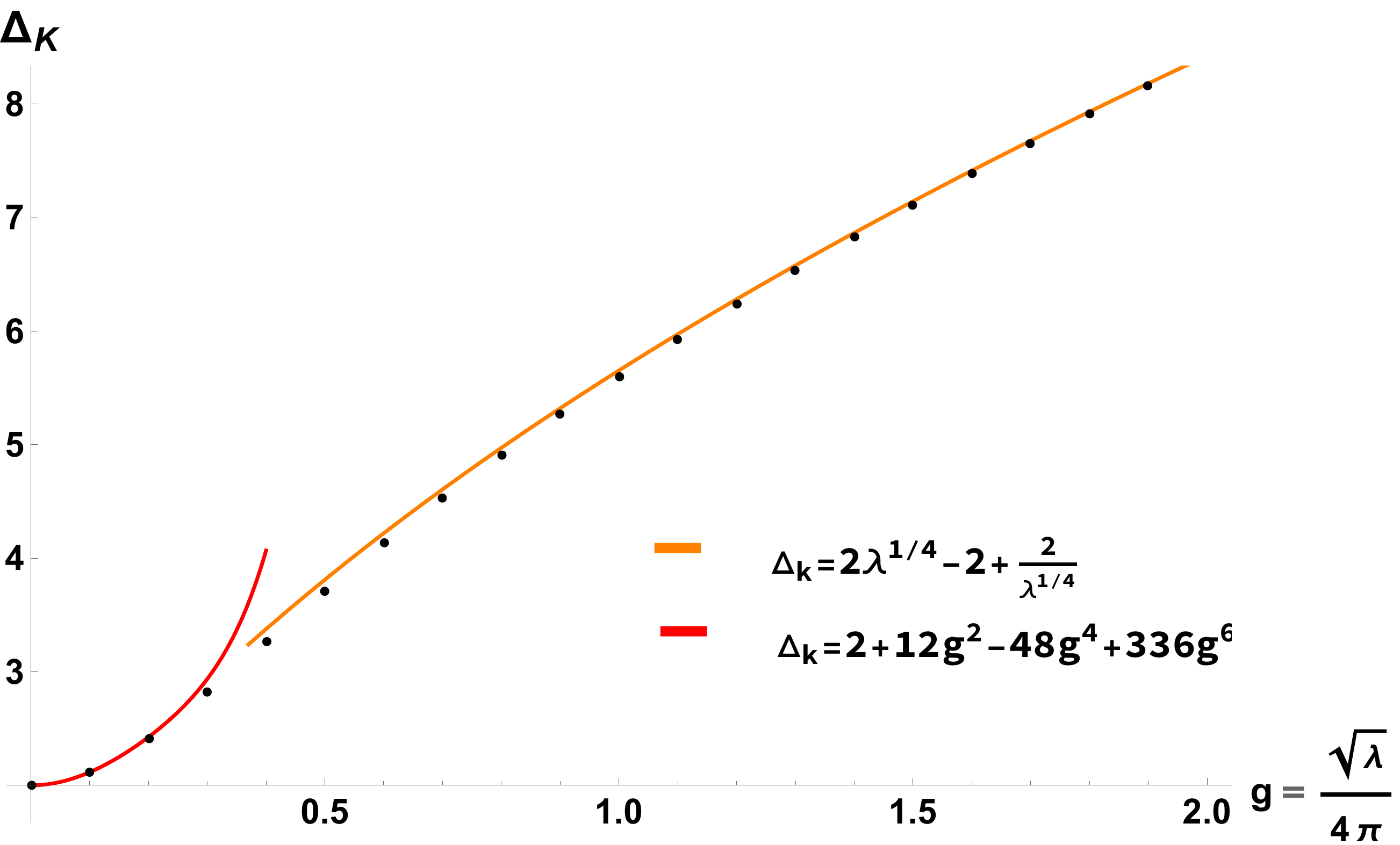}
  \caption{Scaling dimension of the Konishi operator at finite coupling, using data $g\in [0, 2]$ from \cite{Hegedus:2016eop}. This was originally plotted in \cite{Gromov:2009tv} . We also 
include comparison with the first 3 terms at weak and strong coupling, which were originally obtained
in \cite{Eden:2004ua} and \cite{Vallilo:2011fj} respectively.}
  \label{fig:KonishiAllCoupling}
\end{figure}

The Konishi operator will be particularly important in our study since its scaling dimension effectively defines
the 't Hooft coupling, from the point of view of the correlation function we are studying;
bounding its OPE coefficient will be our main focus.

At weak coupling, the scaling dimension of the Konishi operator has been provided to 11 loop orders in \cite{Marboe:2018ugv}
(building on much earlier work referenced there).
We reproduce here the first 5 orders:
\bba
 \Delta_\Konishi &= 2 + 12 g^2- 48 g^4 +336 g^6 +g^8 (576 \zeta_3-1440 \zeta_5-2496)\nonumber\\
 &\qquad+g^{10} \left(6912 \zeta_3-5184 \zeta_3^2-8640 \zeta_5+30240 \zeta_7+15168\right)+\mathcal{O}(g^{12}) .
\label{Konishi weak Delta}
\end{align}
Its OPE coefficient is currently know to 5 loop orders \cite{Georgoudis:2017meq}:
\bba \label{Konishi weak coeff}
\OPE_\Konishi= &\frac{1}{3} - 4 g^2 + g^4 (56 + 24 \zeta_{3}) +
 g^6 (-768 - 128 \zeta_{3} - 400 \zeta_{5}) \nonumber\\
&\qquad + g^8 (9952 + 1312 \zeta_{3} + 288 \zeta_{3}^2 + 3920 \zeta_{5} +
    5880 \zeta_7) \nonumber\\
&\qquad+ g^{10} (-117824 - 28992 \zeta_3 + 6624 \zeta_3^2 - 43008 \zeta_5 -
    13824 \zeta_3 \zeta_5 - 59472 \zeta_7 - 84672 \zeta_9)\nonumber\\
&\qquad    +\mathcal{O}(g^{12}) .
\end{align}
Recall that we removed an overall factor $\frac{1}{c}$ from our OPE coefficients, see eq.~\eqref{OPE scaling}.

\begin{figure}[t]
\centering
    \includegraphics[width=.9\linewidth]{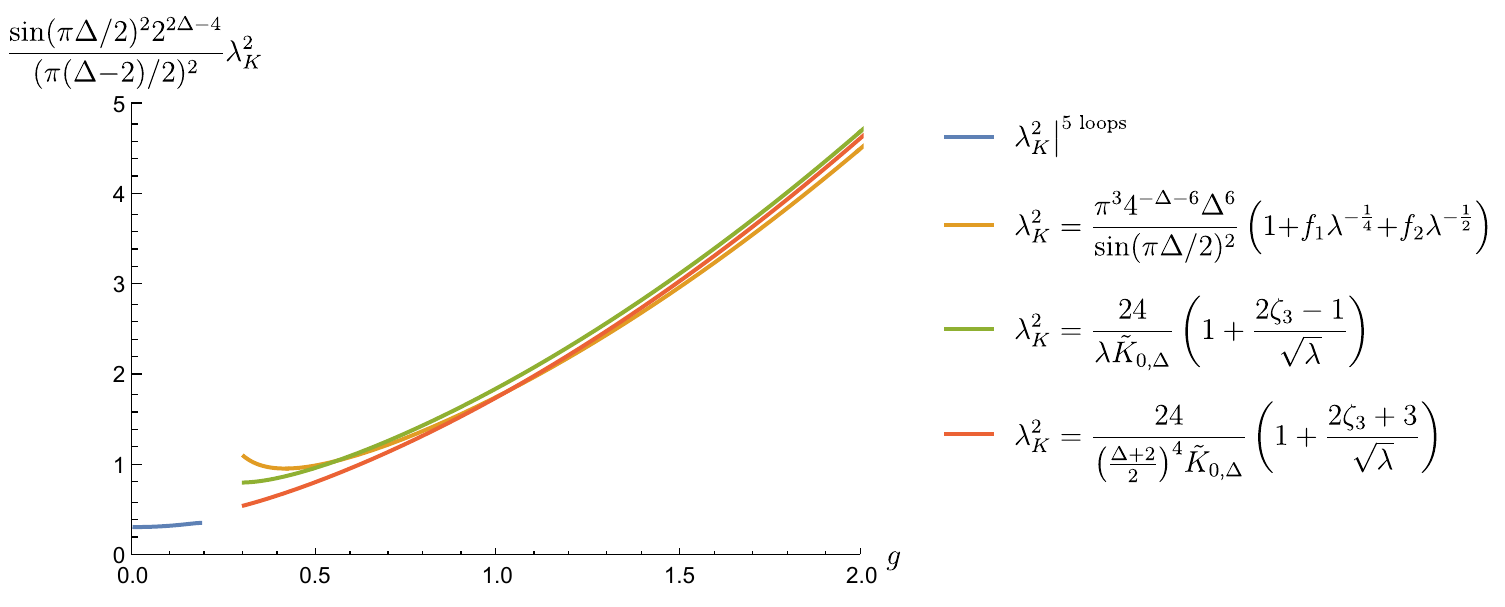}
  \caption{Rescaled OPE coefficient of the Konishi operator to two stress tensor multiplets at weak 
  \eqref{Konishi weak coeff} and strong coupling \eqref{Konishi strong coeff}.
   We display three forms of the strong coupling formula,
   which agree at asymptotically large $g$ but become distinct at intermediate $g$. 
   The rescaling was applied to remove oscillations and the overall exponential trend.}
  \label{fig:KonishiC2}
\end{figure}

At strong coupling, the scaling dimension is known to 3 loop order ($\lambda^{-5/4}$) from \cite{Gromov:2014bva},
\be
\Delta_\Konishi = 2\lambda^{\frac14} -2 + 2\lambda^{-\frac14}+\left(\frac{1}{2} -3 \zeta_3\right)\,\lambda^{-\frac{3}{4}}+\left(\frac{1}{2} +6 \zeta_3+\frac{15}{2}\zeta_{5}\right)\,\lambda^{-\frac54}+ O(\lambda^{-\frac74}),
\ee
while the OPE coefficient is known through its relation to the Virasoro-Shapiro amplitude \cite{Costa:2012cb,Minahan:2014usa,Goncalves:2014ffa}.
Recently, ref.~\cite{Alday:2022uxp} also obtained subleading corrections
by adding spectral information from integrability, together with constraints from localization:
\begin{align}
\OPE_\Konishi &=
\frac{\pi^3 \Delta_\Konishi^6}{4^{\Delta_\Konishi+6}\sin\big(\tfrac{\pi \Delta_\Konishi}{2}\big)^2}
\left(1+f_1\lambda^{-\frac14} + f_2\lambda^{-\frac12} + O(\lambda^{-\frac34})\right)
\nonumber\\&
 =\frac{24}{\lambda\tilde{K}^{N=4}_{0,\Delta_\Konishi}} \left(1+\frac{2\zeta_3-1}{\sqrt{\lambda}}+O(\lambda^{-1}\right)\,, \label{Konishi strong coeff}
\end{align}
where $f_1=\frac{23}{4}$ and $f_2=\frac{405}{32} + 2\zeta_3$ on the first line.\footnote{
The first arXiv version of \cite{Alday:2022uxp} reported a different value for $f_2$.
We are grateful to the authors for communicating sharing with us the corrected value.} 
The second line is an equivalent rewriting of the formula using the factor $\tilde{K}^{N=4}$ defined below \eqref{Kt},
which arises naturally in the derivation of the coefficient.
Note that this rewriting neatly removes the $\lambda^{-\frac14}$ term. We thus
expect the second series to proceed in integer powers of $1/\sqrt{\lambda}$, which would be
interesting to verify.

The quantum spectral curve reviewed above enables to compute the scaling dimension $\Delta_\Konishi$
numerically with arbitrary precision at any $g$.
Figure \ref{fig:KonishiAllCoupling} displays the resulting curve from the original article \cite{Gromov:2015wca},
along with its comparison with weak and strong coupling expansions.

It is amusing to similarly plot the weak and strong coupling predictions for the OPE coefficient,
for which exact results are not yet available.
Since \eqref{Konishi strong coeff} depends strongly on the coupling $g$,
different forms which agree asymptotically become distinct at moderate $g$, as visible in figure \ref{fig:KonishiC2}.
(Without the subleading terms in \eqref{Konishi strong coeff}, the curves would differ from each other much more strongly.)
Below we study the OPE coefficient at $g=0.3$.  It is hard to extract a definite value from the plot,
but it seems reasonable to assume that the true value should lie somewhere between
a linear extrapolation of the weak coupling curve and the lowest of the strong coupling curve, giving
$\OPE_\Konishi(g=0.3)\in [0.24,0.33]$ (corresponding to the range $[0.4,0.55]$ in the rescaled plot).

%%%%%%%%%%%%%%%%%%%%%%%%%%%%%%%%%%%%%%%%%%%%%%%%%%
\section{A menu of functionals} \label{sec:functionals}

All functionals we consider in this work are combinations of the Polyakov-Regge crossing equation
$X_{u,v}$ \eqref{Xuv} and antisubtracted $B_v$-sum rule \eqref{Bv}.
As detailed in the next section, we will seek linear combinations
which are positive on every possible state in the spectrum, and which maximize certain objectives.
Since the space of functionals to explore is infinite-dimensional, different truncations can exhibit different numerical properties.

In this section we define some infinite linear combinations of the $X_{u,v}$ and $B_v$:
the Mellin-transformed functionals $\widehat{X}_{\mS,\mT}$ and $\widehat{B}_\mT$,
and special linear combinations $\Phi_{\ell,\ell+2}$ and $\Psi_\ell$ of $B_v$ which diagonalize its action near twist two.
Although these functionals are infinite linear combinations of each other, no finite relations exist,
and so we will treat them as independent for numerical purposes.

%especially when considering infinite linear combinations (ie. Mellin functionals which are integral transforms of the $X_{u,v}$ and $B_v$).

Generally, a basis of functionals for the numerical bootstrap needs to have the following properties (see for example \cite{Caron-Huot:2020adz}):
\begin{enumerate}
\item Swappability: Each functional commutes with the infinite sum over the CFT spectrum, ie. each gives a valid sum rule.
\item Asymptotic positivity of finite linear combinations: finite linear combinations must exist which are positive
on all but a finite range of $(\Delta,\j)$.
\item Completeness, as the number of elements tend to infinity.
\end{enumerate}
The first requirement is clearly essential, and is rigorously satisfied by the $X_{u,v}$ and $B_v$
thanks to the Regge boundedness of correlators and the analysis of \cite{Caron-Huot:2020adz}.
We discuss convergence for Mellin-transformed functionals below, as well as the asymptotic positivity of various functionals.

Our strategy to fulfill the third requirement is to present the numerical optimization solver with a varied menu of functionals, and see which ones it prefers.

\subsection{Dispersion relations in Mellin space} \label{sec:Mellin functionals}

% problem since $\D_\phi=4 \ \in \mathbb{Z}$ allows for further simplifications.

We now describe the Mellin formulation of the Polyakov-Regge block and collinear functionals corresponding to eq.~\eqref{Xuv} and \eqref{Bv} respectively.
This formulation is also convenient for numerical evaluation.
It was explained in \cite{Caron-Huot:2020adz} how the position space dispersion relation is equivalent to a straightforward dispersion relation in Mellin space.
The Mellin representation for identical-dimension operators takes the form:
\be
\cH(u,v) = \iint\!\!\frac{d\mS\,d\mT}{(4\pi i)^2}\,u^{\tfrac{\mS}{2}-\Df}v^{\tfrac{\mT}{2}-\Df}
\Gamma\!\left(\Df-\tfrac{\mS}{2}\right)^2\Gamma\!\left(\Df-\tfrac{\mT}{2}\right)^2\Gamma\!\left(\Df-\tfrac{\mU}{2}\right)^2
M_{\mS,\mT}\,.
\label{eq:mellinRep2}
\ee 
Here $\mS$, $\mT$, $\mU$ are the Mellin-Mandelstam variables, which are constrained to satisfy $\mS+\mT+\mU=4\Df$,
and from here we set $\Df\equiv 4$ as the effective external dimension of our reduced correlator.
For example, in the $g\to \infty$ limit, the Mellin amplitude goes to
\be
 M_{\mS,\mT}^{\rm strong}  = \frac{1}{(\tfrac{\mS}{2}-3)(\tfrac{\mT}{2}-3)(\tfrac{\mU}{2}-3)}\,.
\ee

The $\mU$-channel Regge behavior of $\cH$ (see section~\ref{sec:Stress tensor})
implies that $\lim_{\mS'\to\infty} M(\mS',\mT')\sim s^{\j_*-4}$, where $\j_*<2$ and $\mT'=\mT+\mS-\mS'$ \cite{Penedones:2019tng}.
In particular the reduced correlator satisfies an unsubtracted dispersion relation:
\be
 M_{\mS,\mT} = \oint \frac{d\mS'}{2\pi i} \frac{M(\mS',\mT')}{\mS-\mS'}  \label{M unsub}
\ee
where again $\mT'=\mT+\mS-\mS'$ and the contour encircles all the poles of $M$ except that at $\mS'=\mS$.
In fact the above only assumes $\j_*<4$.  The stronger expectation $\j_*<2$ at finite coupling implies that an ``anti-subtracted'' dispersion relation also converges,
where we put zeros at some subtraction point:
\be
 M_{\mS,\mT} = \oint \frac{d\mS'}{2\pi i}  \frac{(\mS'-\mS_0)(\mT'-\mS_0)}{(\mS-\mS_0)(\mT-\mS_0)} \frac{M(\mS',\mT')}{\mS-\mS'}.
\ee
A natural choice is $\mS_0=6$, which suppresses the contribution from twist-two operators.
These two relations are not independent, and their equality amounts to the sum rule
\be
 0 = \frac{1}{(\mS-\mS_0)(\mT-\mS_0)}\oint\frac{d\mS'}{2\pi i} (\mS'-\mT) M(\mS',\mT') \label{mellin B pre}
\ee
for any $\mS,\mT$.  This constraint is essentially independent of $\mS_0$.

The poles of the Mellin amplitude, according to the OPE \eqref{OPE}, occur at descendants of primaries
\be
M(\mS,\mT) \sim\frac{\OPE_{\Delta,J}\mathcal{Q}^m_{\Delta+4,J}(\mT)}{\mS-(\Delta+4-\j+2m)}\,,
\ee
where $\mathcal{Q}$ is a Mack polynomial discussed further in appendix~\ref{app:Mack polynomials}, and $m\geq 0$ is an integer.
Assuming the same spectrum in the s- and t- channel, their contribution can be combined in the form
\be
\label{PR Mellin definition}
 M_{\mS,\mT} = M_{\mS,\mT}^{\rm strong} +\sum_{\Delta,\j} \OPE_{\Delta,\j} \widehat{\cP}^{N=4}_{\mS,\mT}[\Delta,\j]\,,
\ee
where the Polyakov-Regge block is defined as
\be \widehat{\cP}^{N=4}_{\mS,\mT}[\Delta,\j]
= \sum\limits_{m=0}^\infty
\mathcal{Q}^m_{\Delta+4,\j}(16 - \mS- \mT)\left[\frac{1}{\mS-(\tau+2m+4)}+\frac{1}{\mT-(\tau+2m+4)}\right] \,.
\label{PR Mellin explicit}
\ee
On the other hand, the constraint \eqref{mellin B pre} amounts to the sum rule
\be
 0= \widehat{B}_{\mT}^{\rm protected} + \sum_{\Delta,\j} \OPE_{\Delta,\j} \widehat{B}_{\mT}[\Delta,\j]
\ee
where (without loss of generality) we focus on the residue at $\mS=\mS_0$ and set $\mS_0=6$:
\begin{align} \label{Bt protected}
 \widehat{B}_{\mT}^{\rm protected} &=\frac{2}{(\tfrac{\mT}{2}-3)(\tfrac{\mT}{2}-2)}, \\
 \widehat{B}_{\mT}[\Delta,\j] &= \sum_{m=0}^\infty \frac{2 (\Delta-\j+2m)+2-t}{t-6} \mathcal{Q}^m_{\Delta+4,\j}(10-\mT).
\label{Bt}
\end{align}
The salient property of these sum rules is that they have double-zeros on all double-trace locations $\Delta=4+2m+\j$,
originating from the Mack polynomials. They are similar but distinct from those used recently in
\cite{Alday:2022uxp} to constrain stringy corrections to double-trace OPE data:
here we concentrate on sum rules which strictly remove all double-traces.

The Polyakov-Regge expansions \eqref{PR coords} and \eqref{PR Mellin definition} are formally similar, and our
nomenclature is not an accident: a result of \cite{Caron-Huot:2020adz} is that $\widehat{\mathcal{P}}$ is precisely the Mellin transform of $\mathcal{P}$ !  This is established using the uniqueness properties of Polyakov-Regge block namely, single-valuedness, Regge boundedness, and the pattern of zeroes on Regge trajectories. 
Similarly, $B_v$ and $\widehat{B}_\mT$ are related by a Mellin transform. Explicitly,
\begin{align}
 \cP_{u,v}^{N=4}[\Delta,\j] &= \int\limits_{5-i\infty}^{5+i\infty} \frac{d\mS\, d\mT}{(4\pi i)^2} u^{\frac{\mS}{2}-4}v^{\frac{\mT}{2}-4}
\Gamma\!\left(4-\tfrac{\mS}{2}\right)^2\Gamma\!\left(4-\tfrac{\mT}{2}\right)^2\Gamma\!\left(4-\tfrac{\mU}{2}\right)^2
\widehat{\cP}_{\mS,\mT}^{N=4}[\Delta,\j], \label{P Mellin} \\
B_v[\Delta,\j] &= \frac{1}{2}\int_{5-i\infty}^{5+i\infty} \frac{d\mT}{4\pi i}v^{\frac{\mT}{2}-4} \Gamma\!\left(4-\tfrac{\mT}{2}\right)^2\Gamma\!\left(\tfrac{\mT}{2}-1\right)^2 \widehat{B}_{\mT}[\Delta,\j].
\label{Bv Mellin}
\end{align}
The physical requirements on these contours is that they run to the left of all $\mS$ and $\mT$-channel poles (so ${\rm Re}(\mS),{\rm Re}(\mT)< \min(8,\Delta-\j+4)$),
and to the right of $\mU$-channel poles (so ${\rm Re}(\mS+\mT)> 8$).
Since all operators considered in this paper have $\Delta-\j\geq 2$, the simple choice ${\rm Re}(\mS)={\rm Re}(\mT)=5$
indicated above works uniformly.

The identities \eqref{P Mellin}-\eqref{Bv Mellin}
are highly nontrivial and give us independent methods to compute functionals numerically.
In appendix~\ref{app:efficient eval of funcs}, we discuss our current best numerical implementations for each.
Roughly, position-space methods seem to scale better with increasing spin,
while our Mellin-space implementations scale better with increasing precision and are generally faster.
The precise numerical agreement between these independent methods is very helpful for debugging.

As simple consistency check, note that $B_v^{\rm protected}$ and $\widehat{B}_\mT^{\rm protected}$ are indeed Mellin-transform of each other:
\be
\frac{1}{2}\int \frac{d\mT}{4\pi i} v^{\frac{\mT}{2}-4}\Gamma\!\left(4-\tfrac{\mT}{2}\right)^2\Gamma\!\left(\tfrac{\mT}{2}-1\right)^2 
\frac{2}{(\tfrac{\mT}{2}-3)(\tfrac{\mT}{2}-2)}=
\frac{v^2-1-2v\log v}{v(1-v)^3} =%f^{(0)}(v)\,.
B_v^{\rm protected}\,.
\ee
For future reference, we also define a $\mS{\leftrightarrow}\mU$ crossing functional in Mellin space as
\be
\widehat{X}_{\mS,\mT} \equiv \widehat{\cP}^{N=4}_{\mS,\mT}-\widehat{\cP}^{N=4}_{16-\mS-\mT,\mT}\,, \label{Xst}
\ee
which is the Mellin transform of $X_{u,v}$ in \eqref{Xuv}.

\subsection{Projection functionals derived from $B_v$}

So far we have two versions of the collinear functional, $B_v$ and $\widehat{B}_{\mT}$, which diagonalize respectively a position cross-ratio or a Mellin moment.
It is natural to try to diagonalize other quantities, for example the action operators of twist close to two and various spins.
Since twist-two operators dominate dispersive sum rules at weak coupling, up to $\sim g^4$ contributions from operators of twists $\tau\geq 4$, these functionals effectively solve the 1-loop problem analytically.
They could also potentially be useful to suppress large-spin contributions.

Such \emph{projection functionals} can be constructed by integrating $\widehat{B}_{\mT}$ against a kernel $W[\mT]$:
\be
W[\D,\j] \equiv \int \frac{d\mT}{4\pi i} W[\mT] \widehat{B}_{\mT}[\D,\j]. \label{projection def}
\ee
These can be thought of as an infinite sums of collinear functionals finely tuned to possess desirable properties.

We constructed the following projectors, labelled by even spins $\ell\geq 0$,
whose details can be found in appendix~\ref{app:projection functionals}.
They are characterized by their zeros near twist two:
\begin{enumerate}
\item The $\Phi_{\ell,\ell+2}$ functional \eqref{Phi kernel}
has simple zeros at $\tau=2$ for $\j=\ell$ or $\ell+2$, and double zeros on all other spins:
\be
\lim_{\tau\to 2} \Phi_{\ell,\ell+2}[\tau+\j,\j] = 0 + (\tau-2) \left( \delta_{\ell,\j} - \frac{\Phi^\infty_\ell}{\Phi^\infty_{\ell+2}}\delta_{\ell+2,\j}\right) + O((\tau-2)^2)\,. 
\label{Phi features}
\ee
\item The $\Psi_\ell$ functional \eqref{Psi summary} has double-zeros for all spins except $\ell=\j$, where it has a nonvanishing intercept and slope:
\be
\lim_{\tau\to 2} \Psi_{\ell}[\tau+\j,\j] = \delta_{\ell,\j}\left( 1+ (\tau-2)\beta_\ell \right)+O((\tau-2)^2)\,.
\label{Psi features}
\ee
\end{enumerate}
All these functionals have double-zeros on double traces with $\tau\geq 4$, as required for our applications: they are saturated by single-trace operators at large $N_c$.
The first was inspired by the $\Phi_\ell$ functional used in \cite{Caron-Huot:2020adz} to prove the existence of operators below the double-twist threshold and fixed spin.

Assuming that the above functionals exist, it is not hard to guess the values of the constants $\Phi^\infty_\ell$ and $\beta_\ell$
by expanding the sum rules at weak coupling.
The anomalous dimensions for the leading family of long operators have been known for some time
(see \cite{Kotikov:2003fb,Kotikov:2004er,Eden:2012rr} for three-loop results):
\begin{align}
 \Delta(\j)-\j &= 2 + 8g^2 S_1(j+2) + O(g^4), \\ %- 16g^4\left(S_{-3}+2S_1S_{-2}+2S_1S_2 +S_3-2S_{-2,1}\right) +O(g^6) \label{konishi twist} \\
  C(\j) &\equiv
   \OPE_{\Delta(\j),\j} = \frac{2\Gamma\big(\frac{\Delta_\j+\j+4}{2}\big)^2}{\Gamma(\Delta_\j+\j+3)} \left( 1-4g^2S_2(j+2) + O(g^4)\right)\,,  \label{protected OPE coeff}
\end{align}
where $\j\geq 0$ and $S_a(m)$ denote harmonic sums $S_a(m)\equiv \sum_{k=1}^{m} \frac{{\rm sign}(a)^k)}{k^|a|}$.
The stress tensor multiplet is formally the $\j=-2$ member of this family, but it is included in the ``protected'' part\footnote{
The protected contribution $\widehat{B}_\mT^{\rm protected}$ in \eqref{Bt protected} is precisely the analytic continuation
of $C^{(0)}_\j \widehat{B}_\mT[\j+2,\j]$ to $\j=-2$.}.
Our focus is on long operators, which have $\j\geq 0$.
Denoting as $Y^{(i)}$ the coefficient of $(g^2)^i$ in the quantity $Y$, the data can be expanded as
\be
 C^{(0)}_\j = \frac{2\Gamma(\j+3)^2}{\Gamma(2\j+5)},\qquad
 \gamma^{(1)}_\j = 8S_1, \qquad
 \frac{C^{(1)}_\j}{C^{(0)}_\j}=  \Big[8S_1(j+2)-8S_1(2j+4)-4S_2(j+2)\Big]\,.
\ee

Generally, each of the above functionals yields a sum rule on single-trace data of the form
\be
 0=W^{\rm protected} + \sum_{(\Delta,\j)\ \rm long} \OPE_{\Delta,\j} W[\D,\j] \label{projected sum rule}
\ee
with $W=\Phi_{\ell,\ell+2}$ or $\Psi_\ell$.
The respective protected parts follow immediately from the protected OPE \eqref{protected OPE coeff}
\be
 \Phi_{\ell,\ell+2}^{\rm protected} = 0, \qquad \Psi_\ell^{\rm protected} = -C^{(0)}_\ell\,. 
\ee
The salient feature of all sum rules we consider is their double zeros at twists 4,6,\ldots.
This means that at one-loop the sum rules are saturated by the twist-two family, which we can evaluate from \eqref{Phi features} and \eqref{Psi features}:
\begin{align}
\sum_{(\Delta,\j)\ \rm long} \OPE_{\Delta,\j} \Phi_{\ell,\ell+2}[\D,\j]& = g^2 \left(C^{(0)}_\ell\gamma^{(1)}_\ell - \frac{\Phi^\infty_\ell}{\Phi^\infty_{\ell+2}}C^{(0)}_{\ell+2}\gamma^{(1)}_{\ell+2}\right) +O(g^4),  \label{Phi one-loop}\\
\sum_{(\Delta,\j)\ \rm long} \OPE_{\Delta,\j} \Psi_{\ell}[\D,\j]& = C^{(0)}_\ell + g^2\left( C^{(1)}_\ell + \beta_\ell C^{(0)}_\ell\gamma^{(1)}_\ell\right) + O(g^4)\,.
\label{Psi one-loop}
\end{align}
Thus, consistency of the bootstrap sum rule \eqref{projected sum rule} with the known perturbative data requires that $C^{(0)}_\ell\gamma^{(1)}_\ell\propto \Phi^\infty_\ell$ with an $\ell$-independent factor,
and that $C^{(1)}_\ell/(C^{(0)}_\ell\gamma^{(1)}_\ell)=-\beta_\ell$.
This is in precise agreement with the constants $\Phi^\infty_\ell$ and $\beta_\ell$ that come out of the derivation, recorded in eqs.~\eqref{boundary term} and \eqref{beta ell result}.
In other words, the $\Phi$ and $\Psi$ sum rules analytically bootstrap the one-loop theory.

The fact that the one-loop data is determined by crossing was first noticed from large spin expansions in \cite{Alday:2015eya},
and later extended to finite spin \cite{Henriksson:2017eej}.
This is a quite generic behavior, which generally works up to a finite number of constants.
For $\mathcal{N}=4$ sYM the one-loop corrections are fixed up to a single overall factor $g^2$. 
The novel feature here is that this is obtained from sum rules with nice sign properties
(the $\Psi_\ell$ are non-negative, see appendix \ref{app:projected functional signs}), thereby uplifting the one-loop approximations
to nonperturbative inequalities.
In general, the action of the projection functionals $\Phi_{\ell,\ell'}[\D,\j]$ and $\Psi_\ell[\D,\j]$
on an arbitrary state can be calculated exactly using the formulas for Mack polynomials \eqref{projection app}.

\subsection{Convergence of Mellin functionals} \label{sec:Mellin convergence}

It is interesting to consider the sum rules $\widehat{X}_{\mS,\mT}$ and $\widehat{B}_\mT$ on their own right, rather than simply
Mellin representations of position-space sum rules.
A crucial fact that is that they are saturated by single-traces in the planar limit, which makes them sensible for our purposes.
This can be seen from the Mack polynomials ${\cal Q}_{\Delta+4,\j}$, which have double zeros
when $\Delta-\j=4+2n$ with $n\geq 0$; physically this happens because the $\Gamma$-functions in the Mellin representation \eqref{P Mellin}, \eqref{Bv Mellin} already account for double-twist operators.
To use Mellin functionals, we need to determine the range of $\mS, \mT$ such that the functionals can be swapped with the OPE.

Regge boundedness ensures that the functionals converge at large twist.
Swappability with respect to the OPE therefore requires that the sum over spin converges.

\begin{figure}[t]
\centering
\begin{subfigure}[b]{0.48 \textwidth}
\centering
\includegraphics[width=\textwidth]{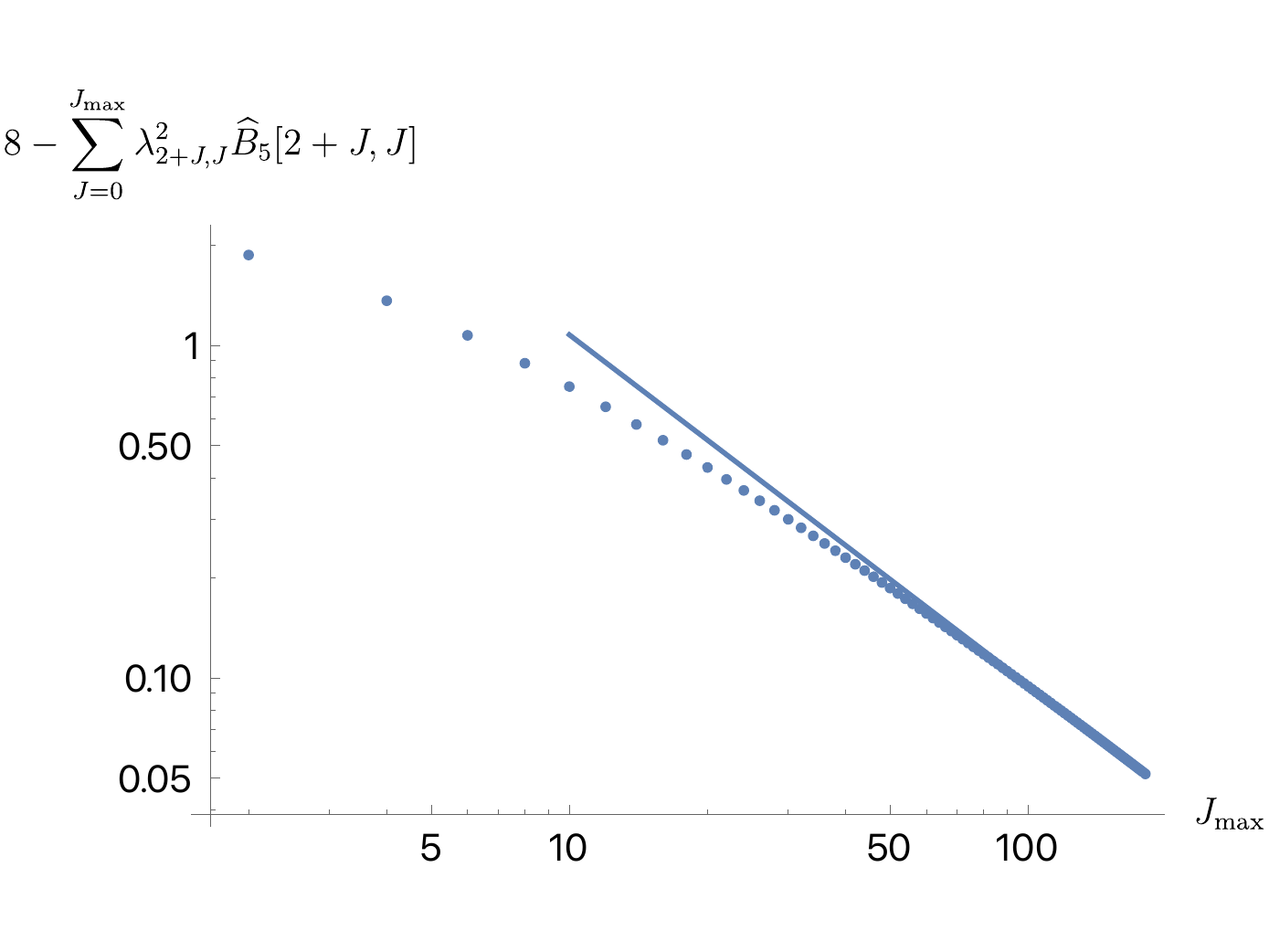}
\end{subfigure}
\begin{subfigure}[b]{0.48 \textwidth}
\centering
\includegraphics[width=\textwidth]{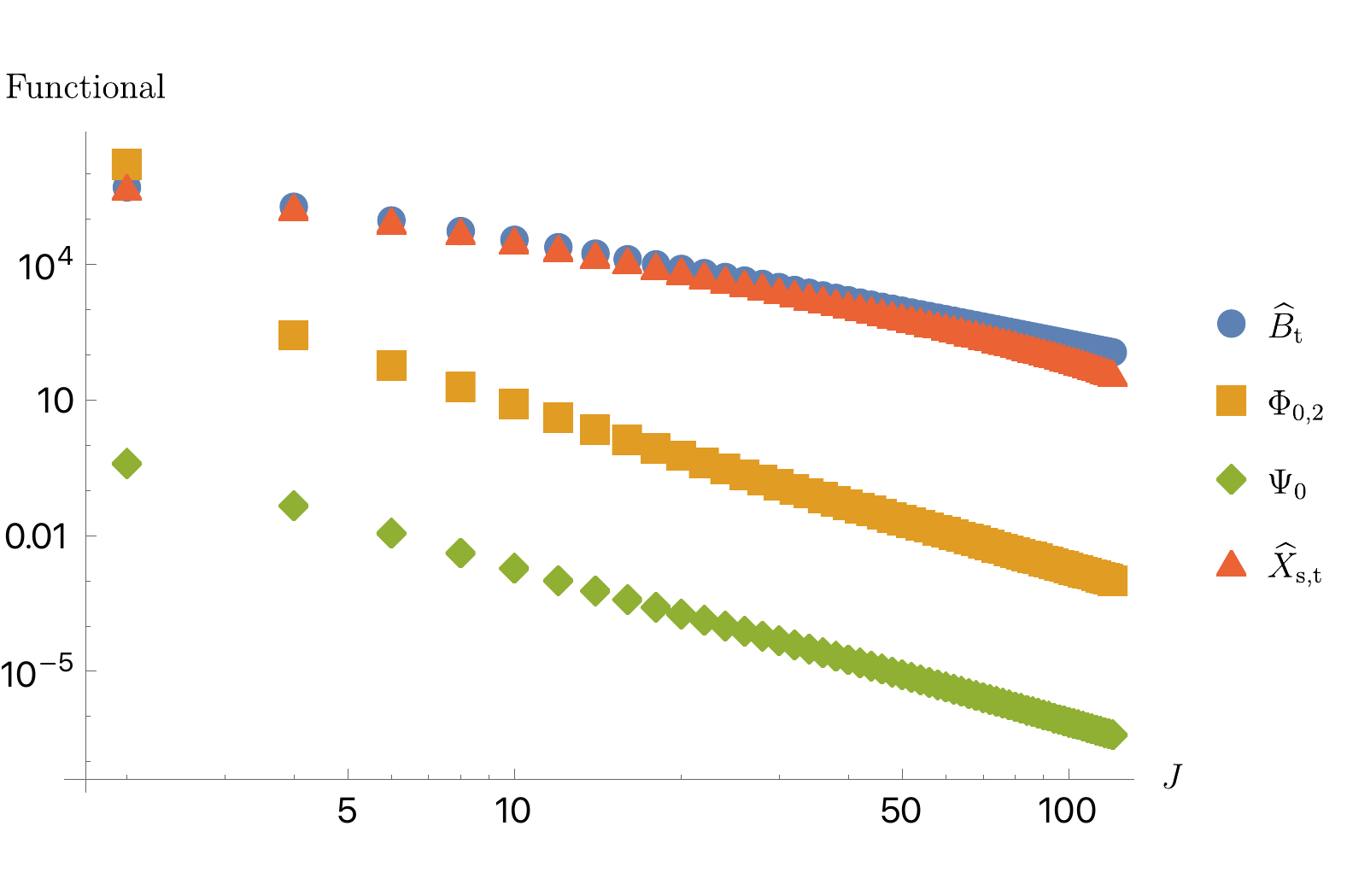}
\end{subfigure}
\caption{Left: Convergence of the $\widehat{B}_{\mT}$ sum rule at zero-coupling. We plot the difference between the sum rule and the protected part for $\mT=5$ at twist $\tau=2.001$. The partial sum converges with $J_{\rm max}^{-1.05}$ as shown by the solid blue line in the plot.
Right: Convergence at large spin of the Mellin-space functionals for fixed twist $\tau=2.001$ without the $2\sin^2(\pi\tau/2)$ factor. We normalized the action of all functionals with respect to the collinear $\widehat{B}_{\mT}$ functional. Furthermore, we evaluated the collinear $\widehat{B}_{\mT}$ functional at $\mT=5$, and the crossing equation sum rule $\widehat{X}_{\mS,\mT}$ at $\mS=\mT=5.01$. The collinear functional $\widehat{B}_{\mT}$ dominates at large spin. The convergence rate $\alpha$ for $J^{-\alpha}$ of the $\widehat{B}_\mT$, $\Phi_{0,2}$, $\Psi_0$, and $\widehat{X}_{\mS,\mT}$ functionals are $2.45$, $3.55$,  $3.54$, and $3.97$ respectively.}
\label{fig:mellin-convergence}
\end{figure}

Let us first consider the $\widehat{B}_{\mT}$ functional.
Given eq.~\eqref{Bt protected}, the protected part of this sum rule, we expect convergence to be bounded by the strip $4 < \mT <6$.
We can verify convergence of the OPE explicitly by taking the large spin and fixed twist limit of the Mack polynomials:
\begin{equation}
\widehat{B}_{\mT}[\tau+J,J] \propto \mathcal{Q}_{\D,J}^m(6-2m+\mT-\tau) \sim J^{-11/2+\mT - \tau} (...) + J^{5/2 -\mT} (...).
\end{equation}
The above suggest that the domain of convergence is saturated by the lowest-twist operator.
Since the OPE coefficients scale as $ \OPE_{\D,J} \sim J^{1/2}$, we conclude that swappability is guaranteed provided that $4< \mT < 6$.
This is further verified numerically at zero-coupling in the left plot of fig.~\ref{fig:mellin-convergence}.
(A similar conclusion was reached in appendix B of \cite{Alday:2022uxp}.)
This exercise further demonstrates that the convergence rate is fixed by the Mack polynomials at large spin,
and therefore it is invariant under Regge boundedness (anti)subtractions.

For $\widehat{X}_{\mS,\mT}$, the arguments of the Mack polynomials are now dependent on both Mellin-Mandelstam variables $\mS$ and $\mT$.
A natural expectation is that
convergence is allowed within the triangle-shaped domain ${\rm Re}\ \mS,\mT,\mU< 6$, which includes the symmetrical point $\mS=\mT=\mU=\frac{16}{3}$.
By evaluating the Mack polynomials in the fixed twist and large spin limit, we find that the forward-channel, the $(\mS \leftrightarrow \mU)$ crossed-channel, and the $(\mT \leftrightarrow \mU)$ crossed-channel blocks scale as follows:
\begin{subequations}
\begin{align}
\widehat{P}_{\mS,\mT}[\tau+J,J] &\propto \mathcal{Q}_{\D,J}^m(-2m+\mS+\mT-\tau) \sim  J^{-23/2+\mS+\mT - \tau} (\ldots) + J^{17/2-\mS -\mT} (\ldots) , \\
\widehat{P}_{16-\mS-\mT,\mT}[\tau+J,J] &\propto \mathcal{Q}_{\D,J}^m(16-2m-\mS-\tau) \sim   J^{9/2-\mS - \tau} (\ldots) + J^{-15/2+\mS} (\ldots).\\
\widehat{P}_{\mS,16-\mS-\mT}[\tau+J,J] &\propto \mathcal{Q}_{\D,J}^m(16-2m-\mT-\tau) \sim   J^{9/2-\mT - \tau} (\ldots) + J^{-15/2+\mT} (\ldots).
\end{align}
\end{subequations}
After accounting for the OPE coefficient and setting $\tau=2$, we conclude that the domain of convergence perfectly matches the triangle-shaped strip with ${\rm Re}\ \mS,\mT,\mU< 6$ !

While we are unable to obtain an analytic expression for the asymptotics of the projected functionals, 
their growth is expected to be bounded by the collinear $\widehat{B}_{\mT}$ functional given the nature of the projection operation.
This is confirmed numerically as shown in the right plot of fig.~\ref{fig:mellin-convergence}.

\subsection{List of functionals we used}

\begin{table}[t]
\centering
\begin{tabular}{c|c|c|c}
Functional & Range & Protected part & Equation(s) \\ \hline
$X_{u,v}$ & $(u,v)$ Euclidean &0 & \eqref{Xuv}, \eqref{cP from Mellin} \\
$B_v$ & $v>0$ real & %$f^{(0)}(v)=$
$\frac{v^2-1-2v\log v}{v(1-v)^3}$ & \eqref{Bv}, \eqref{Bv fast} \\
$\widehat{B}_\mT$ & $4< {\rm Re}(\mT)< 6$ & $\frac{2}{(\frac{\mT}{2}-3)(\frac{\mT}{2}-2)}$ & \eqref{Bt}, \eqref{Bt fast} \\
$\widehat{X}_{\mS,\mT}$ & ${\rm Re}(\mS,\mT,16{-}\mS{-}\mT)< 6$ & 0 & \eqref{Xst}, \eqref{P Mellin fast} \\
$\Phi_{\ell,\ell+2}$ & $\ell\geq 0$ even integer & 0 & \eqref{Phi features}, \eqref{Phi vector} \\
$\Psi_{\ell}$ & $\ell\geq 0$ even integer & $-\frac{2\Gamma(\ell+3)^2}{\Gamma(2\ell+5)}$ & \eqref{Psi features}, \eqref{Psi vector}
\end{tabular}
\caption{Complete list of sum rules we used to bootstrap single-trace operators,
including their allowed parameter ranges and protected contribution.
The Mellin functionals $\widehat{B}_\mT$ and $\widehat{X}_{\mS,\mT}$ allow complex parameters.
We include links to defining equations and to efficient evaluation formulas.
\label{tab:functional-list}
}
\label{tab:all functionals}
\end{table}

To summarize this section, we list the complete set of functionals we use to bootstrap single-trace OPE coefficients.
In particular, we use both position- and Mellin-space sum rules.

Although these functionals encode the same information in their respective spaces, the labelling of these functionals (cross-ratios $(u,v)$ in position-space versus Mellin-Mandelstam $(\mS, \mT)$ in Mellin-space) highlight how this encoding can lead to distinct functionals; generally, we put hats on Mellin functionals.

In each case, we produce a list of functionals by sampling a range of values.  When testing functionals on perturbative data, we observe that
numerical convergence is best achieved near the crossing-symmetric points, which are:
$z=\zb=-1$ for the position-space crossing functionals $X_{u,v}$,
$\mS=\mT=\mU=16/3$ for the Mellin-space Polyakov-Regge blocks $\widehat{X}_{\mS,\mT}$,
or $v=1$ and $\mT=5$ for the Mellin-space collinear functionals $B_v$ and $\widehat{B}_\mT$ respectively.
In principle we could consider derivatives around these points, but for our numerical implementation we find it easier
to sample a random selection of points to high numerical accuracy.
%Therefore, we instead use a basis of derivative functionals near the crossing-symmetric point;
%swappability of the functionals ensures that action of the derivative commutes with the sum.
The list of sum rules with appropriate ranges is recorded in table~\ref{tab:all functionals}.  

\section{Numerical bootstrap} \label{sec:bootstrap}

%In this section we explain how to use the functionals discussed in the preceding sections
%to bound OPE coefficients, given information about the single-trace spectrum.
In this section we describe numerical bounds on OPE coefficients obtained using
the functionals discussed in the preceding sections, given information about the single-trace spectrum.
We focus on the coefficient of the lightest unprotected scalar, the Konishi operator,
at weak and strong(ish) values of the coupling: $g=0.1$ and $g=0.3$.

\subsection{Generalities}

The main concept is similar to  OPE bounds in the numerical conformal bootstrap: we
look for combinations of functionals that have a definite sign on all allowed states.
The main difference is our choice of basis of functionals.
Traditionally, the numerical bootstrap exploits functionals that are derivatives of a crossing equation around
the crossing-symmetric point; as reviewed in introduction, these must be projected out to obtained $N_c$-independent
bounds in the planar limit.
Rather, we rely on the menu of dispersive functionals in table~\ref{tab:functional-list}.
Since we do not a priori know which one are most effective, we use a sample of all of them.
Since the different functionals behave differently in various limits (small twist, large twist, or large spin),
this strategy is intended to help the linear optimization problem and minimize computational resources.

%These parameters define an infinite family of functionals for each type, but we only consider finite linear combinations
%As shown in table~\ref{tab:functional-list}, each type of functional is parametrized by one or two variables: $(u,v)$ in position-space,  $(\mS,\mT)$ in Mellin-space, or $\ell \in 2\mathbb{N} $ for projected functionals.

%with different characteristics  though this set might not satisfy completeness in the limit that its size goes to infinity, given the special characteristic of the functionals inside it, there might still be an optimal linear combination of this finite set that gives us optimal bounds (saturated) on the OPE coefficient to a good accuracy.

Explicitly, let us label as $W_k$ any of the functional shown in table~\ref{tab:functional-list}
for some particular choice of its parameter.
Note that even when functionals allow continuous labels (like $(u,v)$ or $(\mS,\mT$)),
we only consider discrete choices that lie in the allowed ranges.
By swappability, each functional $W_k$ leads to a valid sum rule:
\be
 0=W_k^{\rm protected} + \sum_{(\Delta,\j)\ \rm long} \OPE_{\Delta,\j} W_k[\D,\j], \label{one sum rule}
\ee
and we can take finite linear combinations to get
 \be
 0=\sum_{k}\alpha_{k}\left(W_{k}^{\rm protected} + \sum_{(\Delta,\j)\ \rm long} \OPE_{\Delta,\j} W_{k}[\D,\j]\right).
 \label{many sum rules}
\ee
We then separate the protected part and the target OPE coefficient we want to bound, $\OPE_\Konishi$, and
impose that all other terms are positive.  Namely, we look for linear combinations $\alpha_k$ such that
\be
\hspace{7mm}\sum_{k} \alpha_{k}W_{k}[\Delta',\j']\geq 0\qquad \forall \quad (\Delta',\j') \mbox{ in single-trace spectrum},
\label{pos}
\ee
any of which proves an inequality
\be
%&\text{then}
\hspace{7mm} {-}\OPE_\Konishi\left(\sum_{k}\alpha_{k}W_{k}[\Delta_\Konishi,0]\right)
\geq \sum_{k} \alpha_{k}W_{k}^{\rm protected} \,.
\ee
The prime in $(\Delta',\j')$ indicates that the Konishi operator is omitted.

The optimal bound (for a particular finite list of functionals $\{W_k\}$) is thus found by solving
a standard linear optimization problem on the decision variables $\alpha_k$,
where one maximizes the right-hand-side subject to the inequalities \eqref{pos}
and a suitable normalization condition:
%we choose our decision variables $\alpha_{k_i}$ such that the following conditions are satisfied in addition to the above inequality:
\begin{equation}
\begin{split}
&\text{maximize} \qquad \sum_{k} \alpha_{k}W_{k}^{\rm protected} \\
&\text{such that}\qquad \sum_{k} \alpha_{k}W_{k}[\Delta_\Konishi,0]=\pm 1 \mbox{ and \eqref{pos} holds}.
\end{split}
\end{equation}
The plus (minus) sign yields a upper (lower) bound, respectively.
We use the SDPB solver to efficiently solve this type of problem \cite{Simmons-Duffin:2015qma}.

In principle, the inequality \eqref{pos} needs to be imposed on the infinite list of all single-trace operators of the theory.
In practice, we can only solve the quantum spectral curve for a finite set,
and in any case, we can only solve finite systems of inequalities.
Truncation is necessary.
Our solution is to identify \emph{ranges} of $\Delta$ where operators can be present for each spin,
and to impose positivity on a dense sample of discrete values of $\Delta$ in that range.
Discretization errors can be controlled by plotting the obtained functionals and ensuring that it does not
become
negative between the sampled points.\footnote{As is well known and exemplified in the next subsection,
optimal functionals typically develop double zeros, whose positions often coincide with those of actual operators.
We accept functionals that dip slightly below the real axis between discretization points, as long as their values are small.}

\begin{figure}[t]
\centering
\includegraphics[width=0.8\textwidth]{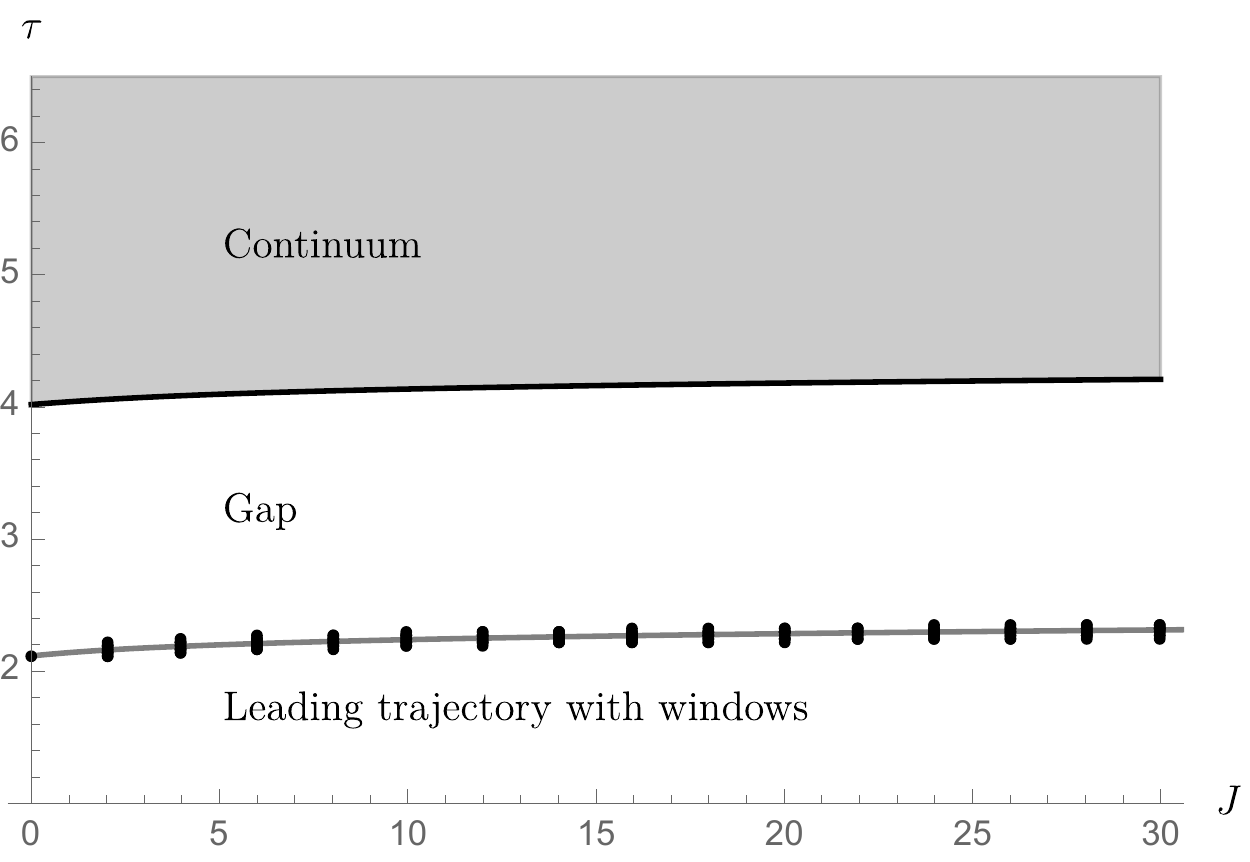}
\caption{Spectral assumptions for $g=0.1$.
For the leading trajectory, we interpolate between results from integrability at low spins and large spin asymptotics,
placing a conservative window around each operator.
For subleading trajectories, we demand that the optimized functional be non-negative in a continuum above a twist gap.
In this instance, we put exaggerated windows to see if the bootstrap discovers the spectrum.
}
\label{fig:weak spectrum}
\end{figure}

Typically, for each spin, we can make good estimates about the leading operator $\Delta_{\rm min}(J)$
and the gap to the next operator, which we call $\Delta_{\rm thresh}(J)$.
Thus, for each spin, we allow to operators in \eqref{pos}
to lie in a small window around $\Delta_{\rm min}(J)$ (with width determined by our error estimates),
or in a continuum at $\Delta>\Delta_{\rm thresh}(J)$.
Our spectral assumptions are further discussed below at weak and strong(ish) coupling.

Having only an incomplete spectrum implies that the inequalities we find, although conservatively valid,
may not be optimal. This discrepancy cannot be removed unless we know the exact spectrum of the theory.
Indeed it was observed in the study of 1d defects in $\mathcal{N}=4$ that including more
operators improves the numerics significantly \cite{Cavaglia:2022qpg}.

%We summarize our procedure in fig.~\ref{fig:bootstrap-approx}.
%Large-spin asymptotics can be computed analytically for low twist, while low-spin and low-twist operators can be computed using the QSC as shown in table~\ref{Frank's table}.
%This allows us to interpolate intermediate-spin operators at low twist.

%In addition, we also have information about the large spin asymptotic of the lowest twist operators, the gap between the first two Regge trajectories at large spin and the gap between.  We interpolate between the large spin asymptotics and the low spin results to obtain an estimate for the intermediate spin.

%Details of the truncation and discretization in twist is discussed in appendix.~\ref{}. 
%This cutoff needs to be large enough so that once we impose positivity up to it, we get positivity for $\tau>\tau_{\text{cutoff}}$ free. This would be part of the criterion for asymptotic positivity that was mentioned earlier. If we make sure that the asymptotic behaviour of the functionals as a function of twist  have kicked in for some $\tau<=tau_{\text{cutoff}}$ then this can be gauranteed to be true.

%We repeat our analysis for various coupling between $g\in[0.1,2.9]$ to obtain upper bound for Konishi at different coupling see fig.~\ref{}. In addition by 
%As we do not predict any error by going to stronger coupling we predict the saturation of our upper bound for stronger coupling as well.

\subsection{Bounds at weak coupling: toy problem with 2 functionals} \label{sec:toy model}

We start our analysis by looking at a toy example with only two functionals, namely $\Psi_0$ and $\Phi_{0,2}$. 
Indeed, in the one-loop approximation, these functionals respectively
compute the Konishi OPE coefficient \eqref{Konishi weak coeff} and the spin-2 anomalous dimension given the Konishi anomalous dimension
defined by eq.~\eqref{Konishi weak Delta}.
It is thus interesting to ask what they prove at finite but small coupling, ie. $g=0.1$.

Our spectral assumptions are shown in fig.~\ref{fig:weak spectrum}.
For the leading twist we impose positivity in a large window close to twist 2 without using any perturbative data.
We also add a conservative gap of $\Delta\tau=1.9$ between the leading twist and a continuum;
this is equal to the asymptotic gap at large spin (see \eqref{twist4}),
while the gap appears to decrease monotonically with spin.

We find that the optimal combination of these two functionals produces an upper bound that
is indeed quite close to the weak coupling OPE coefficient of eq.~\eqref{Konishi weak coeff},
and is almost saturated already by $\Psi_0$:
\be\begin{aligned}
\OPE_\Konishi(g{=}0.1) \leq& 0.30338 \qquad\mbox{from $\Psi_0$ alone} , \label{Psi bounds}\\
\OPE_\Konishi(g{=}0.1) \leq& 0.30255\qquad\mbox{from $\Psi_0$ and $\Phi_{0,2}$}.
\end{aligned}\ee
These can be contrasted with the free theory result $\OPE_\Konishi(g{=}0)=\frac13$ and 
the five-loop prediction $\OPE_\Konishi(g{=}0.1)^{\rm5-loop}=0.30067(1)$.
We recall that our OPE coefficients have $\frac{1}{c}\approx \frac{4}{N_c^2}$ factored out as shown by  eq.~\eqref{OPE scaling}.
Bounds with more functionals are discussed in the next subsection.

\begin{figure}[t]
\centering
\includegraphics[width=0.9\textwidth]{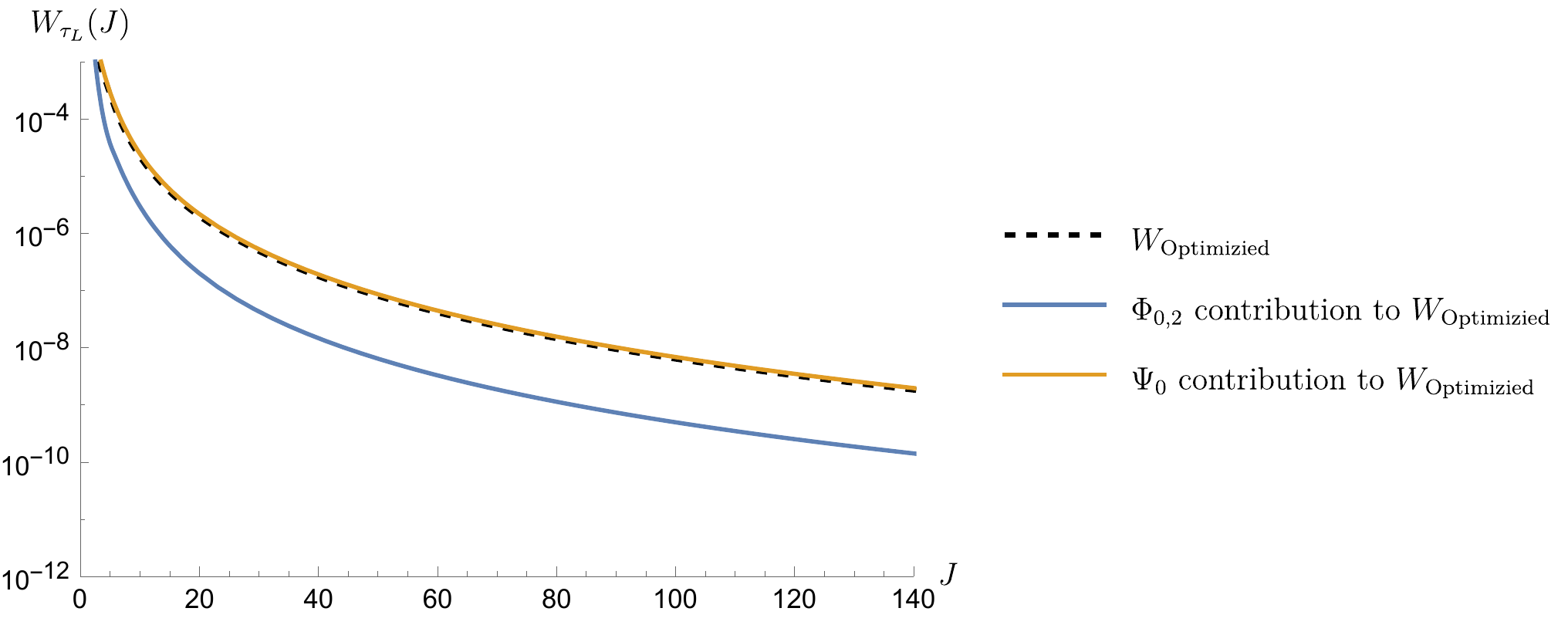}
\caption{Action of the two-term optimized functional $W^{g=0.1}_{\Psi_0+\Phi_{0,2}}$ on leading trajectory states,
showing the relative contributions of  $\Psi_0$ and $\Phi_{0,2}$ (in absolute value). Note the logarithmic scale. 
}
\label{fig:toy-ex}
\end{figure}

The optimal two-functional combination is
\be
W^{g=0.1}_{\Psi_0+\Phi_{0,2}}=0.90764 \Psi_0+0.025498\Phi_{0,2},
\ee
whose action on the leading trajectory states is displayed in fig.~\ref{fig:toy-ex}.
We explicitly see that $\Psi_0$ nearly saturates the action of this functional.

These bounds, which use functionals optimized for the one-loop problem,
improve one-loop perturbation theory in two respects.
First, they are numerically closer to the correct answer (in contrast with $\OPE_\Konishi(g{=}0.1)^{\rm1-loop}=0.29333$).
Second, and perhaps most importantly, they are rigorously valid  at finite $g$.
%In contrast to the one-loop approximation, which gives ,
%the bounds \eqref{Psi bounds} are not only numerically closer to the correct answer, they are rigorously valid independently of a perturbative approximation.
%Adding $\Phi_{0,2}$ to $\Psi_0$ gives the optimizer more degrees of freedom to capture perturbative corrections,
%and the difference between the bound and the actual OPE coefficient is comparable with the 3-loop term, $\sim 0.001$.

\subsection{Bounds at weak coupling: adding more functionals}
\label{sec:pert-num}

It is now interesting to add functionals from table~\ref{tab:functional-list} to see if the improved bounds capture higher-loop effects.
We consider two options: with 20 functionals and with 40 functionals.
Our 20 functionals consist of four $\Phi_{\ell_1,\ell_2}$, $\Psi_0$,
as well as five $\widehat{B}_\mT$ for $\mT$ evenly spread in the $(4,6)$ interval.
Moreover, we use ten $X_{u,v}$ for a set of ${u,v}$ satisfying chosen randomly
with a flat measure above the $s{\leftrightarrow}u$ symmetric curve to avoid redundancy.

The largest set of functionals we were able to use consists of 40 functionals including all 
types from in table~\ref{tab:functional-list}, except for $\widehat{X}_{s,t}$, which we found challenging to stabilize.
In addition to those in the preceding paragraph, we used the $\Psi_{2}$, $\Psi_4$ functionals,
nine more $X_{u,v}$, four $B_v$ with $v$ with 0.95, 1.01, 10 and 50,
and finally, five more $\widehat{B}_\mT$. We find that our bounds are stable with respect to adding more $X_{u,v}$'s,
however adding more $\widehat{B}_\mT$ or $B_v$'s could spoil the convergence with twist and spin
(and would lead to bounds which rules out the theory). Since $X_{u,v}$ are subdominant at large twist and $B$'s dominates in that region, we suspect that this limitation would be removed if we impose a tighter grid in that region or alternatively use an asymptotic formula.

To obtain robust bounds, we impose positivity up to a large $J_{\rm cutoff}\sim 6000$ for the leading trajectory and near the gap in the continuum.\footnote{To approach $J_{\rm cutoff}$ we extrapolate the functionals after they reach their asymptotic behaviour. For $g=0.1$ we exactly calculate up to $J\sim 150$ and then extrapolate.}
However, starting $\tau=20$ we use a smaller cutoff  ($J^{\tau>20}_{\rm cutoff}\sim 250$).
We sample twists up to a cutoff $\tau_{\rm cutoff}\sim 250$. We numerically check that this twist is large enough to ensure positivity on the asymptotic spectrum for the set of functionals considered.

We find that with this increased number of functionals, the upper bound at $g=0.1$ gets reduced to:
\begin{align}
\OPE_\Konishi(g{=}0.1) \leq  0.3018 \qquad \mbox{from 20 functionals}, \\
\OPE_\Konishi(g{=}0.1) \leq  0.3015 \qquad \mbox{from 40 functionals},
\end{align}
again to be compared with the five-loop estimate $0.30067(1)$.
The bound seems stable against increasing the number of $X_{u,v}$ functionals,
however it is not clear whether it has converged yet with respect to $B_v$ and $\widehat{B}_\mT$ functionals,
as mentioned above.

However, as we add more functionals, we gain a higher resolution of the leading-twist spectrum.
This means that despite the small change to the upper bound, the optimized functional is in fact manifestly different than the toy model optimized functional.

\begin{figure}[t]
\centering
\includegraphics[width=0.6\textwidth]{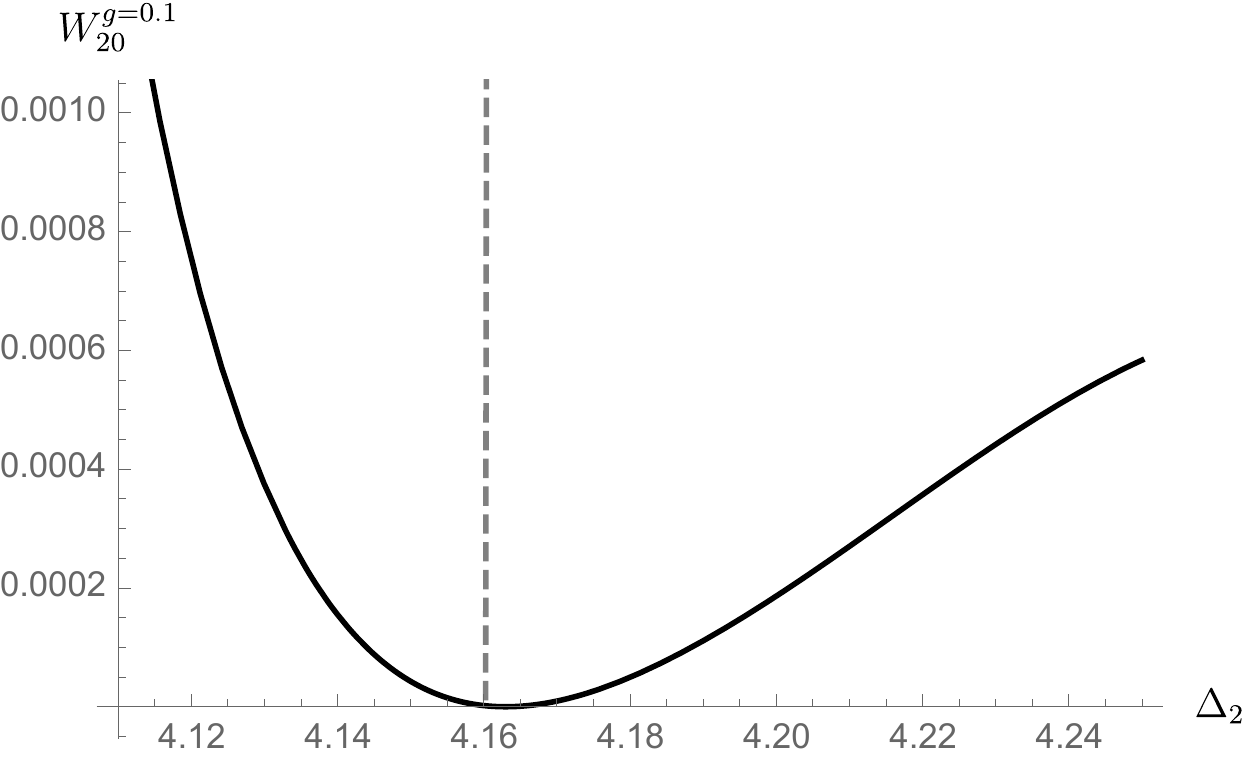}
\caption{The action of the 20-parameter optimized functional inside the window around the leading-twist spin-2 operator for $g=0.1$.
It displays a double zero at $\Delta\approx 4.164$, where the functional has size $W_{20}^{g=0.1}\sim 10^{-13}$; 
the vertical dashed line shows the position of the actual operator at $\Delta_2\approx 4.1603$,
where $W_{20}^{g=0.1}\sim 10^{-6}$.
}
\label{fig:doublezero}
\end{figure}

To investigate our functionals' ability to resolve the spectrum, we put slightly exaggerated windows around the
expected position of single-trace operators (from perturbation theory), requiring positivity within these windows.
We then observe that the optimal functional develops double zeros at these positions.
In fig.~\ref{fig:doublezero}, we show the optimized functional near $(J,\tau)=(2,2)$;
the double-zero of the optimized functional probes the leading spin-2 single-trace operator in the spectrum.  
With 20 functionals, we are able to discover the first $6$ operators in the leading family.
With more functionals one expects of course to discover more states,
but we did not try exaggerated windows with 40 functionals, since our main goal was to bound $\OPE_\Konishi$.
We have not observed stable double-zeros on the subleading families.

\begin{figure}[t]
\centering
\includegraphics[width=0.45\textwidth]{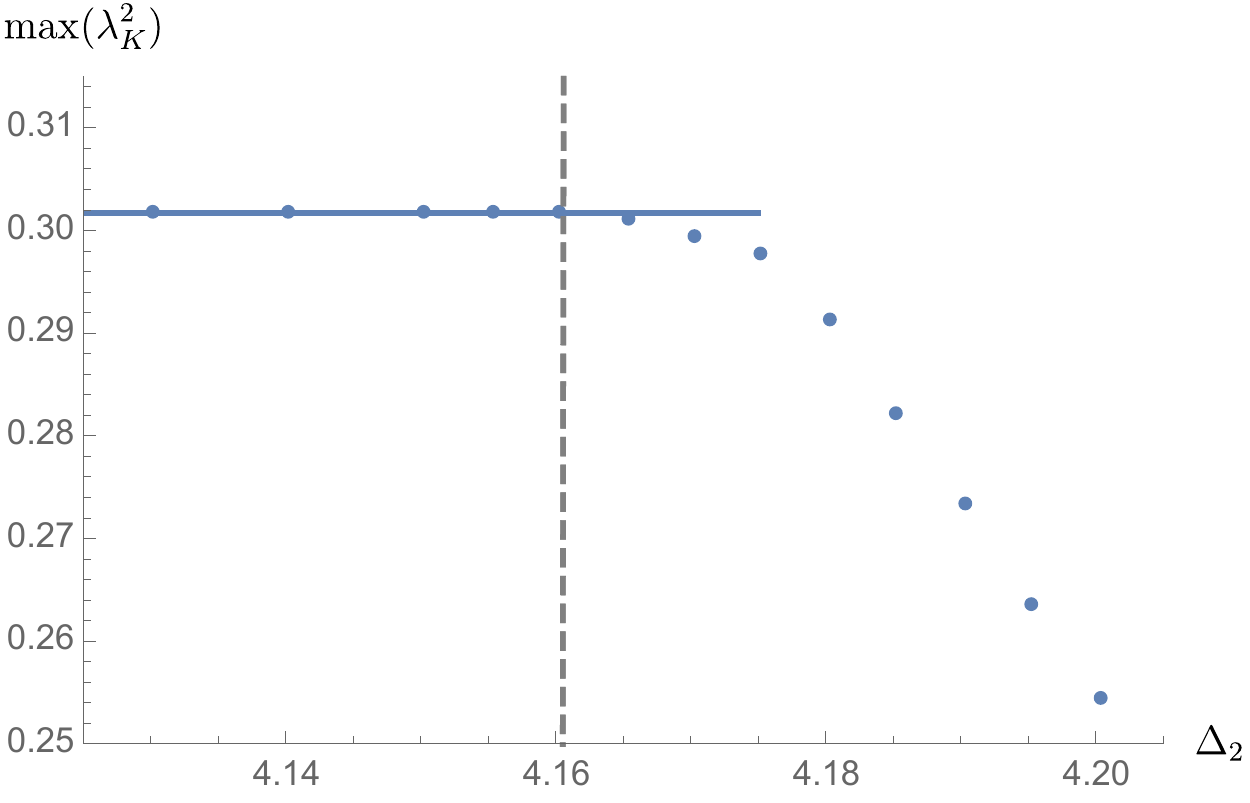}\qquad
\includegraphics[width=0.45\textwidth]{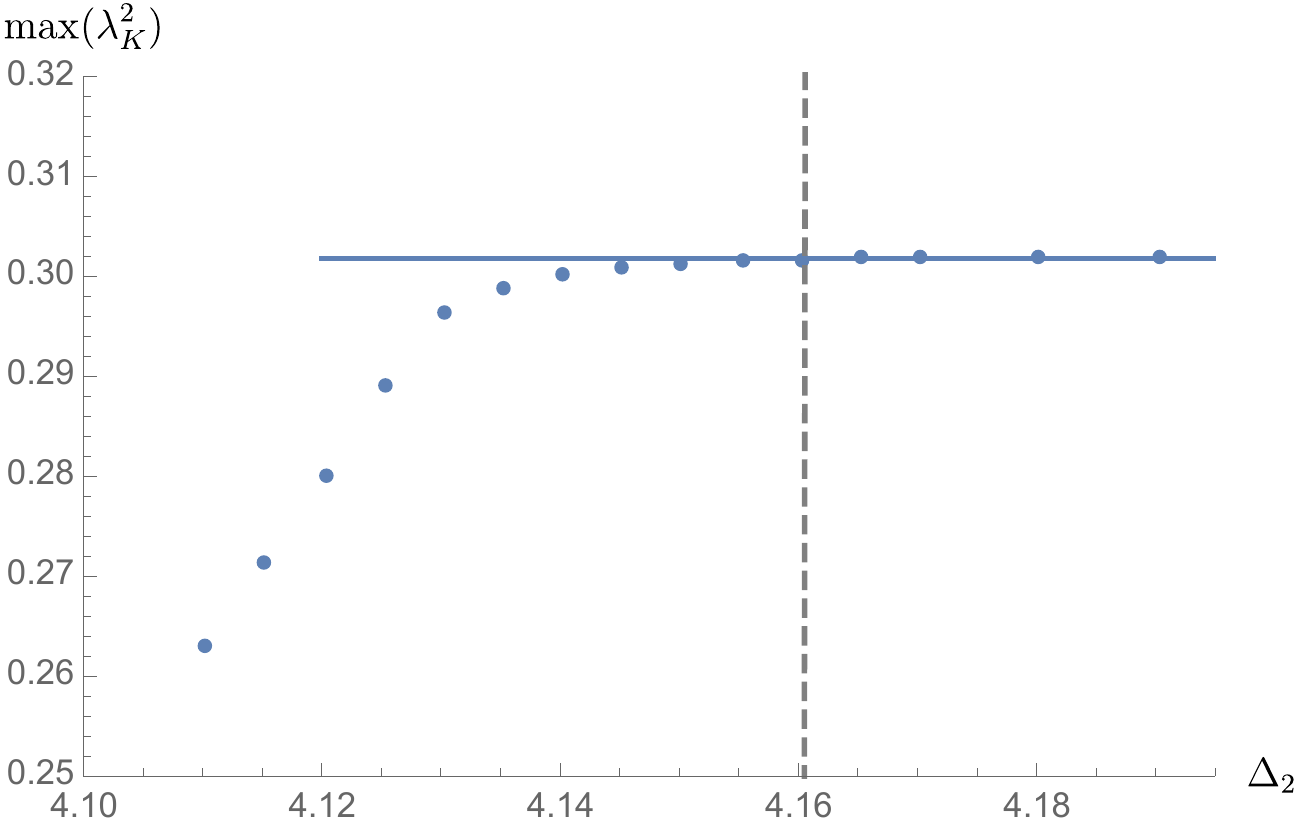}
\caption{Upper bound on the OPE coefficient of the Konishi operator as a function of the lower edge (left plot)
and upper edge (right plot) of the window, using 20 functionals.
The other edge was fixed to 4.56  and 3.76 respectively.
The horizontal line shows the bound when the window contains the physical spin 2 operator, which is exactly $\Delta_2$-independent.  As anticipated from figure \ref{fig:doublezero}, the bounds display kinks
(non-analyticity) near the physical value of $\Delta_2$, shown as a dashed line.}
\label{fig:kink}
\end{figure}

An alternative way to resolve the spectrum is to adjust
the size of the positivity window imposed around a given operator;
as long as the physical operator lies inside the window, the upper bound on $\OPE_\Konishi$ should not vary.
Therefore, by adjusting the upper and lower edges of the positivity window, a kink should form exactly where the operator exits the window.
We observe precisely such kinks as shown in fig.~\ref{fig:kink} for the leading spin-2 operator.
Finally, it is worth mentioning that if the window includes the physical operator in the leading family operators,
its size does not affect our upper bound at all.
Furthermore, omitting windows around operators that are discovered did not change our upper bounds on couplings.

\begin{figure}[t]
\centering
\includegraphics[width=0.8\textwidth]{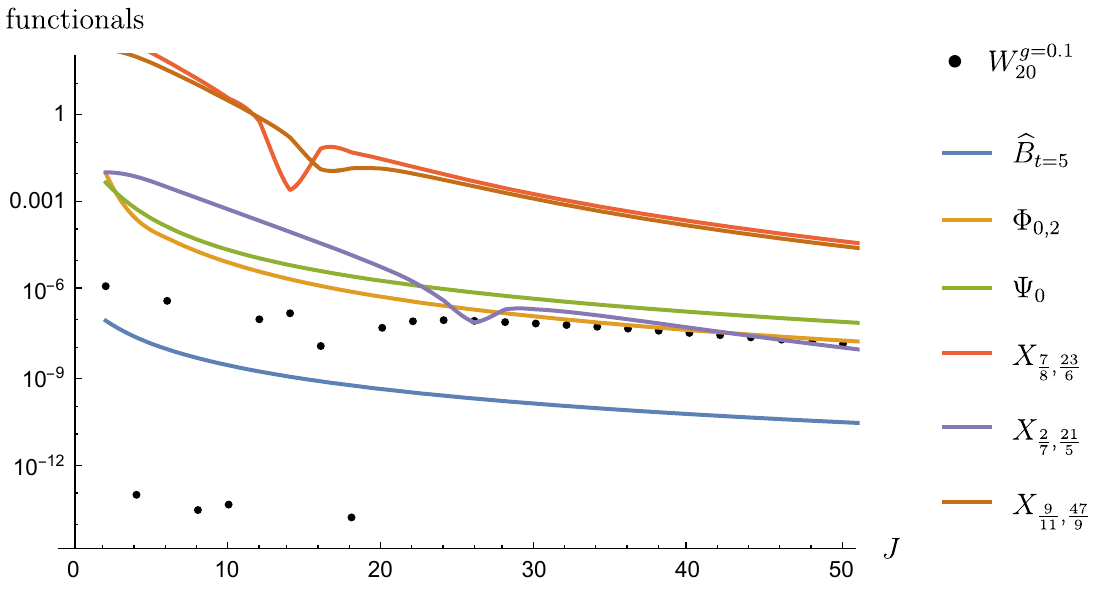}
\caption{The action of the 20-parameter optimized functional on the states belonging to the leading
Regge trajectory for $g=0.1$, as well as the absolute value of 6 (arbitrarily chosen) individual contributions to it.
In contrast to fig.~\ref{fig:toy-ex}, the optimized functional is qualitatively different from any of its individual constituents.} 
\label{fig:20-fnals}
\end{figure}

In fig.~\ref{fig:20-fnals}, we illustrate the action of the individual functionals and the optimized one on the leading-twist states.
This is qualitatively different from the toy model example considered previously, and the optimized functional is distinct from all individual functionals used in the optimization problem.

We repeat our procedure for other weak coupling values such as $g=0.2$,
using the spectral data summarized in table \ref{tab:data}.
For this coupling, we again find a stable upper bound, essentially independent on the number of functionals:
\be
\OPE_\Konishi(g{=}0.2) \leq  0.26,
\ee
which may be compared with the 5-loop estimate $0.25(2)$.

Lastly, let's  briefly discuss lower bounds on the Konishi operator OPE.
%We have performed the numerical implementation of this problem in a similar fashion to the upper bound. 
Proceeding similarly to the above, we were unable to obtain a non-trivial (nonnegative) lower bound %for this spectrum assumption
with the set of functionals at hands. 
We have experimented with including different sets of functionals, as well as with
different spectral assumptions and even tried unphysically large twist gaps. 
We observe that lower bounds converge towards the perturbative value only when we impose an
infinite gap for spin 2 and spin 0 operators.
What improvements are necessary to obtain a lower bound with realistic spectral assumptions is still an open question.

\subsection{Stronger coupling: $g=0.3$} \label{sec:bootstrap strong}

Having analyzed the weak coupling regime extensively, we are now ready to take the next step
and move to values for which perturbation theory is not expected to converge, namely $g>\frac14$.
As we will see, accurate spectral information becomes increasingly essential.

\begin{figure}[t]
\centering
\includegraphics[width=0.8\textwidth]{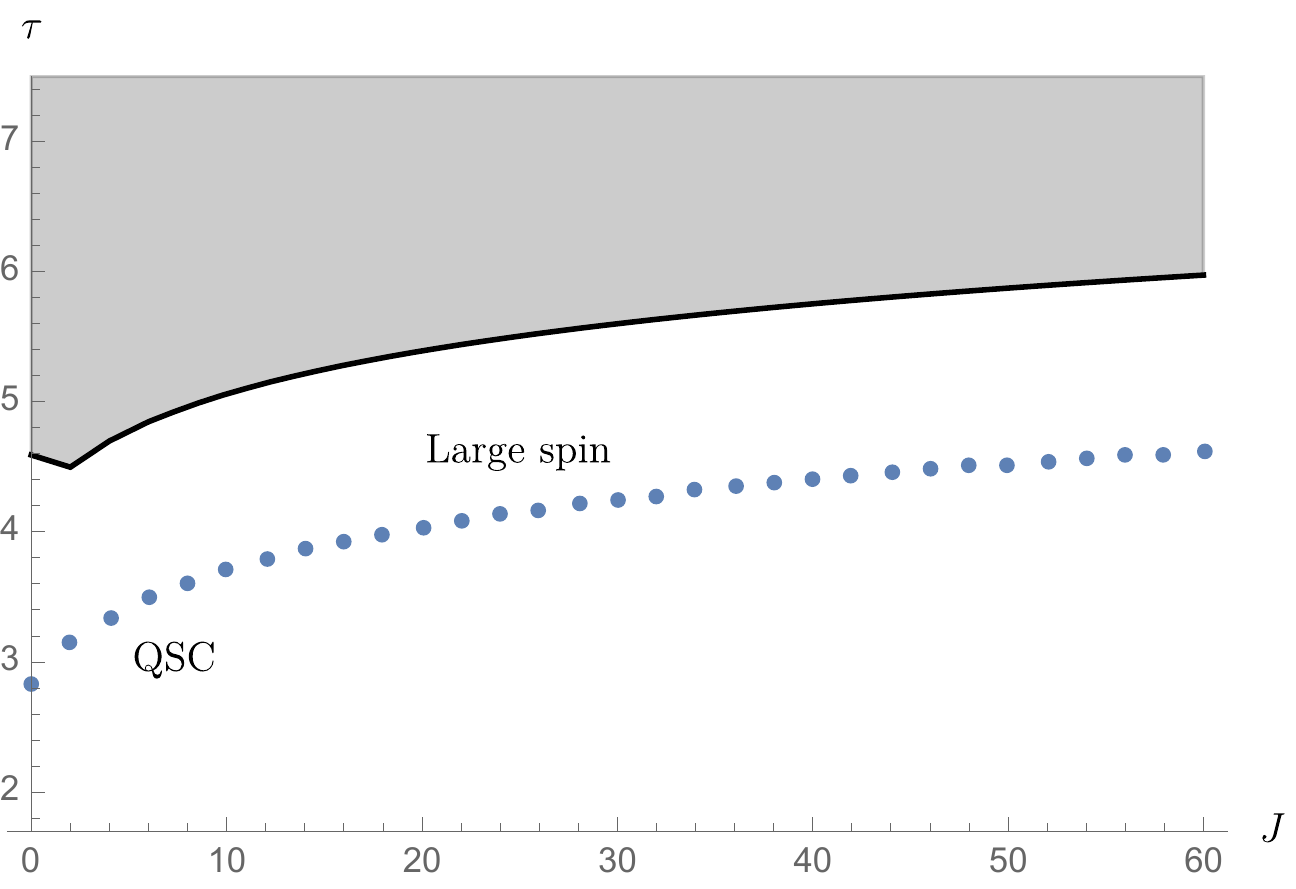}
\caption{Spectral assumption for $g=0.3$.
We use the QSC results up to spin 10, and for higher spins we use the large-spin asymptotics.
Our estimated error is less than 0.0005 for each leading twist operator and is not visible on the plot.
For the subleading trajectory, we conservatively use the large-spin gap for $\j\geq 2$.
}
\label{fig:spectrum-strong}
\end{figure}

Our spectral assumptions are shown in fig.~\ref{fig:spectrum-strong}.
For the leading trajectory, we used precise results from the QSC up to spin $\j=10$, followed by
the improved large-spin asymptotics \eqref{large-spin}.
The latter has an estimated error smaller than $0.0005$
for the twist of higher spin states, and even gives per-mil accuracy for $\j=0$!
To ensure that our bounds are rigorously valid,
we include a window of size $\pm 0.0005$ around each operator up to spin 146.

For the subleading trajectory we use the spin 0 gap derived in table~\ref{tab:data},
and we use the conservative large spin gap for $J\geq2$.
We use the same $J_{\rm cutoff}$ and $\tau_{\rm cutoff}$ as in the preceding subsection.
The only difference is that in order to obtain the asymptotic large spin behavior, we need to go higher in spin
(as can be seen from comparing fig.~\ref{fig:spectrum-strong} and fig.~\ref{fig:weak spectrum}).
We conservatively choose to calculate up to spin 250 and then extrapolate.
We emphasize that we do not have complete control over the large twist, large spin region; for instance, for $20\leq\tau\leq \tau_{\rm cutoff}$ we are unable to reach the large-spin asymptotic region and extrapolate. This could potentially affect the rigour of our bound. Further analysis is needed to completely tame this region.

Accurate spectral information seems important.  In our earliest attempts, we did not use the improved form of
large-spin asymptotics \eqref{large-spin} which led to much larger error estimates
(and larger window sizes ${\sim}0.02$ near $\j=10$).
Using the more accurate spectrum immediately impacted the bounds.

%We vary $\j_{max}$ and $J_{\rm windows}$ to obtain 4 different bounds, as shown in table~\ref{tab:upperbound-jmax}, to illustrate the dependence of the bound on the spectral data.  To obtain these bounds, we use the same $J_{\rm cutoff}$ and $\tau_{\rm cutoff}$. The only difference is that in order to obtain the asymptotic large spin, we need to go higher in spin (as can be seen from comparing fig.~\ref{fig:spectrum-strong} and fig.~\ref{fig:weak spectrum}). We conservatively choose to calculate up to spin 250 and then extrapolate.

\begin{figure}
\centering
\includegraphics[width=0.8\textwidth]{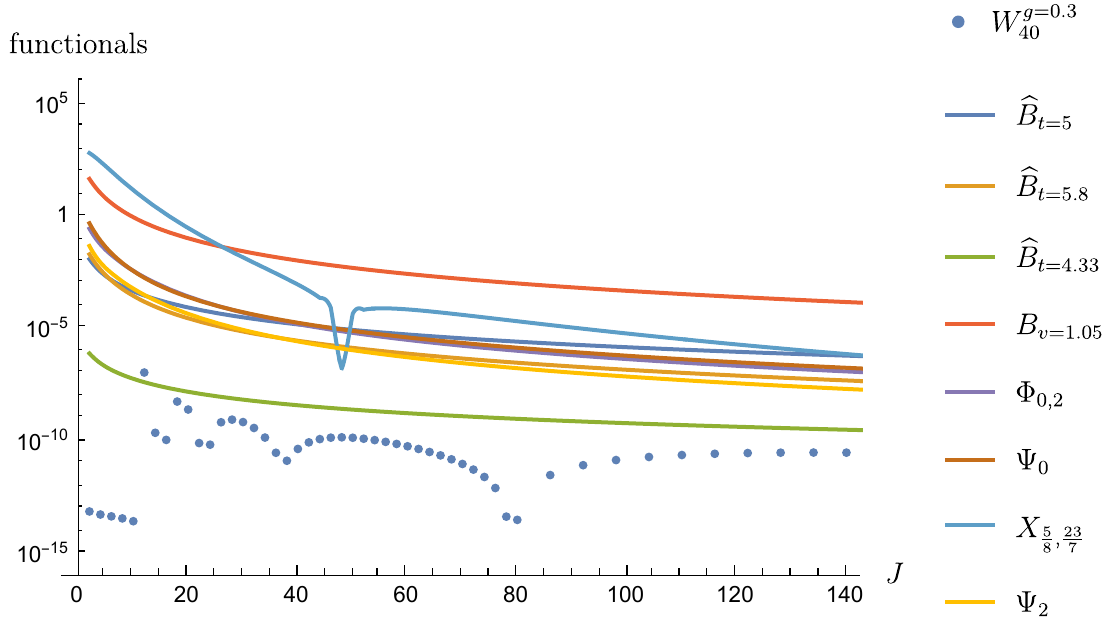}
\caption{
The action of the 40-parameter optimized functional on the states belonging to the leading
Regge trajectory for $g=0.3$, as well as the absolute value of some individual contributions to it.}
\label{fig:40-fnals}
\end{figure}

We use the same set of 40 functionals as described at the top of section \ref{sec:pert-num}.
As in the $g=0.1$ case, the optimal functional does not exhibit zeros near the subleading trajectory,
but it exhibits zeros on several operators on the leading trajectory.\footnote{
We did observe nontrivial dependence on the twist gap when using more $B$ functionals,
but we could not adequately control the large-spin large-twist region with these functionals.
}
This is plotted in fig.~\ref{fig:40-fnals}, along with the contributions from some individual functionals.

%The upper-bounds we get are then
%\begin{table}[h]
%\centering
%\begin{tabular}{c|c|c}
%$J_{\rm max}$ &  $J_{\rm window}$ &Upper-bound \\ \hline
%10 & 40 & 0.20  \\
%10 & 50 & 0.20  \\
%20 & 40 & 0.15  \\
%20 & 50 & 0.15  \\
%\end{tabular}
%\caption{Upper-bounds obtained for different assumption \st{on the unknown of the subleading trajectory}
%}
%\label{tab:upperbound-jmax}
%\end{table}
We quote the result we get with this data to be:
\begin{align}
\OPE_\Konishi(g{=}0.3)&\leq 0.300  \qquad\mbox{from 20 functionals}, \\
\OPE_\Konishi(g{=}0.3)&\leq 0.299  \qquad\mbox{from 40 functionals}. \label{strong bound}
\end{align}
As described at the end of section \ref{ssec:Konishi}, our current best estimates from other methods is
$\OPE_\Konishi(g=0.3)\in [0.24,0.33]$.
The bound \eqref{strong bound} exclude the upper part of that range.

\section{Discussion} \label{sec:discussion}

In this paper, we initiated a numerical study of nonperturbative constraints on
correlators of four stress tensor multiplets in planar four-dimensional $\mathcal{N}=4$ sYM.
Our methodology consists of three main steps.
First, we compute the scaling dimension of low-twist operators using integrability
as described in section~\ref{sec:integrability and QSC}; the resulting data that we used is summarized in table~\ref{tab:data}.
Second, we construct dispersive CFT functionals with distinct characteristics as summarized
in table~\ref{tab:all functionals}; each provides a nonperturbative sum rule on planar OPE coefficients.
Finally, we implement a numerical bootstrap algorithm to find linear combinations
of these sum rules that prove optimal bounds on the desired OPE coefficient, as detailed in section \ref{sec:bootstrap}.

Our main result is that at small but finite coupling, the OPE coefficient of the Konishi operator
satisfies a rigorous upper bound that is nearly saturated by the perturbative series.
As a function of other scaling dimensions, the bound displays kinks (see figure \ref{fig:kink}) which suggest that
with more functionals, the optimal bound will be saturated by the theory.
At stronger coupling $g=0.3$, outside the typical domain of convergence of the perturbative expansion,
we obtain bounds with conservative spectral assumptions.

We focus on the 't Hooft planar limit since exact spectra from integrability are only available in this limit.
On the other hand, this limit is challenging for traditional numerical bootstrap techniques, due to sign-indefinite double-trace contributions. The dispersive functionals described in section \ref{sec:functionals} avoid this problem by formulating
crossing directly at the level of single-trace data, by virtue of having double-zeros that suppress all double-trace operators.
We developed new technology to apply dispersive functionals in a numerical bootstrap context,
summarized in appendix \ref{app:efficient eval of funcs}, which could
be useful for applications to other (non-supersymmetric) conformal field theories.

There is a lot of flexibility in the space of dispersive functionals.
We constructed special combinations,
the projected functionals $\Phi_{\ell, \ell+2}$ and $\Psi_\ell$ (see eqs.~\eqref{Phi features}-\eqref{Psi features}),
which isolate specific twist-two anomalous dimensions and OPE coefficients in the weak coupling limit and determine them at one-loop. Significantly, the $\Psi_\ell$ are nonnegative, which ensure that they prove rigorous bounds at finite coupling.
By studying numerically combinations of more functionals, we find extremal combinations that
develop double zeros at the location of various operators (see section \ref{sec:pert-num}),
demonstrating that the functionals nonperturbatively constrain the spectrum.

%The benefits of these functionals are twofold:
%they decouple the double-trace operators appearing in the OPE, and they possess analytic properties that improves convergence of the numerical bootstrap, notably Regge boundedness, swappability, positivity, and a set of characteristic zero structure.
%The complete list of functionals used in this paper is summarized in table~\ref{tab:functional-list}.

%The role of these projected functionals was on full display  where we detailed our bootstrap analysis at weak coupling.
%By construction, the $\Psi_0$ and $\Phi_{0,2}$ functionals are the analytic results of the Konishi operator's OPE coefficient and anomalous dimension respectively at zero coupling.
%This was confirmed numerically as shown from our two-functional toy model described in section~\ref{sec:toy model}.
%In this setup, we found that the $\Psi_0$ functional saturated the optimized functional.
%Furthermore, we found that the bound is barely sensitive to adding more functionals to the optimizer. Nonetheless, the optimized functional is qualitatively different as evidence from a higher resolution of double-zeros near expected single-trace operators; see fig.~\ref{fig:spingg1over10}.
%This means that the spectrum can be resolved to higher accuracy and for a greater number of operators by providing more functionals to the optimizer.

As we increase the coupling and move beyond the convergence radius of perturbation theory, we observe that
the optimal functionals become qualitatively different from $\Phi$ and $\Psi$.
A main outstanding challenge to reach larger values of coupling is to gain more analytic
control over positivity of the optimal functional in asymptotic regions of large spin and large twist.
This will be needed to make completely rigorous the bounds presented in this paper.
Furthermore, we expect that this will enable the stable inclusion of more $B$-type functionals.
As discussed in subsection \ref{sec:bootstrap strong}, this could
enable the functionals to probe subleading Regge trajectories, whose scaling dimensions can be provided by the QSC.

We only obtain nontrivial upper bounds on OPE coefficients.
This is to be contrasted with the 1D Wilson line defects considered in \cite{Cavaglia:2021bnz,Cavaglia:2022qpg},
where tight lower bounds are also obtained.
It is possible that with more spectral information or more functionals this situation will improve.
An alternative scenario is that, like in the 3D Ising model, bootstrapping the theory down to an isolated island
will require to study more correlation functions \cite{Kos:2014bka}.

Our analysis was performed in the strict planar limit: all the sum rules we used are homogenous in $N_c$.
It could be interesting to compare results with the conventional numerical bootstrap at some large but finite $N_c$.

We anticipate many possible extensions of our analysis.
First, with some optimization it should be possible to find from the QSC the scaling dimension of many more operators,
with higher spin and/or higher twist.
Second, it might be possible to derive polynomial approximations for the $\Delta$-dependence of dispersive functionals,
which would make it possible to include a vastly larger number of functionals.
Such approximations were a crucial step in the development of the modern numerical bootstrap \cite{Poland:2011ey},
and this technology would also be a key step toward applying dispersive functionals to other models.
Third, it may be possible to incorporate sum rules beyond those we considered, for example
the integrated constraints from localization \cite{Binder:2019jwn,Chester:2020dja},
or constraints from mixed correlators involving other half-BPS operators.
It remains to be determined if the method can yield tight two-sided bounds on OPE coefficients.

%as proposed in 
%QSC + conformal bootstrap = solution of SYM
%We could eventually only use this information and try to fix the spectrum and OPE coefficients with the optimal functional. 

\section*{Acknowledgments}
We thank Fernando Alday, Shai Chester, Miguel Paulos, Jo\~{a}o Silva and David Simmons-Duffin for insightful comments.
All authors are supported by the Simons Foundation through the Simons Collaboration on the Nonperturbative Bootstrap.
Work of SCH is additionally supported by the Canada Research Chair program and the Sloan Foundation, while
AT is further supported by the Natural Sciences and Engineering Research Council of Canada.
ZZ is also supported by the Fonds de Recherche du Qu\'ebec--Nature et Technologies.
This research was enabled in part by support provided by Calcul Qu\'ebec and Compute Canada (Narval and Graham clusters).

\begin{appendix}

\section{Additional formulas from integrability}

To obtained the large-spin asymptotics quoted in table \ref{tab:data},
we used the following formulas from the asymptotic Bethe Ansatz, from 
\cite{Basso:2010in,Basso:2013aha}.
The essential step is to build the (infinite) matrix, which effectively inverts the Beisert-Eden-Staudacher (BES) kernel \cite{Beisert:2006ez}:
\beq
\mathcal{K}_{ij} \,=\, (-1)^{(i+1)j}2j \int_0^\infty dt \frac{J_{i}(2 g t)J_{i}(2 g t)}{t\,(e^{t}-1)}
\eeq
By inverting the matrix $(1+\mathcal{K})$ and dotting into suitable vectors, one finds the cusp and virtual anomalous dimensions
\begin{align}
\Gamma_{\text{cusp}} &= 4 g^2 [\,Q\cdot(1+\mathcal{K} )^{-1}\,]_{11} \label{cusp}\\
\Gamma_{\text{virtual}} \,&=\, 4 g\, [\,Q\cdot(1+\mathcal{K})^{-1}\cdot \mathcal{V} ]_{1}\,,
\label{virtual}\qquad
\mathcal{V}_{i} \,\equiv\, \int_0^\infty dt \frac{J_{i}(2 g t)J_{0}(2 g t)-g\,t\,\delta_{i 1}}{t\,(e^{t}-1)} \,.
\end{align}
with $Q_{ij}=j(-1)^{j+1}\delta_{i,j}$.
Similarly, the ground state energy of a scalar excitation is
\be
E_{\phi} = 1+ 4 g\, [\,Q\cdot(1+\mathcal{K})^{-1}\cdot\mathcal{V}^\phi\,]_{1},
\qquad 
\mathcal{V}^{\phi}_i\equiv  \int_0^\infty dt \frac{J_{i}(2gt)(J_{0}(2gt)-e^{\frac{t}{2}})\,}{t(e^{t}-1)}.
 \label{Ephi}
\ee
In practice, these are calculated by truncating the matrix $\mathcal{K}$ to a finite size
and extrapolating the results to infinite size.  For the relatively small values of the coupling that we consider,
convergence is fast.

Since the leading twist-4 operator has not been discussed earlier, we also record weak coupling QSC data corresponding to it,
extracted from the code of \cite{Marboe:2018ugv} solved with quantum numbers $\{\{0,0\},\{2,2,2,2\},\{0,0\}\}$
(after rescaling and shifting $P_3^{\text{there}}$ by a multiple of $P_1^{\text{there}}$ to make explicit the left-right symmetry of the QSC)
\begin{align}
\Delta \,&=\, 4+ 6.60 g^2+ 2.41 g^4 +\mathcal{O}(g^6)\nonumber \\
(gx)^2 P_{1} \,&=\, (-6.7-43 g^2 -100 g^4)+\frac{-18 g^2 -98 g^4}{x^{2}}+\frac{0}{x^4} +\frac{0}{x^6}\nonumber\\
g x P_{2} \,&=\, (-36-170g^2-350g^4)+\frac{-42-260 g^2-960 g^4}{x^2}+\frac{-32 g^2-430 g^4}{x^4}+\frac{0}{x^6}\nonumber\\
P_{3} \,&=\, 1+\frac{\frac{0.34}{g^2}+3.3+4.6 g^2+9.5 g^4}{x^2} +\frac{1.4+6 g^2+3.8g^4}{x^4}+\frac{0.45g^2+8.1g^4}{x^6}\nonumber\\
\frac{P_4}{g x} \,&=\, 1+ \frac{0}{x^2}+\frac{-\frac{0.28}{g^2}-3.1-9.5g^2-19g^4}{x^4}+\frac{0.45-2.7g^2-3.9g^4}{x^6}+\frac{0.27g^2+5.4g^4}{x^8}\,. \label{twist 4 seed}
\end{align}
As mentioned in the main text, we confirmed that this is indeed the twist-four, spin-0 operator which is exchanged between
stress tensor multiplets, by computing its OPE coefficient directly at weak coupling.
For this we worked out the corresponding eigenstate of the 1-loop dilatation operator, using the Hamiltonian from \cite{Beisert:2004ry} (eq. 3.6 therein).
Including double-trace terms, there are four color-singlet operators involving 4 scalars and no derivatives,
\be
 \mathcal{O}_\Delta =
 A \tfrac{1}{N_c^2}{\rm Tr}\big[\phi^i\phi^i\phi^j\phi^j\big]+
 B\tfrac{1}{N_c^2} {\rm Tr}\big[\phi^i\phi^j\phi^i\phi^j\big]
+C \tfrac{1}{N_c^3}{\rm Tr}\big[\phi^i\phi^i\big]\big[\phi^j\phi^j\big]
+D \tfrac{1}{N_c^3}{\rm Tr}\big[\phi^i\phi^j\big]\big[\phi^i\phi^j\big],
\ee
and we find that the eigenfunction corresponding to $\Delta=4+g^2(13-\sqrt{41})+O(g^4)$ is
\be
\left(A,B,C,D\right) \propto \left( \sqrt{41}-5,\ 4,\ \tfrac{1}{2}(35-5\sqrt{41}),\ {-}\sqrt{41}-9\right) + O(g^2,1/N_c^2). \label{eigen}
\ee
By normalizing the two-point function of $\mathcal{O}_\Delta$ and
computing its Wick contractions with two protected operators $\mathcal{O}(x_1,y_1)\mathcal{O}(x_2,y_2)$ of the form
\eqref{def O}, we obtained the tree-level OPE coefficients \eqref{C4}. It was crucial in this calculation to retain the double-trace terms,
which do contribute to the planar OPE coefficient. These are extremal three-point functions in the free-theory
(the sum of twists of two operators is equal to the twist of the third operator)
which are not predicted by the tree-level hexagon formulas \cite{Komatsu:2017buu,Eden:2016xvg}.

\section{Projection Functionals} \label{app:projection functionals}

In this appendix we detail the construction of projection functionals, which diagonalize the action of the $B_v$ (or $\widehat{B}_\mT$)
on twist-two operators.  The idea is exploit orthogonality relations for the polynomials which control
their action on twist-two operators.

\subsection{Action of $\widehat{B}_\mT$ on operators near twist two}

%As discussed in \cite{Caron-Huot:2020adz}, the collinear functionals can be written as an expansion of double-twist analytic functionals:
When acting on an operator with twist close to two, the functional of can be expanded as
\be \label{Bt from a b}
\tfrac12\lim_{\tau\to 2} \widehat{B}_{\mT}[\ell,\ell+\tau]  = a_{\ell}(\mT) + (\tau-2) b_{\ell}(\mT) + O((\tau-2)^2).
\ee
The factor $\tfrac12$ is included for compatibility with the literature.
The two shown terms are insensitive to descendants, thanks to the double-zeros of $\widehat{B}_\mT$,
which suppress operators of twists 4,6\ldots. Thus, the Mellin representation \eqref{Bt} simplifies to
\be \label{Bt near twist 2}
\lim_{\tau\to 2} \widehat{B}_{\mT}[\ell,\ell+\tau]  =\lim_{\tau\to 2} \frac{2\tau+2-\mT}{\mT-6} \mathcal{Q}^0_{\j+\tau+4,\j}(10-\mT) + O((\tau-2)^2)\,.
\ee
Comparing, one thus finds $a_\ell(\mT)$ in terms of the Mack polynomial \eqref{Mack polynomials}, explictly
\begin{equation} \label{a Mack polynomial}
\begin{split}
a_{\ell}(\mT)&= \frac{\Gamma (2 \ell+6)}{\Gamma(3)^2\Gamma (\ell+3)^2} \, _3F_2\left(-\ell,\ell+5,\tfrac{\mT-2}{2};3,3;1\right).
\end{split}
\end{equation}
These polynomials are even under $\mT\mapsto 10{-}\mT$, and are normalized so they
obey the following orthogonality relation for $\ell \geq 0$:
\begin{equation} \label{aMack ortho}
c_{\ell} \int \frac{d\mT}{4 \pi i} \Gamma\!\left(4-\tfrac{\mT}{2}\right)^2\Gamma\!\left(\tfrac{\mT}{2}-1\right)^2 a_{\ell}(\mT)a_{\ell'}(\mT) = \delta_{\ell, \ell'},
\end{equation}
where
\begin{equation} \label{Mellin orthogonality normalization coeff}
c_{\ell} =\frac{\Gamma(\ell+3)^4}{\Gamma(2\ell+6)^2}(\ell+1)_4(2\ell+5)\,.
 %\begin{cases}
%\frac{(5+2\ell)\Gamma(5+\ell) \Gamma(3+\ell)^4}{\Gamma(1+\ell) \Gamma(6+2\ell)^2}.
%& \qquad \text{if } \ell \geq 0.
%\\3 & \qquad \text{if } \ell =-1, \\4 & \qquad \text{if } \ell=-2.\end{cases}
\end{equation}
These properties are similar to those of the leading-twist functional $B_{2,\mT}$ discussed in section~4
of \cite{Caron-Huot:2020adz},
although the details differ slightly since here we are interested in diagonalizing the action on twist-two operators, rather than on the leading double-twists.\footnote{
The definition of $a_\ell$ and $c_\ell$ here coincides with the $\Delta_\phi=3$ case of \cite{Caron-Huot:2020adz}.}

The slope $b_\ell(\mT)$ is more complicated.  For our applications below, it will be useful to know its integrals against $a_j$.
To this aim, we decompose it into odd and even parts.
The odd part can be computed directly using symmetries of \eqref{Bt near twist 2},
while for the even part we adopt an expansion over Mack polynomials:
\begin{align} \label{bEll odd relation}
b_{\ell}(\mT) - b_{\ell}(10-\mT) &= -g(\mT)^{-1} \frac{d}{d\mT} \Big[ a_{\ell}(\mT) g(\mT)\Big],\qquad g(\mT)\equiv (-6+\mT)^2 (-4+\mT)^2,\\
\label{bEll even relation}
b_{\ell}(\mT) + b_{\ell}(10-\mT) &= b_{\ell,-2} a_{-2} (\mT) + \sum \displaylimits_{j \in 2\mathbb{N}}^{\ell} b_{\ell,j} a_{j}(\mT).
\end{align}
In the second term we included a $j=-2$ term to account for the polar part of $b_\ell$:
\be
 a_{-2}(\mT) = \frac{2}{(-6+\mT)(-4+\mT)}\,.
\ee
From investigation of many cases, we find explicitly the coefficients
\begin{align}
b_{\ell,-2} &= \frac{-2 \Gamma(2\ell+6)}{\Gamma(\ell+3)^2(\ell + 1) (\ell + 4)} , \\
b_{\ell,j} &= \bigg( \frac{(\ell+2)^2(\ell+3)^2}{(j+2)(j+3)}-\frac{(\ell+1)^2(\ell+4)^2}{(j+1)(j+4)}\bigg)
\frac{1}{(\ell-j)(\ell+j+5)} \frac{\Gamma(j+3)^2\Gamma(2\ell+6)(j+1)_4}{\Gamma(\ell+3)^2\Gamma(2j+5)(\ell+1)_4},\\
b_{\ell, \ell}&=
%\begin{cases}-43/30, \qquad &\text{if } \ell=0, \\  % SCH: the general formula actually works for j=-2.
4H_{2\ell+4}- 6H_{\ell+2}- 2\frac{23+17\ell+3\ell^2}{(3+\ell)(4+\ell)(5+2\ell)}, % \qquad  & \text{otherwise,} \end{cases}
\end{align}
where the second line is for the generic case $0\leq j<\ell$.
The factor outside the parenthesis matches the unsubtracted dispersion relation (4.45) of \cite{Caron-Huot:2020adz}) evaluated at $\D_\phi=3$.

This representation enables us to find simply the integral of $b_\ell(\mT)$ against an even-spin $a_j$.
The slight subtlety is that is that the non-polynomial function $a_{-2}(\mT)$ is not orthogonal to the polynomial ones, rather:
\begin{equation}  \label{Integral -2 against j}
%I_{-2,j} \equiv
\int %\displaylimits_{5- i \infty}^{5+ i \infty}
\frac{d\mT }{4\pi i} \Gamma\!\left(4-\tfrac{\mT}{2}\right)^2\Gamma\!\left(\tfrac{\mT}{2}-1\right)^2 a_{-2}(\mT) a_j(\mT)  =
\frac{-\Gamma(2j+6)}{\Gamma(j+3)^2(j+1)_4} \qquad\mbox{($\ell\geq 0$ even)}\,.
\end{equation} 
Combining this with the orthogonality relation \eqref{aMack ortho} and expansion \eqref{bEll even relation},
we find nice cancellations such that for even $j$, the integral of a Mack polynomial $a_\ell$ against $b_j$ gives
\begin{align} \label{Iellj integral}
 I_{\ell,j} &\equiv \int
\frac{d\mT }{4\pi i} \Gamma\!\left(4-\tfrac{\mT}{2}\right)^2\Gamma\!\left(\tfrac{\mT}{2}-1\right)^2 c_\ell a_\ell(\mT) b_j(\mT)
\\
&=
\begin{cases}
\frac{1}{(j-\ell)(\ell+j+5)} \frac{(\ell+2)_2}{(j+2)_2}\frac{\Gamma(2j+6)\Gamma(\ell+3)^2}{\Gamma(2\ell+5)\Gamma(j+3)^2},
\qquad
& \mbox{for $0\leq \ell<j$ even},\\
2H_{2\ell+4} -3H_{\ell+2}+\frac{13+8\ell+\ell^2}{(\ell+1)(\ell+3)(2\ell+5)},
& \mbox{for $\ell=j$},\\
\frac{1}{(j+1)(j+4)}\frac{\Gamma(2j+6)\Gamma(\ell+3)^2}{\Gamma(2\ell+5)\Gamma(j+3)^2},
 & \mbox{for $\ell>j$ even}.
\end{cases}
\end{align}
This result, together with \eqref{bEll even relation},
will now be used to define various projectors; we will not further need the $b_{\ell,j}$ coefficients.

\subsection{One-loop anomalous dimensions: the $\Phi_{\ell,\ell+2}$ projection functionals}

In \cite{Caron-Huot:2020adz} a functional $\Phi_\ell$ was constructed analytically, which had double-zeros on all operators
of the first double-twist family of an arbitrary CFT, and single zeros on just the $J=\ell$ one.
The sign properties of that functional, for some $\ell$ and external operator dimensions, established that mean field theory
maximizes the twist gap for that spin.

Here we will construct functionals $\Phi_{\ell_1,\ell_2}$ which analogously has double-zeros near twist $\tau=2$,
except for two operators $J=\ell_1,\ell_2$, where it has a single zero.
This problem is similar, but distinct, from the one studied in \cite{Caron-Huot:2020adz} since here we insist to maintain
double-zeros on every double-twist $\tau\geq 4$.

As a first step, we attempt to construct a functional $\Phi_\ell$ with nonvanishing slope only for $\j=\ell$, by writing
it as an integral over $\widehat{B}_\mT$:
\be
 \Phi_\ell[\Delta,\j] \equiv \int \frac{d\mT}{4\pi i} \Phi_\ell[\mT] \ \widehat{B}_\mT[\Delta,\j]. \label{Phi func ansatz}
\ee
We will find that the kernel $\Phi_\ell[\mT]$ does not lead to a convergent integral unless we allow
a nonvanishing slope on at least two spins.
The construction follows that in \cite{Caron-Huot:2020adz}, although $\Phi_\ell$ here is slightly different.
The conditions on $\Phi_\ell$ are that, for every even $\j\geq 0$,
\be \label{Phi conditions}
 \int \frac{d\mT}{2\pi i} \Phi_\ell[\mT] a_\j(\mT) = 0,\qquad
 \int \frac{d\mT}{2\pi i} \Phi_\ell[\mT] b_\j(\mT) \stackrel{?}{=} \delta_{\ell,\j}\,.
\ee
Note that a factor 2 arose from \eqref{Bt from a b}.
The question mark emphasizes that $\Phi_\ell$ will be obtained by ignoring convergence constraints, which will be addressed below.

First, we observe that first of eqs.~\eqref{Phi conditions} can be satisfied if the kernel is odd: $\Phi_\ell[10{-}\mT]=-\Phi_\ell[\mT]$.
Since the $a_\j$'s form a complete basis for a reasonable function space, we expect this to be the only solution.
For an odd kernel, the second condition, about the slope, can be computed from \eqref{bEll odd relation} as
\be
 \int  \frac{d\mT}{2\pi i} \Phi_\ell[\mT] b_\j(\mT) = 
-\frac12 \int  \frac{d\mT}{2\pi i} \Phi_\ell[\mT] g(\mT)^{-1} \frac{d}{d\mT} \Big[ a_{\j}(\mT) g(\mT)\Big]\,.
\ee
Assuming that $\Phi_\ell[\mT]$ vanishes at infinity faster than any polynomial
(otherwise the functional \eqref{Phi func ansatz} doesn't make much sense), we integrate by parts to
write the second condition as
\be
\delta_{\ell,\j} = \frac12\int \frac{d\mT}{2\pi i} \left(g(\mT)\frac{d}{d\mT} \big[\Phi_\ell[\mT] g(\mT)^{-1}\big]\right) a_{\j}(\mT)\,.
\ee
Comparing with the orthogonality relation \eqref{aMack ortho},
we deduce that the parenthesis should be proportional to $c_\ell a_\ell(\mT)$,
specifically
\be
g(\mT)\frac{d}{d\mT} \big[\Phi_\ell[\mT] g(\mT)^{-1}\big] =
\Gamma\!\left(4-\tfrac{\mT}{2}\right)^2\Gamma\!\left(\tfrac{\mT}{2}-1\right)^2 c_\ell a_\ell(\mT)\,.
\ee
This differential equation admits a unique solution, given that $\Phi_\ell[\mT]$ is odd and so must vanish at $\mT=5$:
\begin{align}
  \Phi_\ell[\mT] &= g(\mT) \int_5^{\mT} d\mT' \frac{c_\ell a_\ell(\mT')}{g(\mT')}\Gamma\!\left(4-\tfrac{\mT'}{2}\right)^2\Gamma\!\left(\tfrac{\mT'}{2}-1\right)^2
  \\ &= \frac{i\pi^2}{16}(y^2+1)^2\int_0^{y} \frac{dy'\ c_\ell a_\ell(5+iy')}{\cosh(\tfrac{\pi y'}{2}\big)^2}
\label{Phi ell kernel}
\end{align}
where we let $\mT=5+iy$ in the second line.
The first few cases can be found analytically, for example
\begin{align}
\Phi_0[5+iy] &= i\frac{\pi}{8}(1+y^2)^2 \tanh(\tfrac{\pi y}{2}), \\
\Phi_2[5+iy] &= i\frac{(1+y^2)^2}{448\pi} \Big( 28\left(\pi y \log(1{+}e^{-\pi y})-\text{Li}_2(-e^{-\pi y})+\tfrac{\pi^2}{12}(3y^2{-}1)\right)
+\pi^2(9{-}7y^2) \tanh(\tfrac{\pi y}{2})  \Big) ,
\end{align}
which are indeed odd functions of $y$.
The only issue with this calculation is that the $\Phi_\ell$ kernels don't actually vanish as $y\to\infty$:
the corresponding functionals do not make sense as the integral \eqref{Phi func ansatz} fails to converge at large $y$.
The solution is to take finite linear combinations.
By investigating the large-$t$ limit of the kernels we find the following simple formula:
\begin{equation} \label{boundary term}
\lim_{t \rightarrow 5+ i \infty} \frac{\Phi_\ell[\mT]}{i \pi g(\mT)} \equiv \Phi_\ell^{\infty} =
\frac{\Gamma(\ell+3)^2}{2\Gamma(2\ell+5)}H_{\ell+2}.
\end{equation}
Since the integrand in \eqref{Phi ell kernel} decays exponentially, the corrections to the limit are proportional to $e^{-\pi y}$.
Thus, to define valid functionals, it suffices to combine any two kernels $\Phi_\ell[\mT]$ so as to cancel the constant:
\be
 \Phi_{\ell_1,\ell_2}[\mT] \equiv \Phi_{\ell_1}[\mT] - \frac{\Phi_{\ell_1}^\infty}{\Phi_{\ell_2}^\infty} \Phi_{\ell_2}[\mT]\,. \label{Phi kernel}
\ee
We have normalized it so it has unit slope as $\tau\to 2$ when acting on operators of spin $\j=\ell_1$.
The $\Phi_{\ell_1,\ell_2}$ are of course not all linearly independent, they are spanned
by the $\Phi_{\ell,\ell+2}$ considered in the main text.
A convenient formula for their evaluation is discussed in \ref{app:PhiPsi eval} below.

As discussed in the main text below \eqref{Phi one-loop},
the combination \eqref{Phi kernel} admits a simple interpretation in terms of one-loop anomalous dimensions,
since the one-loop sum rule forces the following proportionality (with an $\ell$-independent constant):
\be
\Phi_\ell^{\infty}\propto C^{(0)}_\ell \gamma^{(1)}_\ell\,.
\ee
This is indeed satisfied by \eqref{boundary term}.
In other words, the $\Phi_{\ell_1,\ell_2}$ sum rules analytically prove the one-loop formula for anomalous dimensions.
This also makes it physically clear why a functional with a single zero on only one spin could not exist, since that would prove that one-loop anomalous dimensions identically vanish.

\subsection{One-loop OPE coefficients: the $\Psi_{\ell}$ functionals} \label{sec:Psi functional}

The functionals just constructed possess desirable zero structure which makes them highly particularly
sensitive to the scaling dimension of leading-twist operators of a given spin.
However, they lack sensitivity to OPE coefficients, which are the main focus of this paper.
%The previous functional construction possessed a desirable zero structure which suppresses the action of large twist operators.
%However, it lacks sign-definiteness and it still suppresses the leading-twist operator for spin $J=0$.
%In this subsection, we construct a functional that can further help us isolate the contribution of the Konishi operator.

We now construct, for each even $\ell\geq 0$, a functional $\Psi_{\ell}$ which has double zeros around twist two
for all spins except $\j=\ell$, where it has nonvanishing constant term \emph{and} slope.
At weak coupling, $\Psi_\ell$ effectively relates the OPE coefficient and scaling dimension of the leading spin-$\ell$ operator.
In terms of a kernel $\Psi_\ell[\mT]$ defined similarly to \eqref{Phi func ansatz},
these conditions are interpreted as follows:
\begin{align}
\int \frac{d\mT}{2\pi i} \Psi_\ell[\mT] a_{\j}(\mT) &= \delta_{\ell,\j},  \label{aEll condition} \\
\int \frac{d\mT}{2\pi i} \Psi_\ell[\mT] b_{\j}(\mT)  &= \delta_{\ell,\j} \beta_\ell, \label{bEll condition}
\end{align}
for all even $\j\geq 0$, and where $\beta_\ell$ is an a-priori unknown constant.
%We will further find that this functional is positive-definite.
%The derivation of this object is the main goal of this subsection. 

To construct the kernel $\Psi_\ell[\mT]$, we propose to make an ansatz as a sum of its even and odd parts.
The orthogonality relation \eqref{aMack ortho} immediately fixes the even part:
\be\begin{aligned}
 \Psi_\ell[\mT] &\equiv \Psi_\ell^{\rm E}[\mT]+\Psi_\ell^{\rm O}[\mT], 
\\ \Psi_\ell^{\rm E}[\mT]&= \frac12 \Gamma\!\left(4-\tfrac{\mT'}{2}\right)^2\Gamma\!\left(\tfrac{\mT'}{2}-1\right)^2
c_\ell a_\ell(\mT)\,, \label{Psi even}
\end{aligned}\ee
where the odd part satisfies $\Psi_\ell^{\rm O}[10{-}\mT]=-\Psi_\ell^{\rm O}[\mT]$.
This takes care of \eqref{aEll condition}.

We now substitute the ansatz into the second condition \eqref{bEll condition} and evaluate
the even contribution using the integral $I_{\ell,\j}$ in \eqref{Iellj integral}:
\be
 I_{\ell,\j} + \int \frac{d\mT}{2\pi i} \Psi_\ell^{\rm O}[\mT] b_\j(\mT) = \delta_{\ell,\j} \beta_\ell \qquad \mbox{(for all even $\j\geq 0$)}.
\ee
Since we formally diagonalized the $b_\j$ integral using the kernel $\Phi_{\j}[\mT]$ (see \eqref{Phi conditions}),
this system can be inverted as an infinite sum
\be \label{Psi Odd term definition}
  \Psi_\ell^{\rm O}[\mT] \equiv \Phi_\ell[\mT]\beta_\ell -\sum_{n=0}^\infty I_{\ell,2n} \Phi_{2n}[\mT]\,.
\ee
The single parameter left to determine is $\beta_\ell$.
As above, it is fixed by the requirement that the kernel must vanish at large $\mT$
for the functional to make sense, which gives
\begin{align}
0&= \beta_\ell \Phi^\infty_\ell - \sum_{n=0}^\infty I_{\ell,2n} \Phi^\infty_{2n}  \\
\quad\Rightarrow\quad
\beta_\ell H_{\ell+2} &=
I_{\ell,\ell}+\sum_{n=0}^{\frac{\ell}{2}-1}\frac{(4n+5)H_{2n+2}}{(2n+1)(2n+4)}
+\sum_{n=1}^\infty \frac{(\ell+2)_2 (2\ell+5+4n)H_{\ell+2+2n}}{2n(\ell+2+2n)_2(2\ell+5+2n)}
\,, \label{beta sum}
\end{align}
where we plugged in the explicit limit \eqref{boundary term}.
Perhaps surprisingly, these sum up to simple rational numbers.
In fact, as discussed below \eqref{Psi one-loop} in the main text,
there is a simple analytic guess for the result, which is required for the sum rule to be satisfied at one-loop:
\be
 \beta_\ell = H_{2\ell+4}-H_{\ell+2} +\frac{S_2(\ell+2)}{2H_{\ell+2} }\,. \label{beta ell result}
\ee
We verified numerically that this agrees precisely with \eqref{beta sum} for many values of $\ell$.

To summarize starting from \eqref{Psi even}, the action of the $\Psi_\ell$ functional is computed is
\be
 \Psi_\ell[\D,\j]=\int \frac{d\mT}{4\pi i} \Psi_\ell[\mT]\widehat{B}_\mT[\D,\j], \label{Psi summary}
\ee
with the kernel defined by
\be
\Psi_\ell[\mT] = \frac12 \Gamma\!\left(4-\tfrac{\mT'}{2}\right)^2\Gamma\!\left(\tfrac{\mT'}{2}-1\right)^2 c_\ell a_\ell(\mT)
+\beta_\ell \Phi_\ell[\mT] -\sum_{n=0}^\infty I_{\ell,2n} \Phi_{2n}[\mT]\,. \label{Psi kernel}
\ee
The functions $\Phi_\ell[\mT]$ are defined in \eqref{Phi ell kernel} and the coefficients $I_{\ell,2n}$ and $\beta_\ell$ can be found in \eqref{Iellj integral} and \eqref{beta ell result} respectively.

\subsection{Formulas for evaluating $\Phi$ and $\Psi$ functionals} \label{app:PhiPsi eval}

We now describe a practical way to perform the $\mT$ integral that define the projection functionals, starting with $\Phi_{\ell_1,\ell_2}$ in \eqref{Phi func ansatz},
when acting on a generic state.
Following the method in section \ref{app:B from Mack}, it suffices to evaluate the following basic integrals for integer $k\geq 0$:
\be
 \left[\Phi_{\ell_1,\ell_2}\right]_k \equiv \int \frac{d\mT}{4\pi i} \Phi_{\ell_1,\ell_2}[\mT] \frac{(\tfrac{\mT-2}{2})_k}{\mT-6}\,. \label{Phi vector}
\ee
Indeed it is clear from the definition of $\widehat{B}_\mT$ that the action
of $\Phi_{\ell_1,\ell_2}$ on any state can be written as a finite sum of these integrals (see \eqref{Bt}). 
When plugging in the integral representation \eqref{Phi ell kernel}, the $1/(\mT-6)$ denominator neatly cancels out:
\be
 \left[\Phi_{\ell_1,\ell_2}\right]_k = \frac{1}{8i} \int\displaylimits_{-\infty}^\infty \frac{dy}{4\pi} (1+y^2)(\tfrac{1+iy}{2})_{k+1}
 \int_0^y \frac{\pi^2dy'}{\cosh(\tfrac{\pi y}{2})^2}\left[c_{\ell_1} a_{\ell_1}(5+iy')-\frac{\Phi_{\ell_1}^\infty}{\Phi_{\ell_2}^\infty}  c_{\ell_2} a_{\ell_2}(5+iy')\right].
\ee
Again we have set $\mT=5+iy$. This looks daunting, but the trick is to integrate by parts, using that the inner integral vanishes exponentially at infinity.
This yields a difference of two integrals,
\be
\left[\Phi_{\ell_1,\ell_2}\right]_k = \left[\Phi_{\ell_1}\right]_k - \frac{\Phi_{\ell_1}^\infty}{\Phi_{\ell_2}^\infty} \left[\Phi_{\ell_2}\right]_k,
\label{Phi as combin}
\ee
where
\be
 \left[\Phi_{\ell}\right]_k \equiv \frac{1}{16i} \int\displaylimits_{-\infty}^\infty \frac{\pi^2 dy}{4\pi\cosh(\tfrac{\pi y}{2})^2}
c_{\ell} a_{\ell}(5+iy) \int_{i}^y dy' (1+y'^2)\left[\big(\tfrac{1+iy'}{2}\big)_{k+1}-\big(\tfrac{1-iy'}{2}\big)_{k+1}\right]\,. \label{Phi k integral}
\ee
Note that we antisymmetrized in $y'$, which was not strictly necessary but makes each $\left[\Phi_{\ell}\right]_k$ real.
The calculation is thus reduced to integrating a polynomial divided by $\cosh^2$. This can be completed term-by-term using the identity
\begin{equation} \label{yIntegral fast}
\int \displaylimits_{-\infty}^\infty \frac{dy}{4\pi} \frac{\pi^2}{\cosh^2(\pi y/2)} y^q = (-1)^{1+q/2} (2^q-2) B_q,
\end{equation}
where $B_q$ on the right is the Bernouilli number, not to be confused with the position-space functional $B_v$.
In this way $\left[\Phi_{\ell}\right]_k$ can be evaluated as an exact rational number.

The $y'$ lower bound in \eqref{Phi k integral} is arbitrary since any constant added to the integral would cancel
out in the combination \eqref{Phi as combin}, thanks to the definition of $\Phi_{\ell}^\infty$.
The choice made above, which makes the integral proportional to $(1+y^2)$, turns out to ensure that $\left[\Phi_{\ell}\right]_k =0$ for $\ell>k$, which will be convenient below.
This can be interpreted as an orthogonality property of the Mack polynomials, although we were not able to strictly derive it from \eqref{aMack ortho}.
Example values of $[\Phi_\ell]_k$ and $[\Phi_{0,2}]_k$ are given in table \ref{tab:Phi coeffs}.
% need to check against kernels that are quoted.

For the $\Psi_\ell$ functional \eqref{Psi kernel}, we follow the same strategy and integrate by parts in all the terms involving $\Phi_k[\mT]$.
The even contribution is simpler and we could in fact do it analytically; we record only the result,
\be\begin{aligned} \label{Psi vector}
\left[\Psi_{\ell}\right]_k \equiv \int\frac{d\mT}{4\pi i} \Psi_{\ell}[\mT] \frac{(\tfrac{\mT-2}{2})_k}{\mT-6} &=
\frac{\Gamma(\ell+3)^2\Gamma(k+2)}{4\Gamma(2\ell+5)}\left[\frac{(k-\ell)_{\ell+3}}{(k+1)_{\ell+4}}(k+7+\ell(\ell+5))-1\right]
\\&\phantom{=}+ \beta_\ell \left[\Phi_\ell\right]_k -\sum_{n=0}^{\lfloor k/2\rfloor } I_{\ell,2n}\left[\Phi_{2n}\right]_k \,.
\end{aligned}\ee
The crucial fact is that that the $n$ sum terminates, thanks to the vanishing properties just mentioned.
Thus, while we were unable to find a closed-form expression for the kernel $\Psi_\ell[\mT]$ itself,
it is possible to compute its action on a state of arbitrary spin $\j$ as a \emph{finite} sum of $\sim \j$ terms.
The sum gives rational numbers exemplified in table \ref{tab:Phi coeffs}.
We have verified that they agree with the direct numerical integration of \eqref{Psi summary}, with the sum over $n$ truncated to a large order.

\begin{table}[t]\begin{center}
\begin{tabular}{c|cccccccc}
$k$ & 0&1&2&3&4&5&6&7\\\hline
$\left[\Phi_0\right]_k$ & $-\tfrac{1}{30}$&$-\tfrac{1}{15}$&$-\tfrac{59}{315}$&$-\tfrac{44}{63}$&$-\tfrac{1031}{315}$&$-\tfrac{278}{15}$&$-\tfrac{20332}{165}$&$-\tfrac{31072}{33}$%&$-\tfrac{366333988}{45045}$
\\[2pt]
$\left[\Phi_2\right]_k$ & 0&0&$-\tfrac{2}{735}$&$-\tfrac{16}{735}$&$-\tfrac{1228}{8085}$&$-\tfrac{424}{385}$&$-\tfrac{19696}{2275}$&$-\tfrac{1866112}{25025}$%&$-\tfrac{73986784}{105105}$
\\[2pt]
$\left[\Phi_4\right]_k$ & 0&0&0&0&$-\tfrac{5}{11011}$&$-\tfrac{90}{11011}$&$-\tfrac{108}{1001}$&$-\tfrac{14496}{11011}$%&$-\tfrac{2997860}{187187}$
\\[2pt]\hline
$\left[\Phi_{0,2}\right]_k$ & $-\frac{1}{30}$&$-\frac{1}{15}$&$-\frac{37}{225}$&$-\frac{116}{225}$&$-\frac{4943}{2475}$&$-\frac{7658}{825}$&$-\frac{2708092}{53625}$&$-\frac{1536544}{4875}$%&$-\frac{71415404}{32175}$
\\[2pt]
$\left[\Psi_{0}\right]_k$ &
$-\frac{103}{1800}$&$-\tfrac{73}{900}$&$-\tfrac{5996}{33075}$&$-\tfrac{3722}{6615}$&$-\tfrac{74534}{33075}$&$-\tfrac{17492}{1575}$&$-\tfrac{496730}{7623}$&$-\tfrac{84661744}{190575}$%&$-\tfrac{2336521219826}{676350675}$
\end{tabular}
\caption{\label{tab:Phi coeffs}Example results for the $\left[\Phi_\ell\right]_k$ integral in \eqref{Phi k integral},
and for two physical functionals derived from it.}
\end{center}\end{table}

\subsection{Sign properties of $\Phi$ and $\Psi$ functionals} \label{app:projected functional signs}

\begin{figure}[th]
\centering
\begin{subfigure}{\textwidth}
\centering
\includegraphics[width=0.7\textwidth]{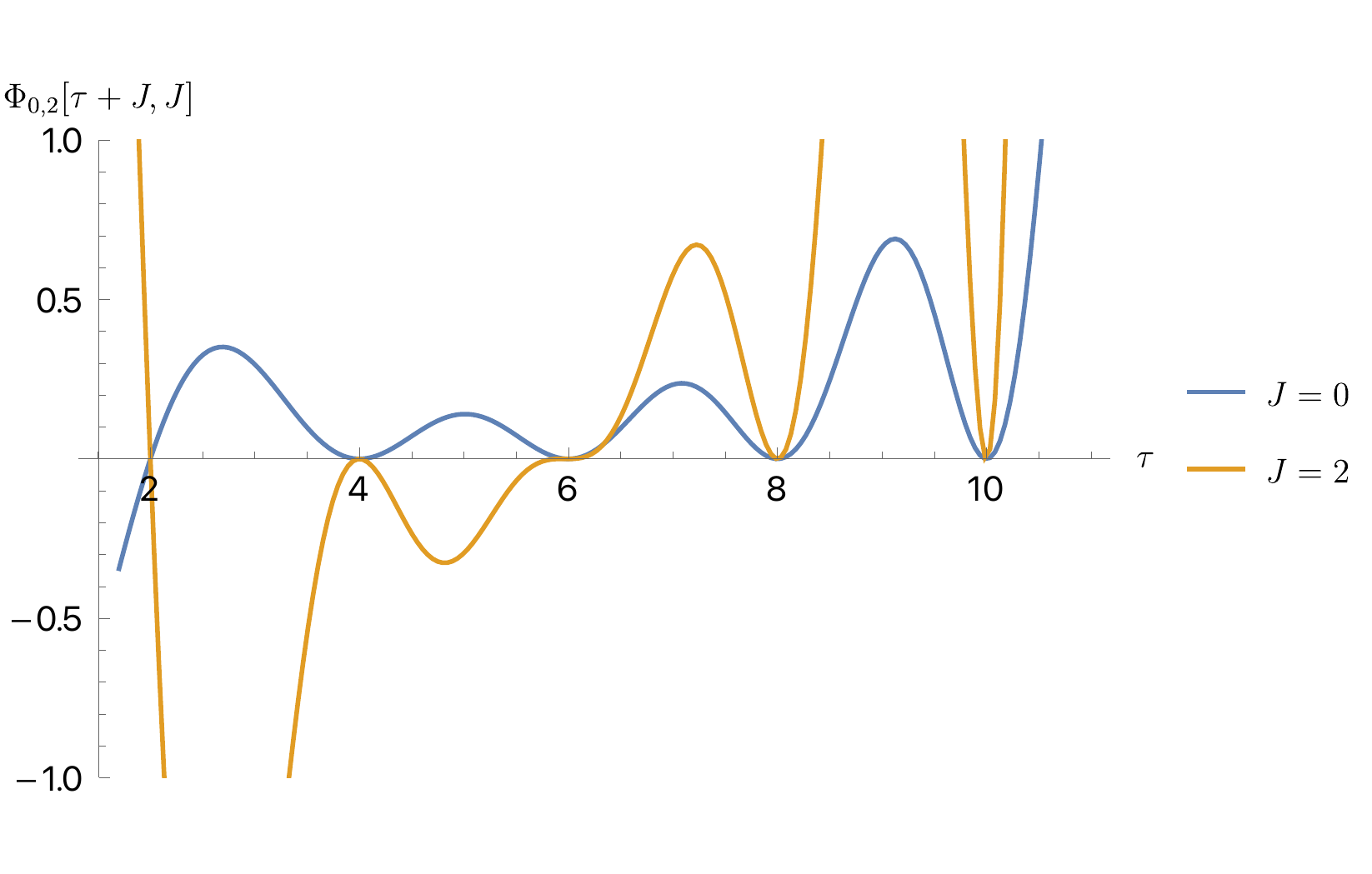}
\end{subfigure}
\begin{subfigure}{\textwidth}
\centering
\includegraphics[width=0.7\textwidth]{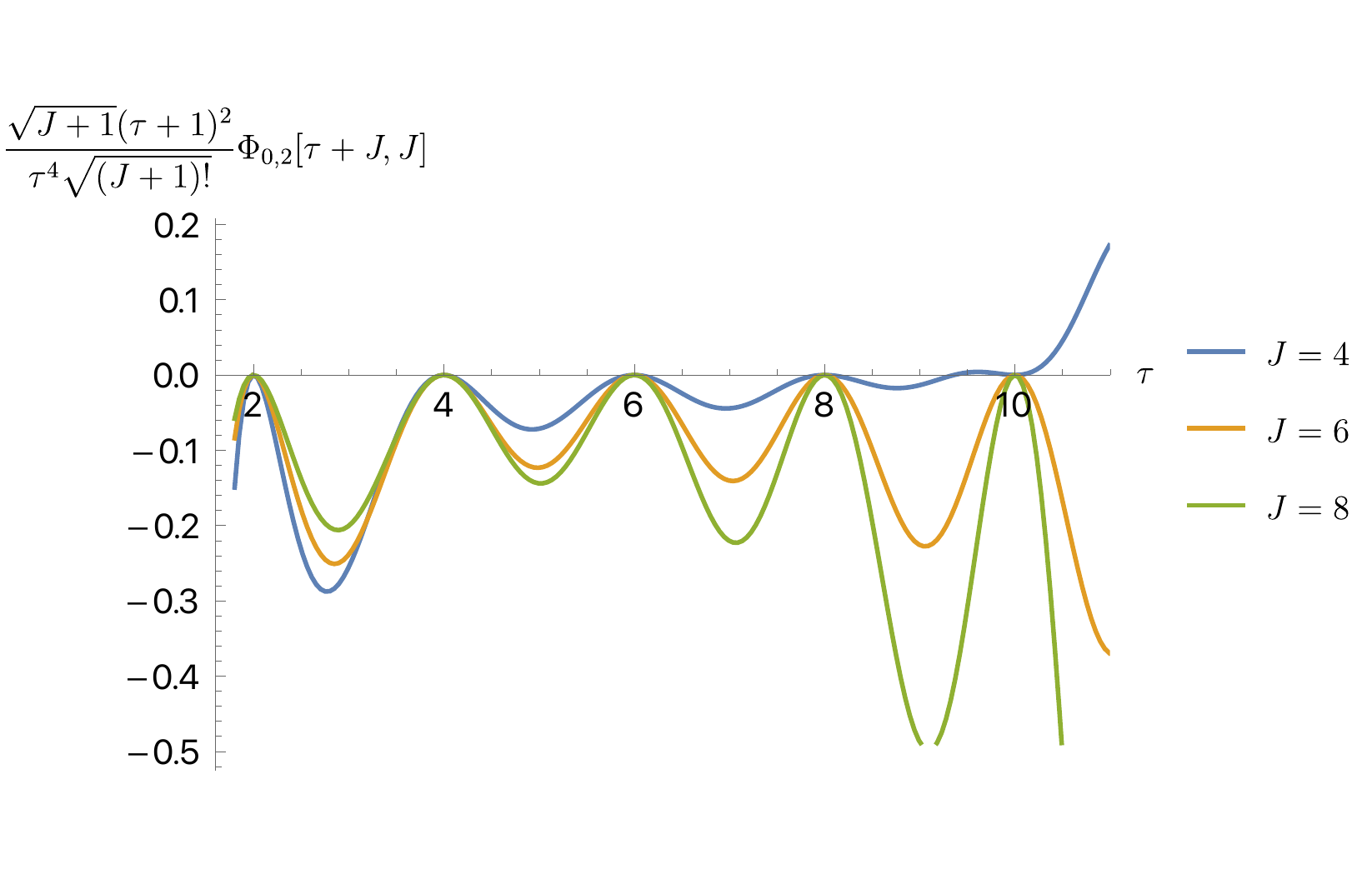}
\end{subfigure}
\caption{Action of the $\Phi_{0,2}$ functional.  We plot the action of $\Phi_{0,2}$ for spins $J=0,2$ in the first panel, and the rescaled functional $\frac{\sqrt{J+1} (\tau+1)^2}{\tau^4 \sqrt{(J+1)!}} \Phi_{0,2}$ for larger spin in the second one for visual clarity.  For $J>0$, the functionals are initially negative for $\tau > 2$, but they become positive at large enough twist.}
\label{fig:Phi02}
\end{figure}

The formulas from the preceding section, combined with the Mack polynomials reviewed in section~\ref{app:Mack polynomials}, enable to rapidly compute the $\Phi_{\ell,\ell+2}$ and $\Psi$ functionals
on generic states.  One readily sees from figure \ref{fig:Phi02} that $\Phi_{0,2}$ indeed has the claimed single zeros at spins $0$ and $2$ and twist 2,
and double zeros on higher spins and all higher twists.
However, since the slopes on spin 0 and 2 have opposite signs, it is not sign definite as also visible from the figure.
In fact, for $\j>0$ the $\Phi_{0,2}$ functional is negative for $\tau< \tau^*$ and eventually becomes positive at large enough twist $\tau$.

\begin{figure}[h]
\centering
\begin{subfigure}{\textwidth}
\centering
\includegraphics[width=0.7\textwidth]{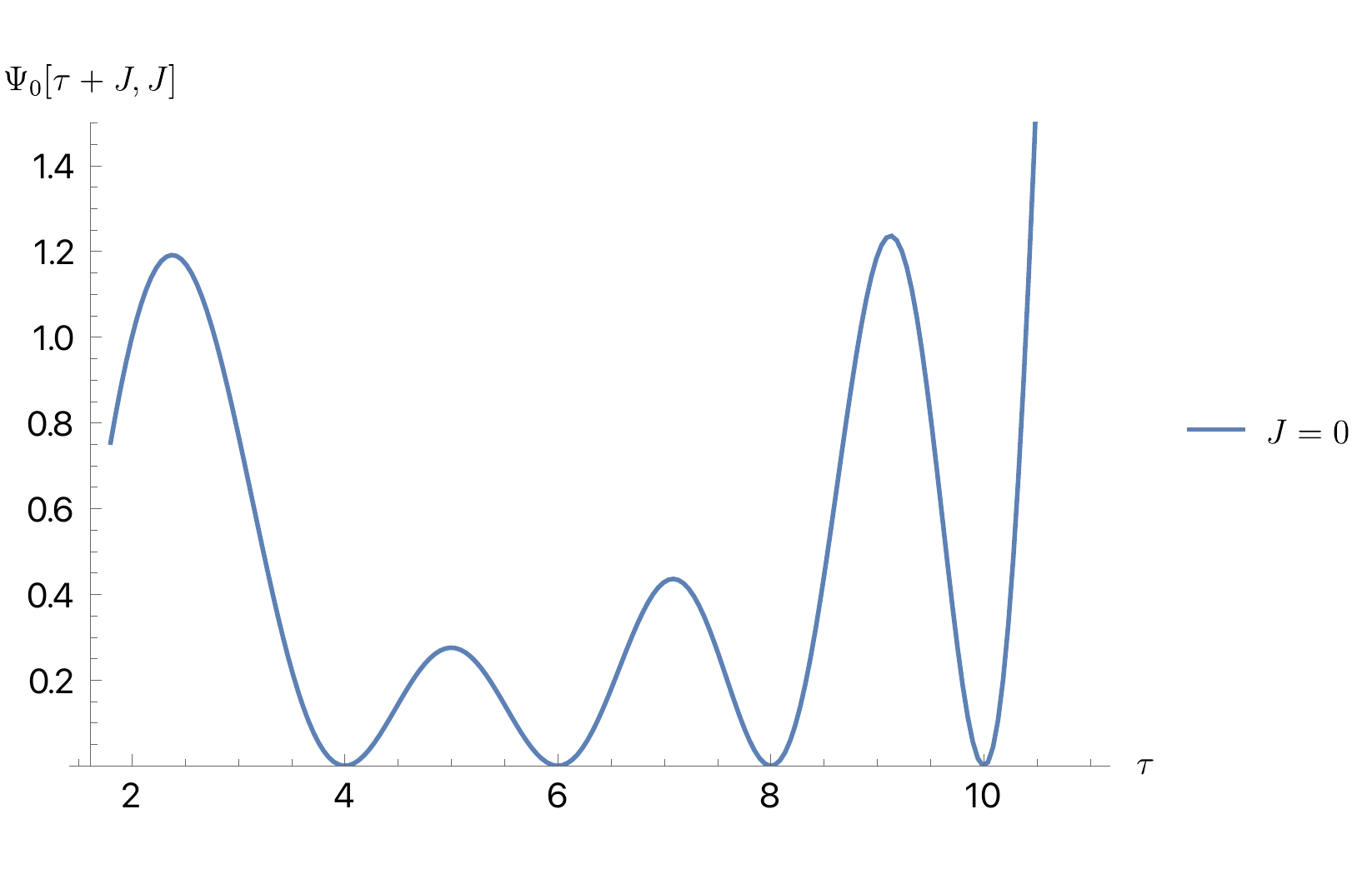}
\end{subfigure}

\begin{subfigure}{\textwidth}
\centering
\includegraphics[width=0.7\textwidth]{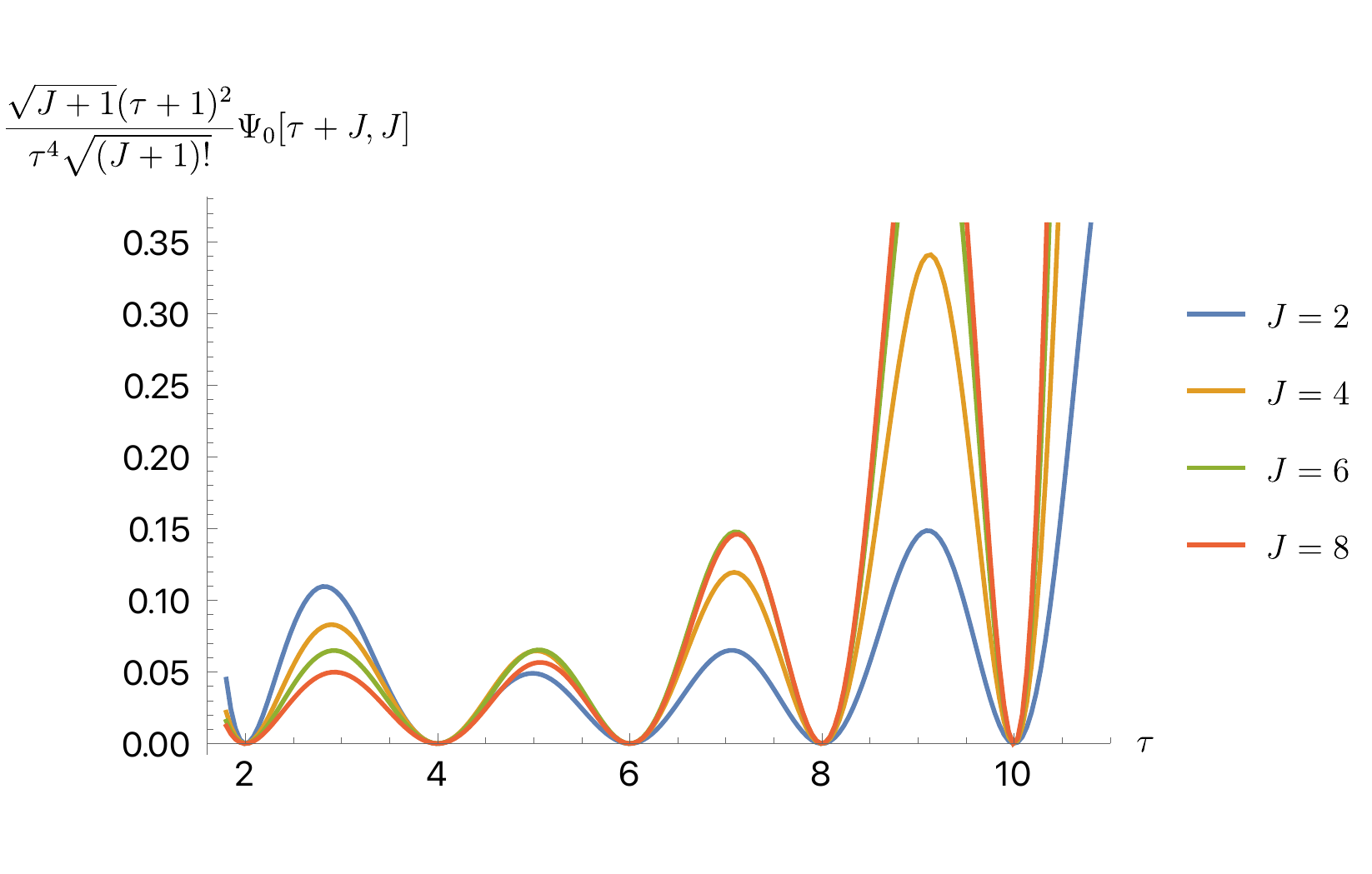}
\end{subfigure}
\caption{Action of the functionals at low twist. We plot $\Psi_0$ in the first panel for $\j=0$, and $\frac{\sqrt{\j+1} (\tau+1)^2}{\tau^4 \sqrt{(\j+1)!}} \Psi_0$ in the second panel for visual clarity.  This functional is positive-definite for all spin and all twist $\tau \geq 2$.}
\label{fig:Psi near twist2}
\end{figure}

Similarly, we show the action of $\Psi_0$ in fig.~\ref{fig:Psi near twist2}, which features double-zeros on all double-twists for $J \geq2$ and it is non-zero when acting on the Konishi operator.
Remarkably, we find that the functional is nonnegative for all $\tau\geq 2$.
The same property is actually shared by all the $\Psi_\ell$:
we observe that $\Psi_\ell[\Delta,\j]$ identically vanishes for $\j<\ell$ and is positive otherwise. 

%%%%%%%%%%%%%%%%%%%%%%%%%%%%%%%%%%%%%%%%%%%%%%%%%%%%%%%%%%%%%%%%%%%
\section{Formulas for efficient evaluation of functionals} \label{app:efficient eval of funcs}

The bootstrap method requires the evaluation of a large menu of trial functionals on a large sample of states $(\Delta,\j)$.
High accuracy is required, since optimal combinations tend to involve large numerical cancellations between different functionals.
Here we detail fast and accurate numerical methods.

%This leads to conflicting requirements of speed and accuracy that are challenging to meet in practice.

\subsection{Dispersion relations in position space} \label{app:position}

The most straightforward method to compute the Polyakov-Regge block $\cP_{\Delta,\j}(u,v)$ is perhaps to
compute the dispersive integral \eqref{PR coords}.
The kernel is that of the unsubtracted dispersion relation of \cite{Carmi:2019cub};
our conventions follow those in (2.9) of \cite{Caron-Huot:2020adz}, where we combine the $s$ and $t$-channels:
\begin{align} \label{K}
 K(u,v;u',v') &= \frac{u'-v'}{64\pi(uvu'v')^{\frac34}} 
 x^{\frac32} {}_2F_1(\tfrac12,\tfrac32,2,1-x)\ \theta(\sqrt{v'}-(\sqrt{u}+\sqrt{v}+\sqrt{u'}))
\nl &\quad +\frac{1}{4\pi(uvu'v')^{\frac14}}
\frac{\sqrt{u'}+\sqrt{v'}}{(\sqrt{u}+\sqrt{u'})(\sqrt{v}+\sqrt{u'})}\ \delta(\sqrt{v'}-(\sqrt{u}+\sqrt{v}+\sqrt{u'})),
\end{align}
where $x$ is a remarkable combination of four cross-ratios:
\be
x= \frac{16\sqrt{uvu'v'}}{
\big[(\sqrt{u}+\sqrt{v})^2-(\sqrt{u'}+\sqrt{v'})^2\big]
\big[(\sqrt{u}-\sqrt{v})^2-(\sqrt{u'}-\sqrt{v'})^2\big]}\,.
\ee
%Thus contribution of an operator with given ($\Delta,\j$) to the crossing equation \eqref{Xuv} with given ($u,v$) can thus be computed as a double integral.
In practice, we perform the double integral in \eqref{disp} by changing to $\rho$ coordinates, in which the kernel also takes a concise form.  By adapting the radial coordinates from \cite{Hogervorst:2013sma} to the $u$-channel, we obtain
\begin{equation}
\rho = \frac{1}{(\sqrt{1-w} + \sqrt{-w})^2}, \qquad \overline{\rho} = \frac{1}{(\sqrt{1-\overline{w}} + \sqrt{-\overline{w}})^2}.
\end{equation}

The choice of numerical integration strategy matters.  In \texttt{Mathematica}, we force
the use of the \texttt{DoubleExponential} method, which essentially computes a simple Riemann sum
after a clever change of variable that makes the integrand decay exponentially near its endpoints.
In theory, for a sufficiently ``nice'' integrand, the error with this method decreases exponentially with effort, however
we only observe a decrease in error if we also force subdivisions using the \texttt{MinRecursion} option.
For example, the two-dimensional $\rho$ integral with
\be\begin{aligned}
 &\texttt{NIntegrate[\ldots, Method -> \{"DoubleExponential", "SymbolicProcessing" -> 0\},}\nonumber\\&\hspace{30mm} \texttt{MinRecursion -> 2, WorkingPrecision -> 25]}
\end{aligned}\ee
will typically achieve near 20 digits of accuracy, which generally
suffices for bootstrap problems with $O(30)$ functionals or fewer.
One way to estimate accuracy is to make simple changes of variable such as $\rho_w\mapsto \rho_w^2$, which theoretically
should not change the integral but in practice do so with this method.
We observe that each increase in MinRecursion typically adds about 10 significant figures at the cost of quadrupling the computation time. Therefore, arbitrary accuracy is in principle achievable with this method,
but only for a limited number of functionals/states.

\subsection{Mack polynomials} \label{app:Mack polynomials}

%An alternative strategy is to use the Mellin representation of Polyakov-Regge blocks using \eqref{P Mellin} from the main text.
In this appendix we detail our evaluation of formulas involving the Mellin representation and Mack polynomials.
For future reference, we keep explicit the dependence on external operator dimensions
$\{\Delta_i\}=\{\Delta_1,\Delta_2,\Delta_3,\Delta_4\}$ and spacetime dimension $d$;
only the case $\{\Delta_i\}=\{4,4,4,4\}$ and $d=4$ is relevant for the main text.

To fix our conventions, we use the following Mellin representation for unequal operators following the convention of \cite{Trinh:2021mll}:
\begin{align}
 \langle \cO(x_1)\cdots \cO(x_4)\rangle&= \frac{1}{(x_{13}^2)^{\tfrac{\D_1+\D_2+\D_3-\D_4}{2}} (x_{24}^2)^{\D_2}} \frac{(x_{34}^2)^{\tfrac{\D_1+\D_2-\D_3-\D_4}{2}} }{(x_{14}^2)^{\tfrac{\D_1+\D_4-\D_2-\D_3}{2}}} \mathcal{G}(u,v), \\
 \mathcal{G}(u,v) &= \iint_{\gamma_C} \frac{d\mS d\mT}{(4\pi i)^2} \Gamma(\tfrac{\D_1+\D_2-\mS}{2}) \Gamma(\tfrac{\D_3+\D_4-\mS}{2}) \Gamma(\tfrac{\D_2+\D_3-\mT}{2}) \Gamma(\tfrac{\D_1+\D_4-\mT}{2}) \nonumber \\
 & \qquad \times \Gamma(\tfrac{\D_2+\D_4 - \mU}{2}) \Gamma(\tfrac{\D_1+\D_3 - \mU}{2}) u^{\tfrac{\mS-\D_1-\D_2}{2}} v^{\tfrac{\mT-\D_2-\D_3}{2}} M_{\mS,\mT}
\end{align}
where $\mS+\mT+\mU=\sum_{i=1}^4 \Delta_i$, and the contour $\gamma_C$ is determined by the Gamma functions.
Then the Mack polynomial ${\cal Q}^{m,\{\Delta_i\}}_{\Delta,\j}(\mT)$ is the residue at $\mS=\Delta-\j+2m$
in the Mellin representation of a conformal block $G_{\Delta,\j}(u,v)$ (or Polyakov-Regge block, which differs by double-twist contributions).

For our applications below, it is useful to explicit the $m$ and $\mT$ dependence of the Mack polynomials.
Up to overall $\Gamma$-functions, the dependence is essentially through a polynomial of total degree $\j$ which is
naturally written using Pochhammer symbols \cite{Mack:2009mi,Penedones:2019tng}:
\def\Gm#1{\Gamma\big({#1}\big)}
\def\Qmat{\left[Q^{a,b}_{\Delta,\j}\right]}
\begin{align}\label{Qcal_rep}
 {\cal Q}^{m,\{\Delta_i\}}_{\Delta,\j}(\mT) =&
 K_{\Delta,j}^{m,\{\Delta_i\}}Q^{m,a,b}_{\Delta,\j}(\mT{-}\Delta_2{-}\Delta_3),
 %\\
 \quad 
 Q^{m,a,b}_{\Delta,\j}(\mT') \equiv  
\sum_{q=0}^\j \sum_{k=0}^{\j-q} (-m)_q  \Qmat_{q,k} \big({-}\tfrac{\mT'}{2}\big)_k\,.
\end{align}
Here and below $a=\frac{\Delta_2-\Delta_1}{2}$, $b=\frac{\Delta_3-\Delta_4}{2}$.
The prefactor, which contains the double-twist zeros mentioned in the main text, is
\be\begin{aligned}
K_{\Delta,\j}^{m,\{\Delta_i\}}=&\frac{1}{m!(\Delta-\tfrac{d}{2}+1)_m\Gamma(\Delta-1)\Gm{\tfrac{\Delta_1+\Delta_2-\Delta+\j}{2}-m}\Gm{\tfrac{\Delta_3+\Delta_4-\Delta+\j}{2}-m}}
\\&\times \frac{2\Gamma(\Delta+\j)\Gamma(\Delta+\j-1)}{\Gm{\tfrac{\Delta+\j+\Delta_1-\Delta_2}{2}}\Gm{\tfrac{\Delta+\j+\Delta_2-\Delta_1}{2}}\Gm{\tfrac{\Delta+\j+\Delta_3-\Delta_4}{2}}\Gm{\tfrac{\Delta+\j+\Delta_4-\Delta_3}{2}}},
\end{aligned}\ee
whereas the coefficients $\Qmat_{q,k}$ are rational functions of $\Delta$.
It is useful to view them as a $(\j{+}1)\times (\j{+}1)$ matrix where $q,k$ range from $0$ to $\j$,
setting to zero the entries with $q+k>\j$.
The matrix can be populated efficiently using the Casimir recursion in appendix A of \cite{Costa:2012cb}.
In the Pochhammer basis, we find
\begin{align} \label{recursion}
&\hspace{-5mm}\big( (2q+k-\j)(\Delta-\j+1-d+q)-(\j+k+d-2)(\j-k-q)\big)\Qmat_{q,k} \nonumber\\
&= (k+1)(\tfrac{\Delta-\j}{2}+a+q+k)(\tfrac{\Delta-\j}{2}+b+q+k)\Qmat_{q,k+1}\nonumber\\
&\phantom{=}+(k+1)(\j-2k-q-a-b+1-d/2)\Qmat_{q-1,k+1}\nonumber\\
&\phantom{=}-2(\j+1-q-k)\Qmat_{q-1,k}\,.
\end{align}
This recursion is seeded with the boundary condition $\Qmat_{0,\j}=(-1)^{\j+1}$ for the top-right element,
which corresponds to the conformal block normalization $\lim_{u\ll 1-v\ll 1} G_{\Delta,\j}(u,v)=u^{\frac{\Delta-\j-\Delta_1-\Delta_2}{2}} (1-v)^\j$.
We fill the matrix row-by-row using \eqref{recursion} to move leftward in $k$.
For example, for the first row ($q=0$) one finds the simple analytic solution
\be
 \Qmat_{0,k} = \frac{(-1)^{\j+1} \j! (\Delta+\j-1)_{k-\j}}{k!(\j-k)! \big(\tfrac{\Delta+\j}{2}+a\big)_{k-\j}\big(\tfrac{\Delta+\j}{2}+b\big)_{k-\j}},
\ee
which resums (see \eqref{Qcal_rep}) to give the standard $m=0$ Mack polynomial \cite{Mack:2009mi,Costa:2012cb}
\be \label{Mack polynomials}
  Q^{0,a,b}_{\Delta,\j}(\mT') = (-1)^{\j+1} \frac{\big(\tfrac{\Delta+\j}{2}+a\big)_{\j}\big(\tfrac{\Delta+\j}{2}+b\big)_{\j}}{(\Delta-1)_\j}
 {}_3F_2\big(\{-\j,\Delta-1,-\tfrac{\mT'}{2}\},\{\tfrac{\Delta-\j}{2}+a,\tfrac{\Delta-\j}{2}+b\},1\big).
\ee
Populating the entire matrix $\Qmat_{q,k}$ requires only $O(\j^2)$ multiplications.
For $\j\sim 100$ and rational values of $\Delta$, this task can be completed with exact rational arithmetic in a
fraction of a second, on a typical laptop.

A slightly upsetting feature of \eqref{recursion} is that it produces spurious poles
at values of $\Delta$ that are not particularly meaningful, where the factor on the left-hand-side vanishes.
We either avoid these values 
or fall back on the following analytic expression for the coefficients (see the formula recorded in appendix of \cite{Trinh:2021mll}),
which is explicitly free of spurious poles:
\begin{align}
\Qmat_{q,k} =&\frac{\j!}{(\j-k-q)!k!}\frac{(\tfrac{\Delta-\j}{2}+b+k+q)_{\j-k-q}}{(\j-q+\tfrac{d}{2}-1)_q}
\nonumber\\ &\times \sum_{p=0}^q
\frac{(-1)^{k+p}}{p!(q-p)!} \frac{(\tfrac{\Delta-\j}{2}+a+k+q-p)_{\j-k-q}}{(\Delta-\j+2-d)_{q-p}(\Delta+k+q-p-1)_{\j-k-q+p}}
\nonumber\\ &\quad\times
\big(\tfrac{\Delta-\j-d+2}{2}-b\big)_{q-p}
\big(\tfrac{\Delta-\j-d+2}{2}+a\big)_{q-p}
\big(\tfrac{\Delta+\j}{2}-b-p\big)_{p}\big(\tfrac{\Delta+\j}{2}+a-p\big)_{p}\,.
\end{align}
Populating the $\Qmat_{q,k}$ matrix with this formula requires $O(\j^3)$ multiplications.

\subsection{Formulas for $B$ functionals using Mack polynomials} \label{app:B from Mack}

The representation \eqref{Qcal_rep} is convenient for our purposes
because the $m$ dependence is isolated in a few Pochhammer symbols.
Consider for example the $\widehat{B}_\mT$ functional, which we recall from \eqref{Bt}:
\be
 \widehat{B}_{\mT}[\Delta,\j] = \sum_{m=0}^\infty \frac{2 (\Delta-\j+2m)+2-t}{t-6} {\cal Q}^{m,\{4,4,4,4\}}_{\Delta+4,\j}(10-\mT).
\label{Bt app}
\ee
The dependence on $m$ is explicit in $K$ and $(-m)_q$, which allows the
sum over the infinite number of descendants to be performed analytically.
The basic formula is
\begin{align} \label{msum}
 \sum_{m=0}^\infty K_{\Delta,\j}^{m,\{\Delta_i\}} (-m)_q &= \tilde{K}_{\Delta,\j}^{\{\Delta_i\}}  %\\&\times
\frac{(\tfrac{\Delta-\j-\Delta_1-\Delta_2+2}{2})_q(\tfrac{\Delta-\j-\Delta_3-\Delta_4+2}{2})_q}{(\tfrac{d+4-\Delta_1-\Delta_2-\Delta_3-\Delta_4}{2}-\j)_q}\,,
\\ \label{Kt}
\tilde{K}_{\Delta,\j}^{\{\Delta_i\}}&\equiv K_{\Delta,\j}^{0,\{\Delta_i\}}\times \frac{\Gm{\Delta+1-\tfrac{d}{2}}\Gm{\tfrac{\Delta_1+\Delta_2+\Delta_3+\Delta_4-d-2}{2}+\j}}{\Gm{\tfrac{\Delta+\j+\Delta_1+\Delta_2-d}{2}}\Gm{\tfrac{\Delta+\j+\Delta_3+\Delta_4-d}{2}}}\,.
\end{align}
From here, we specialize to the $\Delta_i=4$ case of interest
and abbreviate: $\tilde{K}^{N=4}_{\Delta,\j}\equiv \tilde{K}_{\Delta+4,\j}^{\{4,4,4,4\}}$.
The nonnegative factor $\tilde{K}^{N=4}_{\Delta,\j}$ will be present in front of most functionals.

Explicitly, we thus compute $\widehat{B}_{\mT}$ in \eqref{Bt} by forming the vector (which is linear in $\mT$)
\be
\widehat{B}_{\mT}[\Delta,\j]_q = 2(\Delta{-}\j{+}2q{+}1{-}\tfrac{\mT}{2})\frac{(\tfrac{\Delta-\j-2}{2})_q^2}{(-\j-4)_q}
-4\frac{(\tfrac{\Delta-\j-2}{2})_{q+1}^2}{(-\j-4)_{q+1}}, \label{Bt vector}
\ee
which we dot into the Mack coefficients:
\be
 \widehat{B}_{\mT}[\Delta,\j] = \tilde{K}^{N=4}_{\Delta,\j}
 \sum_{q,k=0}^\j \widehat{B}_{\mT}[\Delta,\j]_q \left[Q^{0,0}_{\Delta+4,\j}\right]_{q,k}\frac{(\tfrac{\mT-2}{2})_k}{\mT-6}.
\ee
The result takes the form of a polynomial in $\mT$ of degree $\j+1$, divided by ($\mT-6$).
At this stage we typically express it as a Pochhammer sum,
\be \label{Bt fast}
 \widehat{B}_{\mT}[\Delta,\j]\equiv \tilde{K}^{N=4}_{\Delta,\j}\sum_{k=0}^{\j+1} B[\Delta,\j]_k \frac{(\tfrac{\mT-2}{2})_k}{\mT-6}\,
\ee
where the coefficients $B[\Delta,\j]_k$ can be obtained from the above using simple vector operations.

This representation is useful to compute various functionals related to $B$.
For example, the Mellin transform which gives the position functional $B_v$ \eqref{Bv Mellin}
can be done analytically using
\be
  \frac{1}{2}\int \frac{d\mT}{4\pi i} v^{\frac{\mT}{2}-4}\Gamma\!\left(4-\tfrac{\mT}{2}\right)^2\Gamma\!\left(\tfrac{\mT}{2}-1\right)^2
  \frac{(\tfrac{\mT-2}{2})_k}{\mT-6} =\frac{-\Gamma(k+3)}{2(k+2)_3}{}_2F_1(3,k+3,k+5,1-v) \equiv \left[B_v\right]_k\,,  \label{Bv fast}
\ee
which allows to compute the position space functional as $B_v[\Delta,\j]=\tilde{K}^{N=4}_{\Delta,\j}\sum_k B[\Delta,\j]_k \left[B_v\right]_k$.

The results (and performance) can be compared with the position space integrals described in section \ref{app:position}.
For a sample operator with $\j=150$ and twist $4+\tfrac{1}{19}$, and $v=3/2$, we find for example:
\be\begin{aligned}
B_{3/2}[150,154+\tfrac{1}{19}] &= 2.845137906\ldots\times10^{83} \qquad &\hfill \mbox{(position, 4s)}\\ 
 &= 2.8451379417996980420\ldots\times10^{83} \qquad& \hfill \mbox{(position, 20s)}\\ 
 &= 2.84513794179969806433049731\ldots \times10^{83} \qquad& \hfill \mbox{(Mellin, 1s)}
\end{aligned}\ee
The accurate agreement between methods is a crucial debugging tool which
gives us high confidence in our implementation.
The Mellin space version of $B$ functionals is clearly faster, especially when high accuracy is needed.
This is partly due to the sub-exponential convergence of the $\rho$ integrations discussed above.
In contrast, the Mellin formula boils down to the exact calculation of a matrix of rational numbers,
times a numerical vector of ${}_2F_1$ functions, whose cost increases very slowly with the requested precision.

The projection functionals $\Phi_{\ell,\ell'}$ and $\Psi_\ell$ are computed similarly using respectively
the vectors \eqref{Phi vector} and \eqref{Psi vector}, so that
\be
\Phi_{\ell,\ell'}[\D,\j] =\tilde{K}^{N=4}_{\Delta,\j}\sum_{k=0}^{\j+1} B[\Delta,\j]_k \left[\Phi_{\ell,\ell'}\right]_k,\quad
\Psi_{\ell}[\D,\j] =\tilde{K}^{N=4}_{\Delta,\j}\sum_{k=0}^{\j+1} B[\Delta,\j]_k \left[\Psi_{\ell}\right]_k.  \label{projection app}
\ee
Note that the vectors of rational numbers $[\Phi_{\ell,\ell+2}]_k$ and $[\Psi_\ell]_k$ only need be computed once.
The projection functionals are thus obtained to infinite accuracy in effectively no time,
after the $B[\Delta,\j]_k$ coefficients have been calculated once.

\subsection{Formulas for Polyakov-Regge blocks using Mack polynomials}

A similar strategy works for the Mellin-space Polyakov-Regge block \eqref{PR Mellin explicit},
whose definition we recall:
\be \widehat{\cP}^{N=4}_{\mS,\mT}[\Delta,\j]
= \sum\limits_{m=0}^\infty
\mathcal{Q}^{m,\{4,4,4,4\}}_{\Delta+4,\j}(16 - \mS- \mT)\left[\frac{1}{\mS-(\Delta{-}\j{+}2m{+}4)}+\frac{1}{\mT-(\Delta{-}\j{+}2m{+}4)}\right] \,.
\label{PR Mellin explicit app}
\ee
%The shifts $\Delta\mapsto \Delta+4$ originate from the superconformal block \eqref{}.
The effect of the denominator is to replace the vector of $m$-sums \eqref{Bt vector} by
\be
\widehat{\cP}^{N=4}_{\mS}[\Delta,\j]_{q} \equiv
\sum_{m=0}^\infty \frac{K_{\Delta,\j}^{m,\{\Delta_i\}}}{\tilde{K}_{\Delta,\j}^{N=4}}
\frac{(-m)_q}{\mS-(\Delta{-}\j{+}2m{+}4)}\,. \label{Ps vector}
\ee
It turns out that this sum can also be evaluated analytically,  now in terms of ${}_3F_2$ hypergeometric functions.
Explicitly, for $q=0$ we find:
\be
 \widehat{\cP}^{N=4}[\Delta,\j]_{\mS,0} = 
\frac{\Gm{\tfrac{\Delta+\j}{2}+4}^2}{\Gamma(\Delta+3)\Gamma(\j+5)}
\frac{{}_3F_2\big(\{\tfrac{\Delta-\j-2}{2},\tfrac{\Delta-\j-2}{2},\tfrac{\Delta-\j-\mS}{2}+2\},
\{\tfrac{\Delta-\j-\mS}{2}+3,\Delta+3\},1\big)}{\mS-(\Delta{-}\j{+}4)}. \label{qvec 3F2}
\ee
The terms with $q>0$ admit similar expressions, but a more efficient strategy is to compute them recursively, using combinations
that cancel out the $m$-dependent denominator:
\be\begin{aligned}
\widehat{\cP}^{N=4}[\Delta,\j]_{\mS,q+1}&=(q+\tfrac{\Delta-\j+4-\mS}{2}) \widehat{\cP}^{N=4}[\Delta,\j]_{\mS,q} +
\frac{(\tfrac{\Delta-\j-2}{2}\big)_q^2}{2(-\j-4)_q}\,.
%\left\{\sum_{m=0}^\infty \frac{K_{\Delta,\j}^{m,\{\Delta_i\}}}{2\tilde{K}_{\Delta,\j}^{N=4}}(-m)_q=\frac{(\tfrac{\Delta-\j-2}{2}\big)_q^2}{2(-\j-4)_q}\right\}\,,
\end{aligned}\ee
The inhomogeneous term equals the sum of $K_{\Delta,\j}^{m,\{\Delta_i\}}(-m)_q/(2\tilde{K}_{\Delta,\j}^{N=4})$,
computed using \eqref{msum}.
Once the vector \eqref{Ps vector} is populated, eqs.~\eqref{Qcal_rep} and \eqref{PR Mellin explicit app} readily give the following formula
for Polyakov-Regge blocks:
\be \label{P Mellin fast}
 \widehat{\cP}^{N=4}_{\mS,\mT}[\Delta,\j] = \tilde{K}_{\Delta,\j}^{N=4}
 \sum_{q,k=0}^\j
\left(\widehat{\cP}_{\mS}^{N=4}[\Delta,\j]_{q}+\widehat{\cP}^{N=4}_{\mT}[\Delta,\j]_{q}\right)
\left[Q^{0,0}_{\Delta+4,\j}\right]_{q,k} \big(\tfrac{\mS+\mT-8}{2}\big)_k\,.
\ee
Again this can be rapidly calculated to very high accuracy.

The above expression is also well-suited for performing the Mellin transform to obtain the position-space blocks $\cP_{u,v}$.
For the term with $\cP_\mS$, the Mellin integral over $\mT$ can be done analytically and gives (with $\mU=16-\mS-\mT$):
\be\begin{aligned}
\Gm{\tfrac{\mS}{2}}^{-2}
\int\frac{d\mT}{4\pi i} v^{\frac{\mT}{2}-4}
  \Gamma\!\left(4-\tfrac{\mT}{2}\right)^2\Gamma\!\left(\tfrac{\mS+\mT-8}{2}\right)^2 \big(\tfrac{\mS+\mT-8}{2}\big)_k &= \frac{\Gm{k+\tfrac{\mS}{2}}^2}{\Gamma(k+\mS)}
  {}_2F_1\big(\tfrac{\mS}{2},\tfrac{\mS}{2}+k,\mS+k,1-v\big) \\ &\equiv [\cP_{\mS,v}]_k\,. \label{Pv vec}
\end{aligned}\ee
A rapid way to compute this vector is described below.
%In practice, hypergeometric recursion relations can be used to rapidly construct the $[\cP_{\mS,v}]_k$ vector with $k=0\ldots \j$ in terms of two entries:
%\be
%0 = (k+\tfrac{\mS}{2})^2[\cP_{\mS,v}]_k +((k+1)v-(2k+1+\mS))[\cP_{\mS,v}]_{k+1} +(1-v)[\cP_{\mS,v}]_{k+2}.
%\ee
The outcome is $\cP_{\Delta,\j}(u,v)$ written as a single integral over a matrix product:\footnote{
The factorization of the $q$ and $v$ dependence of the integrand is related to the fact
that we study unsubtracted dispersion relation.}
\be \label{cP from Mellin}
\frac{\cP_{\Delta,\j}(u,v)}{\tilde{K}^{N=4}_{\Delta,\j}}=\int \frac{d\mS}{4\pi i}
\Gamma\big(\tfrac{\mS}{2}\big)^2\Gamma\big(4-\tfrac{\mS}{2}\big)^2 \sum_{q,k=0}^\j
\widehat{\cP}_{\mS}^{N=4}[\Delta,\j]_{q} \left[Q^{0,0}_{\Delta+4,\j}\right]_{q,k}\!
\left(\!u^{\frac{\mS}{2}-4}[\cP_{\mS,v}]_k+v^{\frac{\mS}{2}-4}[\cP_{\mS,u}]_k\!\right)\!.
\ee
There is a single integral left to perform numerically, over $\mS$, in contrast with the two-dimensional $\rho$ integral in the position-space approach.
However, the integrand is a fairly complicated function of $\mS$, especially at large spin $\j$.

Fortunately, it turns out that the $\mS$ integral in \eqref{cP from Mellin} is ``nice'' for numerics.
We use an exponential parametrization $\mS=5+i\sinh(x)$ and simply approximate the integral by a Riemann sum,
sampling $x$ at discrete values uniformly spaced in a range such as $[-6,6]$. Since the integrand decays doubly exponentially with $x$,
it is easy to ascertain that the contribution from $x$ outside the range is smaller than say $10^{-200}$.
Furthermore, since the function is smooth, the Euler-Maclaurin theorem predicts that discretization errors decay nonpertubatively with the spacing $\Delta x$.
We observe empirically that the error decays as $e^{-\#/\Delta x}$ with $\#\sim 4$ for a wide range of spins, twists, and cross-ratios.

Convergence is extremely fast.  The Riemann sum using just 300 sampling points is typically accurate to 50 digits.
Thus, our method for evaluating $\cP_{\Delta,\j}(u,v)$ boils down to evaluating the
vectors $\widehat{\cP}_{\mS}^{N=4}[\Delta,\j]_{q}$ and $[\cP_{\mS,u/v}]_k$ on a few hundreds values of $\mS$, and dotting into the Mack coefficient matrix $\left[Q^{0,0}_{\Delta+4,\j}\right]_{q,k}$.
The result is highly accurate and stable under changes in the parametrization or in the real part of $\mS$.

This method agrees precisely with the position space integrals described above.
Again, it is instructive to compare the results (and performance on one of the authors' laptop)
of the two methods:
\be \begin{aligned} \label{X numerics}
X_{2/3,5/4}[150,154+\tfrac{1}{19}] &= 1.947607331\ldots{\times} 10^{78} \hfill\mbox{(position, 30s)}\\
%X_{2/3,5/4}[150,154+\tfrac{1}{19}] &= 1.947607348863473773996\ldots\times 10^{78}  \hfill \mbox{(position, 130s)}\\
&= 1.9476073488634737739526678531298896\ldots{\times} 10^{78}  \hfill \mbox{(position, 1800s)}\\
&= 1.9476073\ldots{\times} 10^{78}  \hfill \mbox{(Mellin, 700s)}\\
&= 1.9476073488634737739526678531298896365\ldots{\times}10^{78}  \hfill
\mbox{(Mellin, 1200s)}.
%X_{2/3,5/4}[150,154+\tfrac{1}{19}] &= 1.94760734886347377395266785312988963652567826516\ldots\times 10^{78}  \hfill \mbox{(Mellin, 1500s)}.
\end{aligned}\ee
Again, the precise agreement gives us high confidence in the validity of our codes.
Generally, our position space implementation tends to be faster for getting a small number of figures,
but its cost increases rapidly with the requested accuracy.
On the other hand,  the Mellin method requires some effort to get any significant figure at all (due to strong numerical cancellations in the matrix product), but it scales much better with requested accuracy.
The above timings reflects a naive implementation of the ${}_3F_2$ function \eqref{qvec 3F2}, which is the most expensive step in the calculation; faster timings are achieved using the optimizations below.

A significant advantage of the Mellin approach is that the same ingredients can be recycled for many functionals.
This makes the average evaluation time per functional much smaller than the above numbers suggest.
Most of our intensive runs were computed using this method.

%The bottleneck is the ${}_3F_2$ function that enters the $m$-sum in \eqref{},
%so computing an arbitrary number $\cP_{u,v}[\Delta,\j]$'s take about the same resources
%as computing $\widehat{\cP}_{\mS,\mT}[\Delta,\j]$ for one hundred values of $(\mS,\mT)$.

\subsection{Some algorithmic improvements}

For the reasons just mentioned, we invested significant effort to optimize Mellin-based formulas.
The most expensive ingredient in the preceding subsection is to perform the sum over descendants
in \eqref{Ps vector} for $q=0$, which we expand here for convenience:
\be
\widehat{\cP}^{N=4}_{\mS}[\Delta,\j]_{0} \equiv
\frac{\Gamma\big(\tfrac{\Delta+\j}{2}+4\big)^2}{\Gamma(\Delta+3)\Gamma(\j+5)}
\sum_{m=0}^\infty \frac{\big(\tfrac{\Delta-\j-2}{2}\big)_m^2}{(\Delta+3)_m m!}\ \frac{1}{\mS-(\Delta{-}\j{+}2m{+}4)}\,.
\label{Ps 2}
\ee
We focus here on the sum which appears in $\mathcal{N}=4$ sYM, but we expect
similar techniques to work for the more general dispersive sum rules discussed in \cite{Trinh:2021mll}.
A relevant fact is that for each $(\Delta,\j)$, we need the above sum for several hundred values of $\mS$.
Instead of using \texttt{Mathematica}'s \texttt{HypergeometricPFQ}, 
we find it advantageous to compute the sum by combining exact evaluation of
the summand at small $m$ with its asymptotic series at large $m$:
\be
 \widehat{\cP}^{N=4}_{\mS}[\Delta,\j]_{0}  = \sum_{m=0}^{m_{\rm max}}  \frac{X_m}{\mS-(\Delta{-}\j{+}2m{+}4)}
 + \sum_{m_{\rm max}+1}^\infty \left[\sum_{n=0}^{n_{\rm max}} \frac{Y_n}{m^{\j+6+n}} \frac{1}{\mS-(\Delta{-}\j{+}2m{+}4)}\right]\,,
\ee
where $X_m$ collects all other factors in \eqref{Ps 2} and the $Y_n$ parametrize its $1/m$ expansion.
In the last term, the $m$-sum can be rapidly computed as a series in $1/m_{\rm max}$ up to order $n_{\rm max}$.
For calculations aiming for $O(100)$ digits, we typically choose $n_{\rm max}\sim 100$ or more, and
increase it and $m_{\rm max}$ until the error in the formula becomes smaller than the requested precision.
Errors are readily estimated using the observation that the $X_m$ sum up to exactly 1 (see \eqref{msum}).

The crucial point is that the expensive ingredients in this formula, $X_m$ and $Y_n$, only need to be evaluated
once for each operator $(\Delta,\j)$: the $\mS$-dependent factor is very simple.
The sum can thus be computed for multiple $\mS$ values for essentially the price of one,
easily reducing the timings quoted in \eqref{X numerics} by two orders of magnitude.

The second expensive ingredient is the vector of ${}_2F_1$ functions in \eqref{Pv vec}.
We evaluate it by replacing it by linear combinations that involve $\rho$-like variables
and computing those recursively. Specifically, define the two-vector:
\be \label{Pv prime}
\cP_{\mS,v}^\prime \equiv \frac{\Gamma\big(\tfrac{\mS}{2}\big)^2}{\Gamma(\mS)}(1+\rho_v)^{\mS}
\begin{pmatrix}
{}_2F_1\big(\tfrac12,\tfrac{\mS}{2},\tfrac{\mS+1}{2},\rho_v^2\big) \\
{}_2F_1\big(\tfrac32,\tfrac{\mS}{2},\tfrac{\mS+3}{2},\rho_v^2\big) \end{pmatrix}, \qquad
\rho_v\equiv \frac{1-\sqrt{v}}{1+\sqrt{v}}.
\ee
The desired integral \eqref{Pv vec} can be expressed in terms of those using
\begin{align}
&\Gm{\tfrac{\mS}{2}}^{-2}
\int\frac{d\mT}{4\pi i} v^{\frac{\mT}{2}-4}
 \Gamma\!\left(4-\tfrac{\mT}{2}\right)^2\Gamma\!\left(\tfrac{\mS+\mT-8}{2}\right)^2
\begin{pmatrix}
\big(\tfrac{\mS+\mT-8}{2}\big)_r^2 \\
\big(\tfrac{\mS+\mT-8}{2}\big)_r\big(\tfrac{\mS+\mT-8}{2}\big)_{r+1}
\end{pmatrix}
\\ &\hspace{20mm}=
\big(\tfrac{\mS}{2}\big)_r^2
\begin{pmatrix}
1&0\\
\tfrac{1+\rho_v}{4}(\mS+2 r){-}\tfrac{\rho_v}{4} & \tfrac{\rho_v}{4}
\end{pmatrix} \cdot \cP_{\mS+2r,v}^\prime  \label{Pv from Pv'}\,.
\end{align}
More precisely, the integral $[\cP_{\mS,v}]_k$ is equal to a sum of the left-hand-side
of the preceding equation with $0\leq r\leq \lfloor\tfrac{k}{2}\rfloor$ and integer coefficients.
%\Gm{\tfrac{\mS}{2}}^{-2}
%\int\frac{d\mT}{4\pi i} v^{\frac{\mT}{2}-4}
%  \left(4-\tfrac{\mT}{2}\right)^2\Gamma\!\left(\tfrac{\mS+\mT-8}{2}\right)^2 \big(\tfrac{\mS+\mT-8}{2}\big)_k
The two-vectors $\cP_{\mS+2r,v}^\prime$ which appear on the right-hand-side
can be populated recursively in terms of the one with the largest $r$, using
\be
\cP_{\mS,v}^\prime = \frac{4}{\mS (1+\rho_v)^2}
\begin{pmatrix} \mS{+}1 & -\rho_v^2 \\ 1 & (\mS{-}1)\rho_v^2\end{pmatrix} \cdot  \cP_{\mS+2,v}^\prime\,.
\ee
(We avoid using the recursion in the opposite direction because it is not numerically stable for $v\approx1$.)
Even without the change of basis \eqref{Pv from Pv'},
hypergeometric relations would allow the vector $[\cP_{\mS,v}]_k$ to be populated
with only two hypergeometric evaluations, a significant speedup over $O(\j)$ evaluations.
The special combinations \eqref{Pv prime} further optimize the computation
of two seeds by making the hypergeometric argument numerically smaller (see \cite{Hogervorst:2013sma}).
%: $\rho_v^2<|1-v|$ for $v>0$.

\end{appendix}

\bibliographystyle{JHEP}
\bibliography{references}

\end{document}